\documentclass[letterpaper,11pt]{article}
\pdfoutput=1

\usepackage{jheppub}
\usepackage{multirow}
\usepackage{shuffle}

\usepackage{breqn}

\usepackage{subfig}
\usepackage{xspace}
\usepackage[countmax]{subfloat}
\usepackage{enumitem}
\usepackage{amssymb}
\usepackage{amsmath}
\usepackage{cancel}
\usepackage{color}
\usepackage{braket}
\usepackage{graphicx}
\usepackage{multirow}
\usepackage{verbatim}
\usepackage{amsthm}
\usepackage{slashed}
\usepackage{wasysym}
\usepackage{simplewick}
\usepackage{mathtools}
\usepackage{soul}
\usepackage{xspace}

\usepackage{mathrsfs}

\def\beq{\begin{equation}}
\def\eeq{\end{equation}}
\def\bea{\begin{eqnarray}}
\def\eea{\end{eqnarray}}

\def\Ord{{\cal O}}

\def\cB{\mathcal{B}}

\def\cM{\mathcal{M}}
\def\cN{\mathcal{N}}
\def\cO{\mathcal{O}}

\def\cP{\mathcal{P}}

\def\nn{{\nonumber}}

\newcommand{\Eq}[1]{Equation~\eqref{#1}}

\DeclareRobustCommand{\Sec}[1]{Sec.~\ref{#1}}
\DeclareRobustCommand{\Secs}[2]{Secs.~\ref{#1} and \ref{#2}}
\DeclareRobustCommand{\App}[1]{App.~\ref{#1}}

\DeclareRobustCommand{\Fig}[1]{Fig.~\ref{#1}}

\DeclareRobustCommand{\Eq}[1]{Eq.~(\ref{#1})}
\DeclareRobustCommand{\Eqs}[2]{Eqs.~(\ref{#1}) and (\ref{#2})}

\def\be{\begin{equation}}
\def\ee{\end{equation}}

\newcommand{\Sl}[1]{\slashed{#1}}

\allowdisplaybreaks[3]

\newcommand{\e}{\epsilon}

\newcommand{\momf}{\xi}
\newcommand{\psthree}{\Phi_c^{(3)}}

  \newcount\hour \newcount\minute
  \hour=\time \divide \hour by 60 \minute=\time
  \newcommand{\todaytime}{\today \ -- \number\hour :\ifnum \minute<10 0\fi\number\minute}

\title{Three Point Energy Correlators in the Collinear Limit:\\
Symmetries, Dualities and Analytic Results}

\author[1]{Hao Chen,}
\author[1]{Ming-Xing Luo,}
\author[2]{Ian Moult,}
\author[1]{Tong-Zhi Yang,}
\author[1]{Xiaoyuan Zhang,}
\author[1]{and Hua Xing Zhu}

\affiliation[1]{Zhejiang Institute of Modern Physics, Department of Physics, Zhejiang University, Hangzhou, Zhejiang 310027, China}
\affiliation[2]{SLAC National Accelerator Laboratory, Stanford University, CA, 94309, USA\vspace{0.5ex}}

\emailAdd{chenhao201224@zju.edu.cn}
\emailAdd{mingxingluo@zju.edu.cn}
\emailAdd{imoult@slac.stanford.edu}
\emailAdd{yangtz@zju.edu.cn}
\emailAdd{xyzhang0314@zju.edu.cn}
\emailAdd{zhuhx@zju.edu.cn}

\abstract{Energy Correlators measure the energy deposited in multiple detectors as a function of the angles between the detectors. In this paper, we analytically compute the three particle correlator in the collinear limit in QCD for quark and gluon jets, and also in $\cN=4$ super Yang-Mills theory. We find an intriguing duality between the integrals for the energy correlators and infrared finite Feynman parameter integrals, which maps the angles of the correlators to dual momentum variables. In $\cN=4$, we use this duality to express our result as a rational sum of simple Feynman integrals (triangles and boxes). In QCD our result is expressed as a sum of the same transcendental functions, but with more complicated rational functions of cross ratio variables as coefficients. Our results represent the first analytic calculation of a three-prong jet substructure observable of phenomenological relevance for the LHC, revealing unexplored simplicity in the energy flow of QCD jets. They also provide valuable data for improving the understanding of the light-ray operator product expansion. }

\begin{document} 

\maketitle

\section{Introduction} \label{sec:intro}

Observables which probe the flow of energy in quantum field theories are interesting both from a theoretical perspective for studying the Lorentzian limits of field theories, as well as from a practical perspective, where they can be used as event shape or jet substructure observables to measure properties of QCD, and search for new physics with jets (see e.g. \cite{Larkoski:2017jix,Marzani:2019hun} for a review).

From the theoretical perspective, one of the simplest observables is the Energy-Energy Correlator (EEC), defined as \cite{Basham:1978bw,Basham:1978zq}
\begin{align}
  \label{eq:EECdef}
  \frac{d\sigma}{dz}= \sum_{i,j}\int d\sigma\ \frac{E_i E_j}{Q^2} \delta\left(z - \frac{1 - \cos\chi_{ij}}{2}\right) \,.
\end{align}
Here $E_i$ and $E_j$ are the energies of final-state partons $i$ and $j$ in the center-of-mass frame, $\chi_{ij}$ is their angular separation, and $d\sigma$ is the product of the squared matrix element and the phase-space measure.
The EEC can also be defined in terms of correlation functions of ANEC operators \cite{Sveshnikov:1995vi,Korchemsky:1999kt,Lee:2006nr,Hofman:2008ar,Belitsky:2013xxa,Belitsky:2013bja,Belitsky:2013ofa}
\begin{align}
\mathcal{E}(\vec n) =\int\limits_0^\infty dt \lim_{r\to \infty} r^2 n^i T_{0i}(t,r \vec n)\,,
\end{align}
where the limit is taken with the retarded time $u = t - r$ fixed, and we integrate over the advanced time $v = t + r$. In terms of these ANEC operators, the EEC is given by the four-point Wightman correlator
\begin{align}
\frac{1}{\sigma_{\rm tot}} \frac{d\sigma}{dz}=\frac{ \int \! d^4 x \, e^{i q \cdot x} \langle \cO (x) \mathcal{E}(\vec n_1)  \mathcal{E}(\vec n_2) \cO^\dagger (0) \rangle }{\int \! d^4 x \, e^{i q \cdot x} \langle \cO (x)  \cO^\dagger (0) \rangle}\,,
\end{align}
for some source operator $\cO$ that produces the localized excitation.
This provides a connection between event shape observables and correlation functions of ANEC operators, allowing the study of event shapes to profit from recent developments in the study of ANEC operators. Conversely, the EEC provide a concrete situation for studying the behavior of ANEC operators using jets at the LHC.

\begin{figure}
\begin{center}
\subfloat[]{
\includegraphics[scale=0.45]{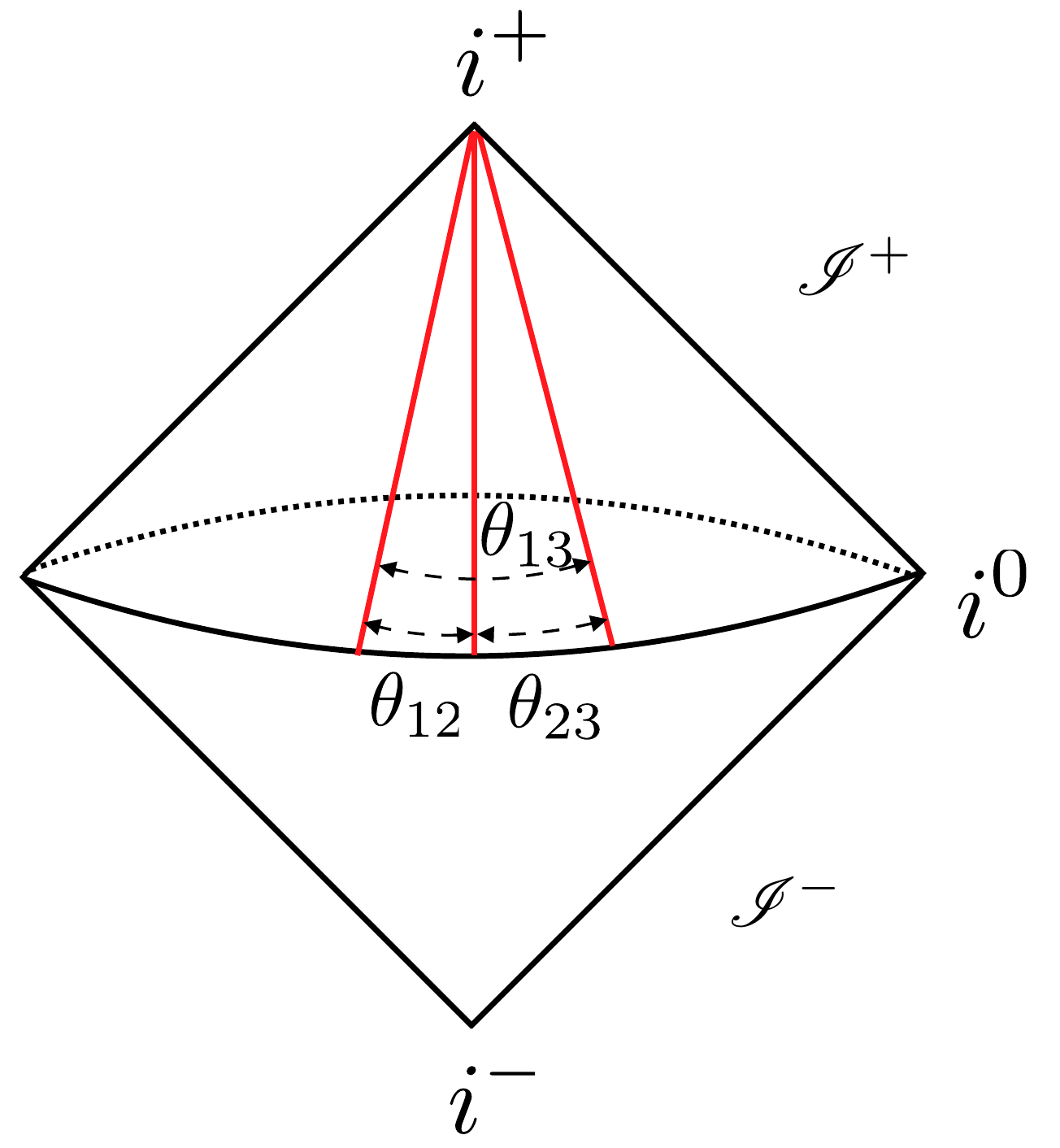}\label{fig:jet_EEC_a}
}\qquad
\subfloat[]{
\includegraphics[scale=0.45]{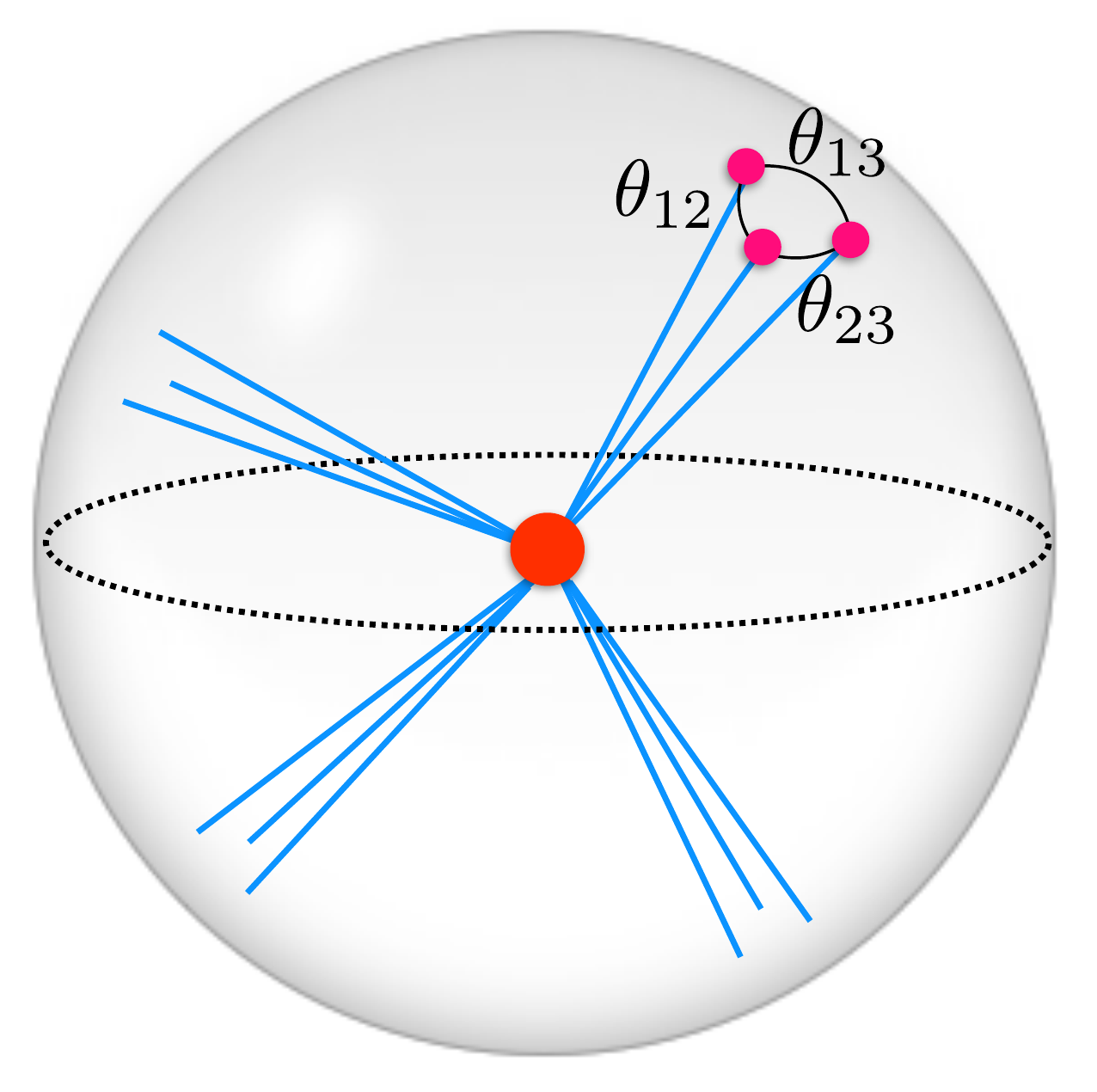}\label{fig:jet_EEC_b}
}\qquad
\end{center}
\caption{(a) Penrose diagram for the EEEC showing the three energy flow operators. One dimension of the $S_2$ has necessarily been suppressed. (b) The EEEC as measured on a jet in a four jet event. The pink dots denote energy flow measurement operators.}
\label{fig:jet_EEC}
\end{figure}

There has recently been interesting progress in the understanding of the EEC from a number of different directions.  The EEC has been computed for generic angles at next-to-leading order (NLO) in QCD \cite{Dixon:2018qgp,Luo:2019nig}, and up to NNLO in $\cN=4$ SYM \cite{Belitsky:2013ofa,Henn:2019gkr}. It has also been computed numerically in QCD at NNLO \cite{DelDuca:2016csb,DelDuca:2016ily}. There has also been progress in understanding the singularities of the EEC, which occur as $z\to 0$ (the collinear limit) and $z\to 1$ (the back-to-back limit). In the back-to-back limit, the EEC exhibits Sudakov double logarithms, whose all orders logarithmic structure at leading power is described by a factorization formula \cite{Moult:2018jzp,Gao:2019ojf}, and whose subleading power logarithms have been resummed in $\cN=4$ SYM \cite{Moult:2019vou} using recent developments in understanding the structure of subleading power corrections for this class of observable \cite{Ebert:2018gsn}.  In the $z\to 0$ limit, which will be studied in this paper, the EEC exhibits single collinear logarithms, originally studied at leading logarithmic order in  \cite{Konishi:1978yx,Konishi:1978ax,Konishi:1979cb,Kalinowski:1980wea,Richards:1982te}. Formulas describing the behavior of the EEC in the collinear limit were recently derived in \cite{Dixon:2019uzg} for a (not necessarily conformal) gauge theory, and in \cite{Kravchuk:2018htv,Kologlu:2019bco,Kologlu:2019mfz,Korchemsky:2019nzm} for a generic conformal field theory (CFT). The $z\to 0$ limit is of theoretical interest for studying the operator product expansion (OPE) in Lorentzian signature, and is of phenomenological interest as a jet substructure observable for the LHC.

The two-point correlator is particularly simple since it depends on a single variable, $z$. Indeed, in a conformal field theory (CFT), its behavior in the collinear limit is fixed to be a power law \cite{Dixon:2019uzg,Kologlu:2019mfz,Korchemsky:2019nzm}
\begin{equation}
\Sigma(z) = \frac{1}{2} \, C(\alpha_s) \, z^{\gamma_1^{{\cal N}=4}(\alpha_s)} \,,
\label{Neq4NNLLcumulant}
\end{equation}
where $C(\alpha_s)$ is a constant independent of $z$, $\gamma_1^{{\cal N}=4}(\alpha_s)$ is the universal anomalous dimension of local twist-2 spin-3 operator, and we use $\Sigma(z)$ to denote the cumulant~\cite{Dixon:2019uzg}. In a non-conformal theory, such as QCD, one can derive a \emph{time-like} factorization formula~\cite{Dixon:2019uzg}
\begin{align}
\label{eq:fact}
 \Sigma(z, \ln \frac{Q^2}{\mu^2} , \mu)
= \int_0^1 dx\, x^2 \vec{J} (\ln\frac{z x^2 Q^2}{\mu^2},\mu)
   \cdot  \vec{H} (x,\frac{Q^2}{\mu^2},\mu) \,,
\end{align}
where $\vec{H}$ is the coefficient functions for semi-inclusive fragmentation, and $\vec{J}$ is the jet function for the EEC.
One finds that instead of involving a single fixed moment, the jet function involves with the timelike splitting kernel
\begin{align}
  \label{eq:jetRG}
\frac{d \vec{J}(\ln\frac{z Q^2}{\mu^2}, \mu) }{d \ln \mu^2} = \int_0^1 dy\, y^2 \vec{J} (\ln\frac{z y^2 Q^2}{\mu^2}, \mu) \cdot \widehat P_T(y,\mu) \,.
\end{align}
Therefore, in both conformal and non-conformal theories, the dependence on the angle $z$ is completely fixed by symmetry or renormalization group arguments, as expected for a two-point observable.

In this paper, we study the three point energy correlation function, which we will refer to as the triple correlator, or EEEC. The EEEC is defined as
\begin{equation}
\frac{1}{\sigma_{\rm tot}} \frac{d^3\Sigma}{dx_1 dx_2dx_3}=	\sum_{i,j,k}\int \frac{E_iE_jE_k}{Q^3}d\sigma \delta \left(  x_1- \frac{1-\cos \theta_{ij}}{2} \right)   \delta \left(  x_2- \frac{1-\cos \theta_{jk}}{2} \right)    \delta \left(  x_3- \frac{1-\cos \theta_{kl}}{2} \right)\,,
\end{equation}
where $i$,$j$ and $k$ run over all the final-state massless partons. We will focus on the collinear limit of the EEEC, where the angles between detectors are small, $x_1\,, x_2\,, x_3 \ll 1$, but we will not assume that there is any hierarchy between the angles, so that the result will have a dependence on the ratio of angles that is not (yet) predicted by symmetries or renormalization group arguments. The complete calculation of the triple correlator for generic angles is also interesting, and will be considered in future work.

There are a number of motivations for extending the understanding of the energy correlators beyond the simplest case of the two point correlator. First, the three point correlator is the first time that a non-trivial dependence on the angles can occur, and so the result provides a more non-trivial probe of the structure of the OPE in Lorentzian signature. Secondly, the triple correlator directly probes multi-particle correlations within a jet, which is extremely interesting from the perspective of jet substructure. Observables capturing multi-particle ($>2$) correlations have not been analytically computed before, and we believe that this provides a significant advance for jet substructure that will have phenomenological applications at the LHC.

In this paper we directly compute the triple correlators in QCD for both quark and gluon jets, and also in $\cN=4$ super Yang-Mills (SYM) theory. We show that the result can be compactly written in terms of polylogarithmic functions of a complex variable that encodes the two conformal cross ratios, and we highlight a number of interesting features of the results.  We also find an intriguing duality between the integrals associated in computing the EEEC in the collinear limit, and Feynman parameter integrals for loop amplitudes. This duality maps energies to Feynman parameters, and the angles of the energy correlators to dual coordinates. Due to the well established techniques for the integration of Feynman parameter integrals, this allows the result to be computed relatively easily, and we believe that it will also enable calculations of higher point correlators.

For the case of quark jets in QCD, we are able to compare our analytic result with a numerical calculation using the full $e^+e^-\to 4$ parton matrix elements using \textsc{Nlojet++} \cite{Nagy:2001fj,Nagy:2003tz} and \textsc{Event2} \cite{Catani:1996vz}. We find good agreement for all partonic channels in the singular limit, providing strong evidence both for the correctness of our results, and for the factorization properties of the EEEC in the collinear limit.

An outline of this paper is as follows. In \Sec{sec:calculation}, we describe our calculation of the EEEC using timelike factorization and the triple collinear splitting functions. In \Sec{sec:sym} we discuss the symmetries of the observable, the parametrization of the result, and the functions that will appear. In \Sec{sec:feyn_parameter} we illustrate an intriguing duality between the integrals encountered in our calculation for the energy correlator, and Feynman parameter integrals for loop amplitudes. In \Sec{sec:results} we present results for the EEEC in $\cN=4$, and for both quark and gluon jets  in QCD, and discuss their structure. In \Sec{sec:numerics}, we compare our analytic predictions with numerical calculations using the full $e^+e^-\to 4$ parton matrix elements. We conclude in \Sec{sec:conclusions} and discuss many future directions for further understanding the energy correlators in the collinear limit, and for phenomenological applications.

\section{Factorization in the Collinear Limit} \label{sec:calculation}

\begin{figure}
\begin{center}
\includegraphics[scale=0.2]{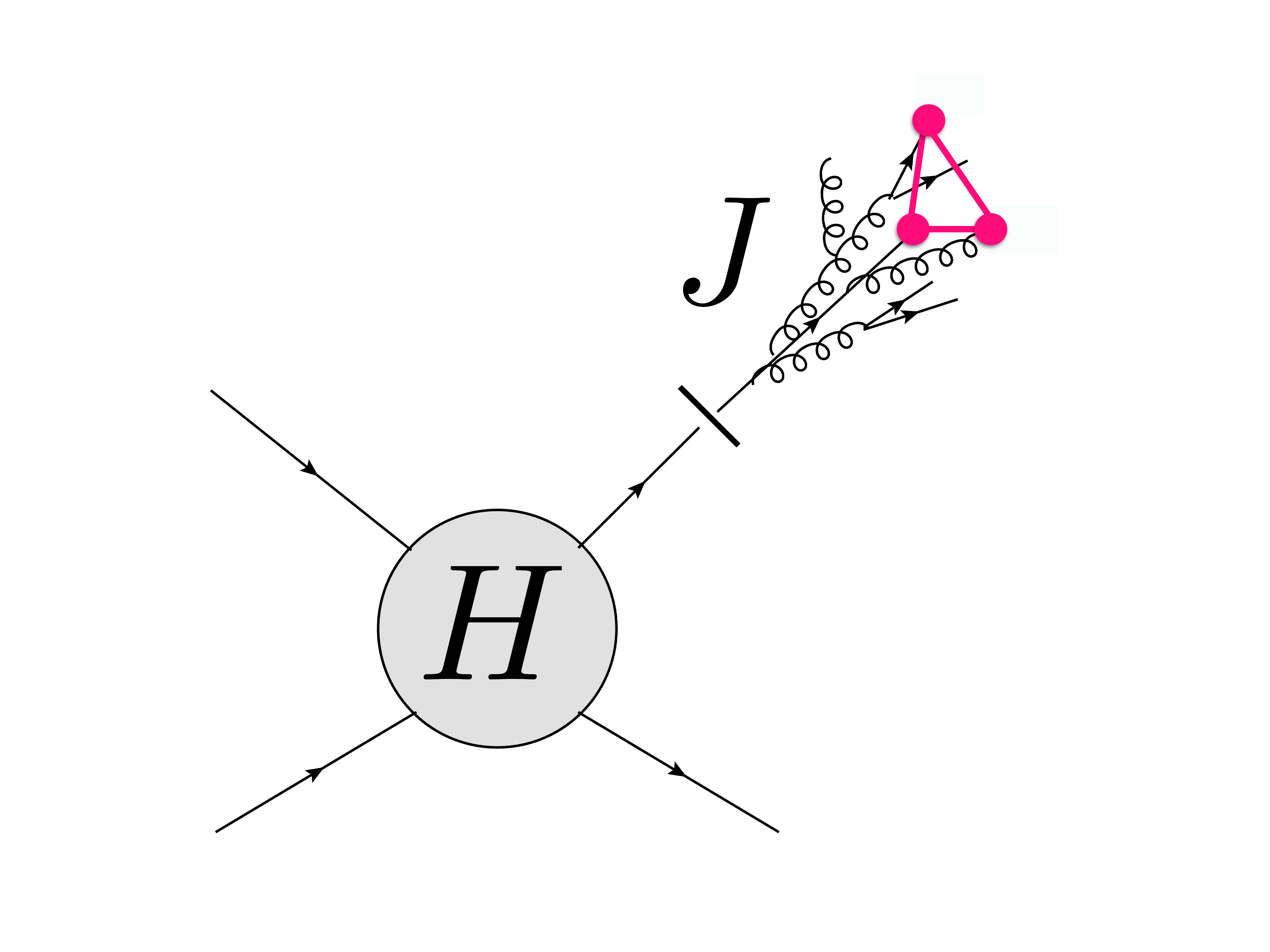}
\end{center}
\caption{A schematic of the factorization for the triple correlation function in the collinear limit. In the collinear limit, we can factorize on a single partonic state produced by the hard function, $H$. The measurement is then imposed on the jet function, which can be computed perturbatively for different partonic channels.}
\label{fig:factorization}
\end{figure}

To compute the behavior of the EEEC in the collinear limit, we will use timelike collinear factorization, and extend the factorization formula presented in \cite{Dixon:2019uzg} for the EEC. Following \cite{Dixon:2019uzg}, we factorize the EEEC in the collinear limit into a hard function, $H$, and a jet function, $J$, both of which are vectors in flavor space. This factorization is shown schematically in \Fig{fig:factorization}. The hard function depends on the source, but is independent of the measurement, and is therefore identical for the EEEC and EEC (and more generally for an arbitrary number of correlators). The jet function, $J$, describes the dependence on the EEEC measurement, and will be the focus of this paper. Corrections to this formula are suppressed by powers of the angles, and are described by higher twist jet functions.  In this paper we will only study the fixed order properties of the jet function, leaving a study of the all orders resummation structure of the factorization formula to a future publication.

The jet function for the EEEC for a quark (or antiquark) jet is defined as
\begin{align}\label{eq:jet_func_quark}
J_q(x_1,x_2,x_3,Q,\mu^2)&=\\
&\int \frac{dl^+}{2\pi}\frac{1}{2N_C} \text{Tr} \int d^4x e^{i l\cdot x} \langle 0 | \frac{\Sl{\bar n}}{2} \chi_n(x) \widehat\cM_{\text{EEEC}} ~ \delta (Q+\bar n \cdot \cP) \delta^2(\cP_\perp) \bar \chi_n(0) |0\rangle\,, \nn
\end{align}
and for a gluon jet, as
\begin{align}\label{eq:jet_func_gluon}
J_g(x_1,x_2,x_3,Q,\mu^2)&=\\
&\hspace{-1cm}\int \frac{dl^+}{2\pi}\frac{1}{2 (N^2_C-1)} \text{Tr} \int d^4x e^{i l\cdot x} \langle 0 |  \cB^{a,\mu}_{n,\perp}(x) \widehat\cM_{\text{EEEC}} ~ \delta (Q+\bar n \cdot \cP) \delta^2(\cP_\perp) \cB^{a,\mu}_{n,\perp}(0) |0\rangle\,.\nn
\end{align}
Here $\chi_n$ and $\cB^{a,\mu}_{n,\perp}$ are gauge invariant collinear quark and gluon fields in SCET \cite{Bauer:2000ew, Bauer:2000yr, Bauer:2001ct, Bauer:2001yt}. The operator $\widehat\cM_{\text{EEEC}}$ implements the EEEC measurement in the collinear limit,
\begin{align}
\widehat\cM_{\text{EEEC}}=\sum_{i,j,k} \frac{E_iE_jE_k}{Q^3} \delta \left(  x_1- \frac{\theta^2_{ij}}{4} \right)   \delta \left(  x_2-  \frac{\theta^2_{jk}}{4} \right)    \delta \left(  x_3- \frac{\theta^2_{kl}}{4} \right)\,.
\end{align}
Technically speaking, it can be written as an operator in terms of the stress energy tensor \cite{Sveshnikov:1995vi,Korchemsky:1997sy,Korchemsky:1999kt,Belitsky:2001ij,Hofman:2008ar,Belitsky:2013bja}, however, here we will only need its expression in momentum space. For the EEEC, there must be three partons in the jet to have a non-trivial functional dependence (beyond contact terms).

To compute the jet function of \Eqs{eq:jet_func_quark}{eq:jet_func_gluon}, we use its equivalence to a calculation of the full QCD splitting functions with the EEEC measurement function inserted. For an extensive discussion of the equivalence between jet functions and results integrated against collinear splitting functions, see \cite{Ritzmann:2014mka}. We must  therefore simply integrate the triple collinear splitting functions  against the measurement function for the EEEC, over the triple collinear phase space. 

The triple collinear splitting functions $P_{ijk}$ were computed in \cite{Campbell:1997hg,Catani:1998nv}, and are summarized in \App{app:A} for convenience. The triple collinear phase space expressed in terms of the Mandelstam invariants $s_{ij}$, $s_{123}$ and the momentum fractions $\momf_i = 2 E_i/Q$ of the partons is given by \cite{GehrmannDeRidder:1997gf}
\begin{align}
d\psthree &=ds_{123}ds_{12}ds_{13}ds_{23} \delta(s_{123}-s_{12}-s_{13}-s_{23}) d\momf_1 d\momf_2 d\momf_3 \delta(1-\momf_1-\momf_2-\momf_3) \nn \\
&\times \frac{4 \Theta(-\Delta)  (-\Delta)^{-1/2-\epsilon}}{(4\pi)^{5-2\epsilon}\Gamma(1-2\epsilon)}\,,
\end{align}
where
\begin{align}
\Delta=(\momf_3 s_{12}-\momf_1 s_{23}-\momf_2 s_{13})^2-4 \momf_1 \momf_2 s_{13}s_{23}\,.
\end{align}
The angles between the partons are related to the Mandelstam variables and energy fractions used in the phase space parametrization as
\begin{align}
s_{ij}=2 E_i E_j (1-\cos \theta_{ij}) \stackrel{\theta_{ij} \to 0}{=} \momf_i \momf_j Q^2 \frac{\theta_{ij}^2}{4}\,,
\end{align}
where in the second equality we use the collinear approximation.
We can therefore write the result for the EEEC jet function at the lowest non-trivial order as
\begin{align}
J_{\widehat{ijk}}=\int d\psthree \left(\frac{\mu^2 e^{\gamma_E}}{4\pi}   \right)^{2\epsilon}   \frac{4 g^4}{s_{123}^2}   \sum_{i,j,k} P_{ijk} \widehat{\cM}_{\text{EEEC}}\,.
\end{align}
This expression is also true for a generic measurement function, $\cM$,  but in general it is hard to perform the integral analytically. This is due both to the constraints of the measurement function, and from the $\Theta(-\Delta)$ constraint appearing in the collinear phase space. However, here we will find that a remarkable simplification occurs. First, we can change variables to express the integration over the Mandelstam variables in terms of integrals over the $x_i$, using
\begin{align}
ds_{12} ds_{13}ds_{23}=(\momf_1 \momf_2 \momf_3)^2 Q^6 dx_1 dx_2 dx_3\,,
\end{align}
and
\begin{align}
\Delta=(\momf_1 \momf_2 \momf_3)^2  (x_1^2 +x_2^2+x_3^2 -2x_1 x_2 -2x_1 x_3 -2x_2 x_3)\,.
\end{align}
Now, we note that in the collinear limit, we have that
\begin{align}
x_1=\frac{\theta_{23}^2}{4}\,, \qquad x_2 =\frac{\theta_{13}^2}{4}\,, \qquad x_3 =\frac{\theta_{12}^2}{4}\,,
\end{align}
and $\sqrt{x_1}$, $\sqrt{x_2}$, $\sqrt{x_3}$ are subjected to the constraint that they must form the sides of a triangle.  Using Heron's formula, which expresses the area of a triangle, $A$, in terms of the lengths of its sides, we find that
\begin{align}
A^2=\frac{-\Delta}{(\momf_1 \momf_2 \momf_3)^2} \geq 0\,.
\end{align} 
This implies that the measurement function automatically guarantees that the $\Theta(-\Delta)$ constraint is satisfied. This enables the integrals over $s_{123}$, and the energy fractions $\momf_i$, to be performed analytically, while the integrals over $x_i$ are fixed by the measurement function. This can be compared with other examples, for example planar flow considered in \cite{Field:2012rw}, where analytic integration is extremely difficult due to the $\Theta(-\Delta)$ constraint.

In \Sec{sec:feyn_parameter} we will show that the particular structure of the measurement function for the energy correlators which completely fixes the angle, but integrates over the energies allows the phase space integrals appearing in its calculation to be mapped to Feynman parameter integrals for loop amplitudes. This particular factorization of the measurement function into energies and angles is non-standard in jet substructure, but we find that it is extremely convenient and leads to many of the nice properties of the result.

\section{Parametrization, Symmetries, Functions and Constraints} \label{sec:sym}

In this section we discuss our parametrization of the EEEC and its associated symmetries, and present arguments for the classes of functions that will appear. In \Sec{sec:feyn_parameter} we give another way of understanding the functions that appear in the result by illustrating a relationship between the integrals for the energy correlators in the collinear limit and Feynman parameter integrals for loop amplitudes. 

\subsection{Symmetries on the Celestial Sphere and Parametrization}

To study the structure of the energy correlators, it is convenient to switch from a parametrization in terms of energies and angles, to coordinates on the celestial sphere. We therefore parametrize the momentum of particles as
\begin{align}
k^\mu_i = \omega_i \frac{Q}{2} (1+|z_i|^2, z_i+\bar z_i, -i (z_i-\bar z_i), 1-|z_i|^2 )\,.
\end{align}
Here $\omega_i = \momf_i/(1 + |z_i|^2)$ is a rescaled energy fraction, and the complex coordinate $z_i$ are coordinates on the $S_2$ celestial sphere. In terms of these coordinates, the standard Mandelstam invariants are
\begin{align}
s_{ij}= Q^2 \omega_i \omega_j |z_i-z_j|^2 \equiv Q^2 \omega_i \omega_j |z_{ij}|^2\,.
\end{align}
Note that in terms of standard angles, we have
\begin{align}
|z_{ij}|^2=\frac{1-\cos \theta_{ij}}{2} \simeq \frac{\theta^2_{ij}}{4}=x_k\,,
\end{align}
where the final equality holds in the small angle limit. 
The interesting feature of the energy correlator observables is that the measurement function fixes the coordinates on the sphere, but integrates freely (up to momentum conservation) over the energies. The result is therefore a function of the complex variables $z_i$, and it is therefore interesting to begin by studying the symmetries acting on the $z_i$ variables.

Lorentz transformations in four dimensional Minkowski space act as the global conformal group, SL$(2,\mathbb{C})$, on the $z_i$ coordinates on the celestial sphere \cite{Terrell:1959zz,Penrose:1959vz,Penrose:1987uia}. Explicitly, 
\begin{align}\label{eq:transforms}
z_i \to z_i'=\frac{a z_i+b}{cz_i+d}\,, \qquad
\omega_i \to \omega_i' =|cz_i+d|^2 \omega_i\,.
\end{align}
For a pedagogical review, see e.g. \cite{Oblak:2015qia}. However, for energy correlators at generic angles, the Lorentz symmetry is broken by the timelike vector $Q^\mu$ of the source from SO$(3,1)\to$ SO$(3)$. For SO$(3)$, this leaves three generators, which act on the coordinates on the celestial sphere as translations, $z\to z+b\,, b\in \mathbb{C}$ and rotations, $z\to e^{i\theta}z, \theta \in \mathbb{R}$ (see also the discussion in \cite{Kologlu:2019mfz}).

Due to collinear factorization, an additional symmetry is restored at small angles, which corresponds to boosts along the jet direction. Physically this arises since the jet function in \Eqs{eq:jet_func_quark}{eq:jet_func_gluon} depends only on the source through the momentum conserving delta function for the momentum along the jet direction $\delta[\bar n \cdot Q- Q (\omega_1+\omega_2+\omega_3) (1+|\tilde z|^2)]$. Here we have used that $\tilde z \simeq z_1 \simeq z_2 \simeq z_3$ at leading power in the collinear limit, where $\tilde z$ denote the jet direction on $S^2$. Without loss of generality, in the rest of this paper we use the $SO(3)$ symmetry to set the jet direction to $\tilde z = 0$, so that $\omega_i = \xi_i$ at leading power. 
If we apply a general SL$(2,\mathbb{C})$ transform to this, then we see from \Eq{eq:transforms} that if $c=0$, then $\omega_i \to |d|^2 \omega_i$, which can then be pulled out of the delta function.\footnote{We also use that $Q^\mu = (Q, 0,0,0)$ can be decomposed as $Q^\mu = n^\mu (\bar{n} \cdot Q)/2 + \bar{n}^\mu (n \cdot Q)/2$, where $n^\mu (\bar{n} \cdot Q)/2$ transforms as a null momenta.} This boost, which acts as $\omega_i \to |d|^2 \omega_i$, rescales the coordinates on the sphere as $z_i\to z_i/d$, and therefore acts as a dilatation. Therefore, in the collinear limit, in terms of the coordinates on the sphere, we have the following symmetry group that will act on the energy correlators
\begin{align}
z&\to z+b\,, \qquad b\in \mathbb{C} \,,\nn \\
z&\to e^{i\theta}z\,, \qquad \theta \in \mathbb{R} \,, \nn \\
z& \to e^{-\chi} z\,, \qquad \chi \in \mathbb{R} \,.
\end{align}
These symmetries can be interpreted in two ways. First, this is the conformal group of $\mathbb{C} \cup \{\infty\}$, with the point at infinity fixed. This has a fairly natural interpretation with the point at infinity being the source for the jet function in the collinear limit. The only missing generator for the SL$(2,\mathbb{C})$ transformations on the sphere that is broken in the collinear limit is inversions. We can therefore interpret the EEEC as a three point function with these symmetries. Unlike when inversions are included, in this case the three point function is no longer completely fixed and is a non-trivial function of scale invariant cross ratios.

\begin{figure}[t]
\begin{center}
\includegraphics[scale=0.35]{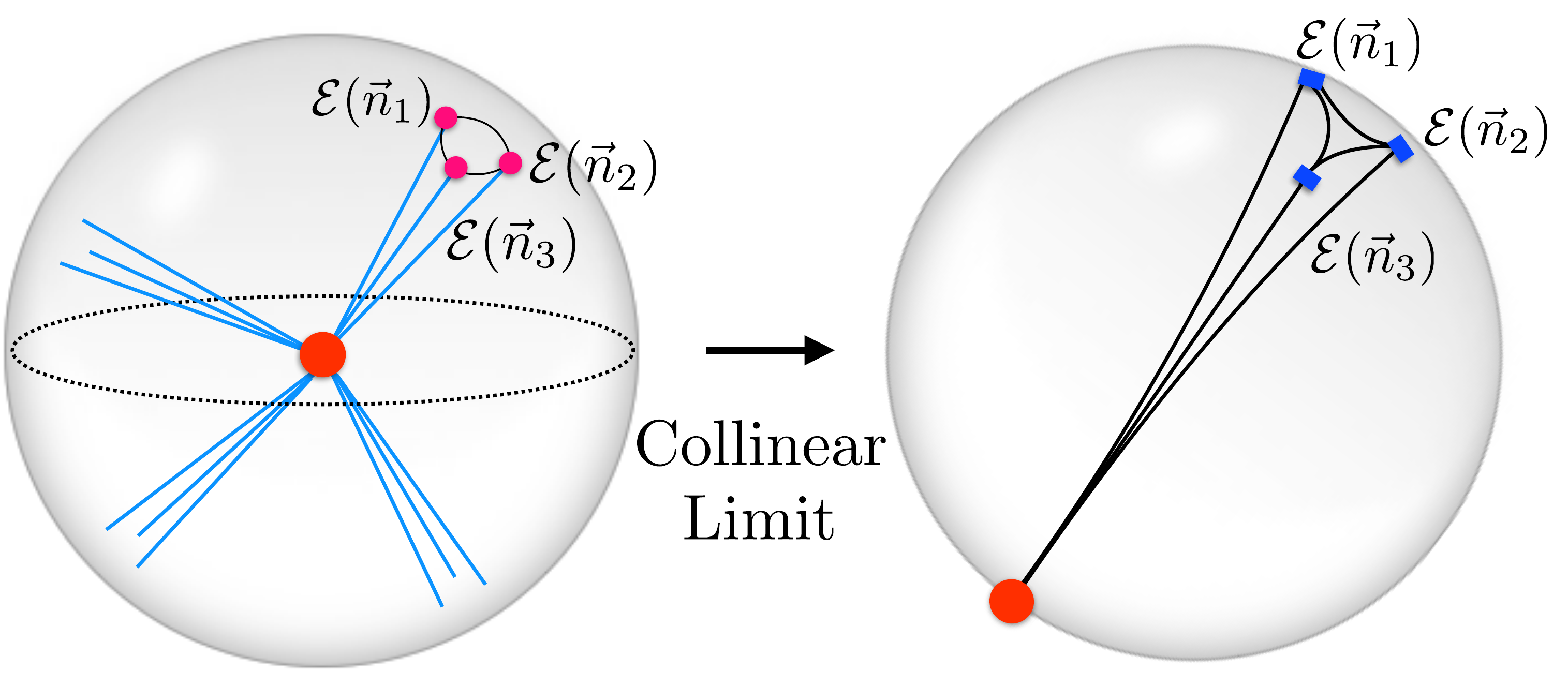}
\end{center}
\caption{In the collinear limit, the EEEC factorizes onto a massless state, as shown in \Fig{fig:factorization}. The source therefore moves to the $S_2$ celestial sphere, and the EEEC exhibits the symmetries of a four point function on $S_2$.}
\label{fig:holographic}
\end{figure}

Alternatively, and perhaps more promising from the speculative perspective of trying to develop a field theory on the celestial sphere describing the energy correlators, we can view the EEEC as a four point correlation function. For generic angles, the source momentum $Q^\mu$ satisfies $Q^2>0$, and therefore lives as a point in the bulk of $\mathbb{H}_3$.\footnote{See e.g. \cite{Almelid:2017qju} for a discussion of the parametrization of a massive particle as a point in $\mathbb{H}_3$, as well as a discussion of the massless limit.} When one factorizes in the small angle limit,  as in \Fig{fig:factorization}, one expands to leading power in the limit that the invariant mass of the intermediate state $q^2\to 0$, and therefore the source for the jet also moves to the $\partial \mathbb{H}_3=S_2$ boundary where the other massless final state particles live. This is shown schematically in \Fig{fig:holographic}. We can therefore interpret the EEEC as a four point correlator on the $S_2$ where we have already put one point at infinity (or at the very least, it possesses the symmetries of such a four point correlator).

\begin{figure}[t]
\begin{center}
\subfloat[]{
\includegraphics[scale=0.30]{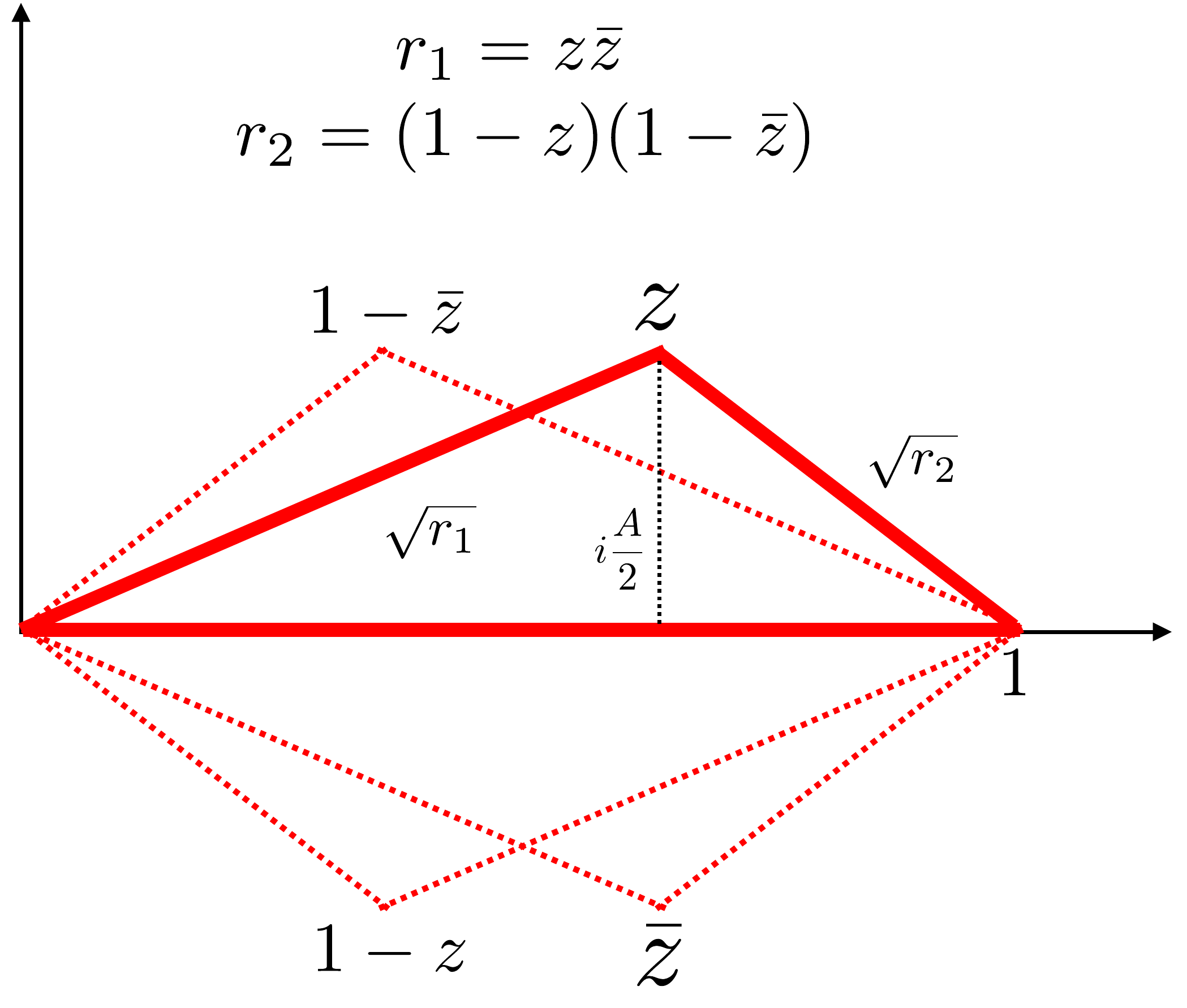}\label{fig:param_a}
}\qquad
\subfloat[]{
\includegraphics[scale=0.30]{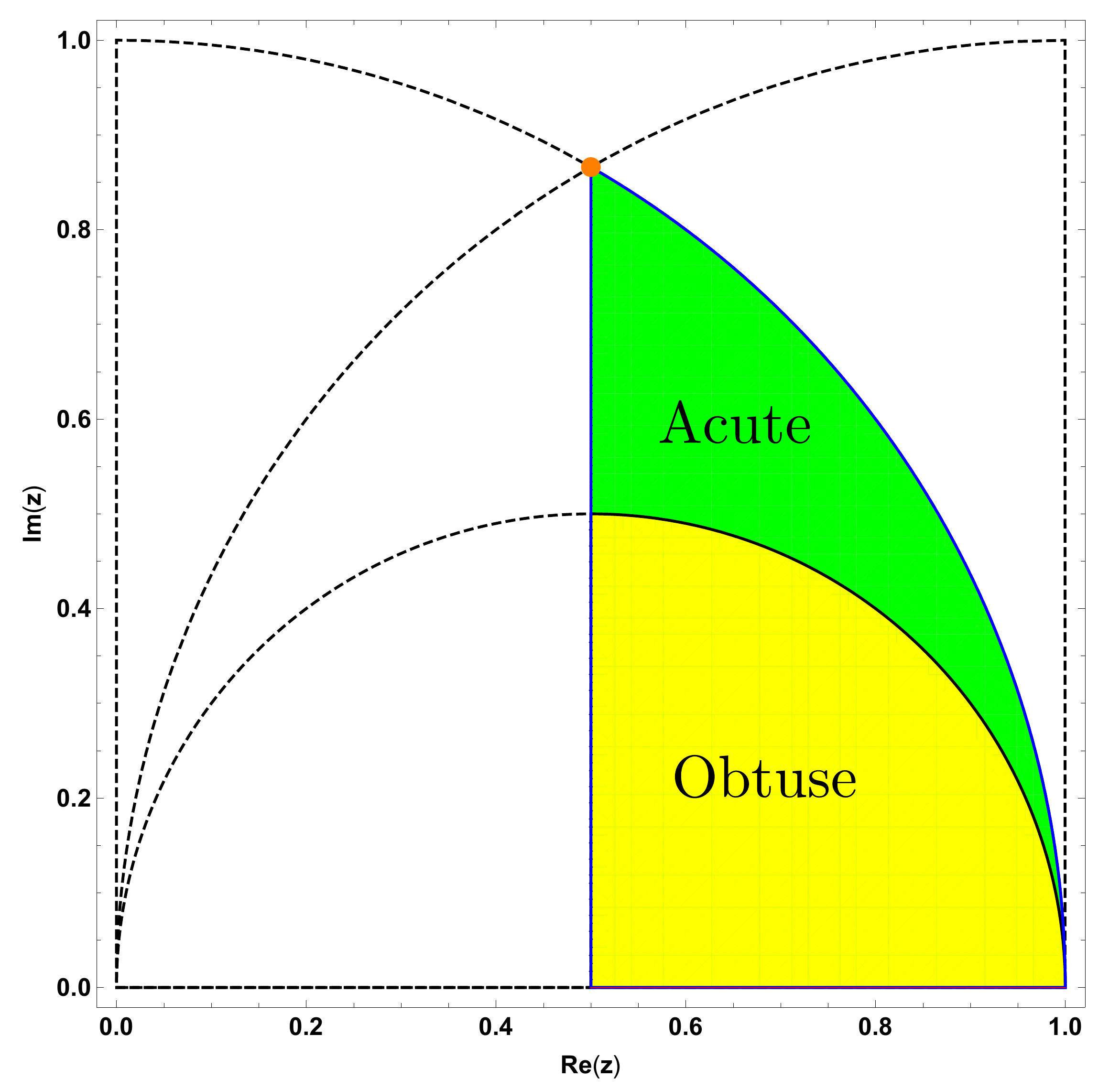}\label{fig:param_b}
}\qquad
\end{center}
\caption{(a) Parametrization of unit triangle in terms of a complex variable $z$. (b) Enforcing that the unit side is the longest side, and removing triangles that are identical by parity, restricts the triangle to lie in the shaded region. Regions corresponding to obtuse and acute triangles are shown in yellow and green respectively.}
\label{fig:param}
\end{figure}

Therefore, interpreting the EEEC as a four point correlator on $S_2$, we use the standard parametrization of a conformal four point correlator, where we put the ANEC operators at $\{0,1,z\}$, and the fourth point at $\infty$. This is shown in \Fig{fig:param_a}. Algebraically, we  then have
\begin{equation}
	\begin{aligned}
	r_1=\frac{x_1}{x_3} = z \bar{z}&,\quad
	r_2=\frac{x_2}{x_3}  = (1-z)(1-\bar{z})\,.
	\end{aligned}
\end{equation}
A technically convenient property of the $z$ parametrization is that it rationalizes the area of the triangle, which we will see appears as an argument of the transcendental functions in the result. The area of the rescaled triangle can be written as 
\begin{align}\label{eq:def_A}
i A=(z-\bar z)=i \sqrt{-r_1^2-(-1+r_2)^2+2r_1(1+r_2)}\,.
\end{align}
Note that for a triangle, the expression in the square root is guaranteed to be positive by Heron's formula.

Finally, there is also a discrete $S_3 \times \mathbb{Z}_2$ symmetry that acts on the result. The $\mathbb{Z}_2$ acts as complex conjugation and is equivalent to parity. The final result is invariant under the $\mathbb{Z}_2$,\footnote{Here we assume that we are factorizing on unpolarized states. We thank Jesse Thaler for discussions on this point.} but individual transcendental functions can be classified as being either even, or odd under this symmetry. The $S_3$ acts as a permutation symmetry on the three identical energy detectors. In the $z$ variables, the $S_3$ acts as 
\begin{align}\label{eq:S3_action}
S_3 : z\to 1-z\,, \qquad z \to 1-\frac{1}{z}\,, \qquad z\to \frac{1}{z}\,, \qquad z\to \frac{z}{1-z}\,, \qquad z \to \frac{1}{1-z}\,.
\end{align}
This is simply the anharmonic group, which is the stabilizer of $\{0,1,\infty\}$. It is also the standard action of $S_3$ on the arguments of harmonic polylogarithms, which will be a fact that will be useful later when understanding the function space.

To restrict to a region in the complex plane that uniquely specifies a triangle without redundancies, we must restrict to the region shown in \Fig{fig:param_a}. This region can be determined by enforcing that the sides of the triangles are ordered, $x_3>x_2>x_1$. This is the region that is relevant for e.g. an experimental analysis. However, when expressing the result as a function of $z$, it is useful to view the function as defined globally in the complex plane, using the $S_3$ action.

With this set of coordinates in mind, we can now perform a change of variables to write the cross section as a function of $x_L = x_3$, and $z$, $\bar z$. We find\footnote{Note that the Jacobian appearing here is important, and cancels a spurious singularity in the collapsed limit of the triangle where $z=\bar z$. This spurious singularity occurs in the $x_1$, $x_2$, $x_3$ variables because they degenerate in this limit, but does not occur in the $x_L$, $z$, $\bar z$ variables. }
\beq
\begin{split}
\frac{1}{\sigma_{\rm tot}} \frac{d^3\Sigma}{dx_L \, d\text{Re}(z)\, d\text{Im}z} &= 2 x_L \sqrt{2 x_1 x_2+ 2 x_2 x_3 +2 x_3 x_1-x_1^2-x_2^2-x_3^2} \frac{1}{\sigma_{\rm tot}} \frac{d\Sigma}{dx_1 dx_2 dx_3}\\
 &=\frac{g^4}{16 \pi^5} \frac{1}{x_L} G(z)\,.
\end{split}
\eeq
The goal is therefore to determine the non-trivial function $G(z)$.

\subsection{Transcendental Functions}\label{sec:transcendental}

The function $G(z)$ is obtained by integrating a rational function of $z$ (namely the tree level splitting functions), and can only have singularities at $\{0,1,\infty\}$, which correspond to the squeezed (or OPE) limits (see \Secs{eq:squeeze_short}{sec:squeeze-limit-three}). This places strong constraints on the possible functions that can appear. In particular, it is a known mathematical result that such functions are harmonic polylogarithms \cite{2006math......6419B,Almelid:2017qju} multiplied by rational functions of $z$. Furthermore, from the structure of the integrals the transcendentality of the functions appearing in the result is at most two. 

Due to the action of the anharmonic group, the result is defined on the entire plane, and it must be single valued.  There are two ways that this can occur. The first way is if the functions dependend on $|z|^2$ or $|1-z|^2$.  The second is if the polylogarithms are single valued harmonic polylogarithms (SVHPLs) \cite{Brown:2004ugm}. These functions are well studied, and have appeared elsewhere in the physics literature, for example in the study of the Regge limit \cite{Dixon:2012yy}, in anomalous dimensions of cusped Wilson lines \cite{Almelid:2017qju}, or in the multi-loop soft gluon emission amplitude~\cite{Dixon:2019lnw}. In particular, it is well known that at weight 2, there is a unique SVHPL, the Bloch-Wigner dilogarithm, which is defined as
\beq
2 i D_2^{-}(z)=\text{Li}_2(z)-\text{Li}_2\left(\bar z\right)+\frac{1}{2}
   \left(\log (1-z)-\log \left(1-\bar z\right)\right) \log \left(z \bar z\right)\,.
\eeq
This function has an odd parity under the $\mathbb{Z}_2$ symmetry, and also has a very simple action under the $S_3$ action discussed before, namely~\cite{Zagier:2007}
\beq
D_2^-(z)=D^-_2\left(1-\frac{1}{z}\right)=D_2^-\left(\frac{1}{1-z}\right)=-D_2^-\left(\frac{1}{z}\right)=-D_2^-(1-z)=-D_2^-\left(\frac{-z}{1-z}\right)\,.
\eeq
Another useful identity relating to the Bloch-Wigner dilogarithm is that it can be written in terms of a unit argument
\begin{align}
D_2^-(z) =\frac{1}{2} \left[  D_2^-\left( \frac{z}{\bar z} \right)+  D_2^-\left( \frac{1-1/z}{1-1/\bar z} \right)  +D_2^-\left( \frac{1/(1-z)}{1/(1-\bar z)} \right)  \right]\,.
\end{align}
In our particular case, the arguments are interpreted as the angles of the triangle. This shows that one will find polylogarithms of roots of unity when the result is evaluated for particular triangles. For example, for an equilateral triangle, one will find that the result is expressed in terms of polylogarithms of the third root of unity. 

It is also well known that $D_2^-(z)$ is the area of the ideal tetrahedron in hyperbolic three space, $\mathbb{H}_3$, with points at $\{0,1,z,\infty\}$. The celestial sphere is the boundary of $\mathbb{H}_3$, and therefore this can be interpreted as the volume of the tetrahedron formed by the three points of the energy correlators, and the point at infinity which originated the jet in the collinear limit, as shown in \Fig{fig:holographic}. The appearance of $D_2^-(z)$ is therefore natural since the SL$(2,\mathbb{C})$ symmetry on the $S_2$ acts as diffeomorphisms in the bulk $\mathbb{H}_3$. It would be interesting to understand this point better, and if a similar interpretation extends to higher points.

We will additionally need transcendental functions of $|z|^2$ or $|1-z|^2$. Our result is not of uniform transcendentality, and therefore we must describe functions of both weight $1$ and $2$. At weight 1, we have simply the logarithms $\log|z|^2$, $\log|1-z|^2$. These functions are even under $\mathbb{Z}_2$. We will also need weight $2$ even functions.  It is convenient to introduce the following particular variant of the dilogarithm (and its $S_3$ permutations)\footnote{This function is remarkably similar to the Roger's dilogarithm, and it would be interesting to better understand its appearance from symmetries.}
\begin{align}
D^+_2(z)=\left(\text{Li}_2\left(1-|z|^2\right)+\frac{1}{2} \log \left(|1-z|^2\right) \log \left(|z|^2\right)\right)\,,
\end{align}
where the logarithms are fixed by demanding that the result is a pure functions of either $|z|^2$ and $|1-z|^2$ that has no double logs as $z\to 0,1,\infty$. This is a property of the result at this order, since there are no logarithms appearing in the OPE limit. In both QCD and $\cN=4$ we will find that $D^+_2(z)$ is sufficient to describe the even weight 2 part of our result, and that the functions $\{ \log^2 \left(|1-z|^2\right) , \log^2 \left(|z|^2\right) \}$ do not appear independently.

We therefore expect to find the following independent functions in our result. 
\begin{align}
1\,,\quad \zeta_2\,,\quad \log(|z|^2)\,, \quad D_2^+(z)\,,\quad D_2^-(z)\,.
\end{align}
These functions will also appear in the permutations under the $S_3$ action, and will be multiplied by rational functions of $z$, $\bar z$.

Here we have argued for these functions based purely on the form of the functions being integrated, and using known mathematical results.  In \Sec{sec:feyn_parameter}, we will provide another way of arriving at these same functions by showing explicitly that all integrals appearing in the calculation of the $\cN=4$ EEEC can be directly mapped to one-loop Feynman integrals for amplitudes. Here the side lengths, $|z|^2$ and $|1-z|^2$ will be associated with Mandelstam invariants. Since the symbol~\cite{Goncharov:2010jf} of amplitudes obey a first entry condition \cite{Gaiotto:2011dt}, namely that the first entry is a Mandelstam invariant, this guarantees that the first entry of our result is either $|z|^2$ or $|1-z|^2$. This is true for both $D_2^+(z)$, $D_2^-(z)$.

\subsection{Squeezed Limits and Behavior at Infinity}\label{eq:squeeze_short}

A particularly strong constraint on our result comes from the squeezed (OPE) limits, namely $z\to 0$ or $z\to 1$. By the $S_3$ symmetry, these OPE limits are also related to the limit $z\to \infty$. For the tree level three point correlator that we are computing, we have the following OPE limits
\begin{align}
G(z)\Big|_ {z\to 0}\simeq \frac{1}{|z|^2}\,, \qquad G(z)\Big|_ {z\to 1}\simeq\frac{1}{|1-z|^2}\,, \qquad G(z)\Big|_ {z\to \infty}\simeq \frac{1}{|z|^2}\,.
\end{align}
The coefficient of proportionality depends on the theory, and a more detailed discussion of the OPE limit, and the calculation of the relevant constants, will be given in \Sec{sec:squeeze-limit-three}.
In particular, this result implies that all weight $1$ and weight $2$ functions must cancel in these limits.
This power law scaling remains true beyond tree level, but it is modified by logarithms.  

The behavior at infinity also strongly constrains the form of the polynomials in the result.  To have higher order polynomials that describe power corrections to these limits, one must also have higher order denominators, and intricate cancellations are required to respect the required constraints.  Similar cancellations were also seen in the case of the two point energy correlator in \cite{Dixon:2018qgp,Luo:2019nig}.

\subsection{Collapsed Triangles}

Finally, we find that another limit which plays an important role in understanding the EEC is the limit $z\to \bar z$, namely where the triangle collapses to a line.\footnote{In the study of cosmological correlators, this limit is referred to as the flattened, or collapsed triangle (see e.g. \cite{Arkani-Hamed:2015bza}). There the correlator is also finite in this limit for an adiabatic vacuum. } This limit is non-singular for a generic value of $\Re(z)$, i.e. away from the OPE limit. Interestingly, we find that the approach to this limit  largely controls the complexity of the answer, and is the primary difference between the QCD and $\cN=4$ results. 

To understand why the $z\to \bar z$ limit plays an important role, we note that one of the two weight two functions that appears in our result, $D_2^{-}(z)$, vanishes as $z\to \bar z$, since it is odd. Since the full result for the cross section is even, $D_2^{-}(z)$ must be multiplied by an odd power of $(z-\bar z)$. If this power is positive, then this function completely vanishes on the line $z=\bar z$. However, if the power is negative, then one must Taylor expand $D^{-}(z)$ in powers of  $(z-\bar z)$ allowing this function to contribute on the line $z=\bar z$.

In the $\cN=4$ case, we find that only the following functions appear
\begin{align}
\frac{D_2^{-}(z)}{(z-\bar z)^{3}}\,, \qquad \frac{D_2^{-}(z)}{(z-\bar z)}\,, \qquad (z-\bar z) D^{-}(z)\,. 
\end{align}
On the other hand, in QCD, we find powers up to $(z-\bar z)^{-11}$. This leads to a very intricate set of cancellations as $z\to \bar z$. Furthermore, as soon as one has these large denominators in one part of the result, they immediately appear in other parts of the result. For example, assume that the function $D_2^{-}(z)/(z-\bar z)^{3}$ appears in the answer. Taylor expanding the function as
\begin{align}
\frac{D_2^{-}(z)}{(z-\bar z)^{3}}\to   \frac{ \partial_{z-\bar z} D_2^{-} (z)|_{z=\bar z}}{ (z-\bar z)^2 }+ \frac{1}{3!}\partial^3_{z-\bar z} D_2^{-} (z)|_{z=\bar z} +\cdots\,,
\end{align}
we see that the first term, which is now a weight $1$ function must be cancelled as $z\to \bar z$ to have a well behaved answer. This requires an equal complexity for the polynomials of the weight $1$ terms. With 11 inverse powers, this generates an extremely complicated pattern of cancellations. It would be interesting to understand if one can predict before hand the powers of $(z-\bar z)$ that appear and to understand what generates them. In \Sec{sec:feyn_parameter} we give some progress towards this for the case of $\cN=4$, where we are able to write the result as a rational sum of Feynman parameter integrals.

\section{Relation to Feynman Parameter Integrals} \label{sec:feyn_parameter}

In this section we discuss an interesting relation between the integrals that are encountered when computing energy correlators in the collinear limit, and standard one loop integrals for scattering amplitudes. We will show that, at least for the three point correlator, we can interpret the energy integrals as Feynman parameter integrals, and the angles between the energy correlators as dual coordinates of Feynman graphs. This relationship is particularly interesting due to the fact that techniques for the calculation of one-loop integrals are highly developed, and their structure is much better understood.

The integrals that arise when considering the EEEC in the collinear limit take the form\footnote{We note that in QCD it is conventional to pull out a factor of $1/(s_{123})^2$ from the splitting functions. Due to the simpler structure of the splitting functions in $\cN=4$, this is no longer convenient there. Here we have written it without the factor of  $1/(s_{123})^2$ pulled out.}
\begin{equation}
\frac{1}{\sigma_{\rm tot}}\frac{d^3\Sigma}{dx_1dx_2dx_3}=\mathcal{N}\int d\omega_1d\omega_2d\omega_3\delta (1-\omega_1-\omega_2-\omega_3) \frac{(\omega_1 \omega_2 \omega_3)^2}{16}\times P_{1\to 3}\,,
\end{equation}
where we have set the jet direction to $\tilde z = 0$ such that $\omega_i = \xi_i$ is the parton momentum fraction, and
\begin{equation}
	\mathcal{N} = \frac{g^4}{32\pi^5\sqrt{-\Delta}}\,.
\end{equation}
Here we will work in the coordinates on the celestial sphere, where
\begin{align}
s_{ij}=Q^2 \omega_i \omega_j |z_i-z_j|^2\,.
\end{align}
To illustrate the correspondence with Feynman parameter integrals, we choose first a particularly simple term in the $\cN=4$ splitting function (the complete splitting functions can be found in the Appendix)
\begin{align}
P_{1\to 3} \supset \frac{1}{\omega_1 \omega_3 s_{12} s_{123} } \sim \frac{1}{\omega_1^2 \omega_2 \omega_3 |z_{12}|^2 s_{123} }\,.
\end{align}
Writing
\begin{align}
s_{123}=Q^2 (\omega_1 \omega_2 z_{12}^2+\omega_1 \omega_3 z_{13}^2+\omega_2 \omega_3 z_{23}^2)\,,
\end{align}
we then find that the integral we must consider in the collinear limit is
\begin{equation}
	F_{1}=\mathcal{N} \frac{1}{2 |z_{12}|^2}\times\int d\omega_1d\omega_2d\omega_3 \delta(1-\omega_1-\omega_2-\omega_3) \frac{\omega_2 \omega_3}{\omega_1 \omega_2 z_{12}^2+\omega_1 \omega_3 z_{13}^2+\omega_2 \omega_3 z_{23}^2}\,.
\end{equation}
We immediately see that this resembles a Feynman parameter integral for the three mass triangle.  To make this relation precise, we consider the Feynman parametrization of the integral
\begin{equation}
\begin{aligned}
	\mathcal{J}^{(d)}(\nu_{1},\nu_{2},\nu_{3})
	&=\int \frac{d^d l}{i \pi^{\frac{d}{2}}}\frac{1}{(l^2)^{\nu_1}((l+p_1)^2)^{\nu_2}((l+p_1+p_2)^2)^{\nu_3}}\\
	&=(-1)^{\frac{d}{2}} \frac{\Gamma(\sum_i\nu_i-\frac{d}{2})}{\prod_{i=1}^{3}\Gamma(\nu_i)}\int d\omega_1 d\omega_2 d\omega_3 \delta(1-\omega_1-\omega_2-\omega_3)\\
	& \times \omega_2^{\nu_1-1} \omega_3^{\nu_2-1} \omega_1^{\nu_3-1}(\omega_1\omega_2(p_1+p_2)^2+\omega_1\omega_3p_2^2+\omega_2\omega_3p_1^2)^{ \frac{d}{2}-\sum_{i}\nu_i}\,,
\end{aligned}
\end{equation}
We are then able to identify
\begin{equation}
	F_{1}=\mathcal{N} \frac{1}{2 |z_{12}|^2}\times \mathcal{J}^{(d=8)}(2,2,1)\,,
\end{equation}
where we have the association
\begin{equation}
	p_1^2\rightarrow |z_{23}|^2 = x_1\,,\qquad p_2^2\rightarrow |z_{13}|^2 = x_2\,,\qquad (p_1+p_2)^2\rightarrow |z_{12}|^2 = x_3\,.
\end{equation}

\begin{figure}
\begin{center}
\subfloat[]{
\includegraphics[width=0.3\textwidth]{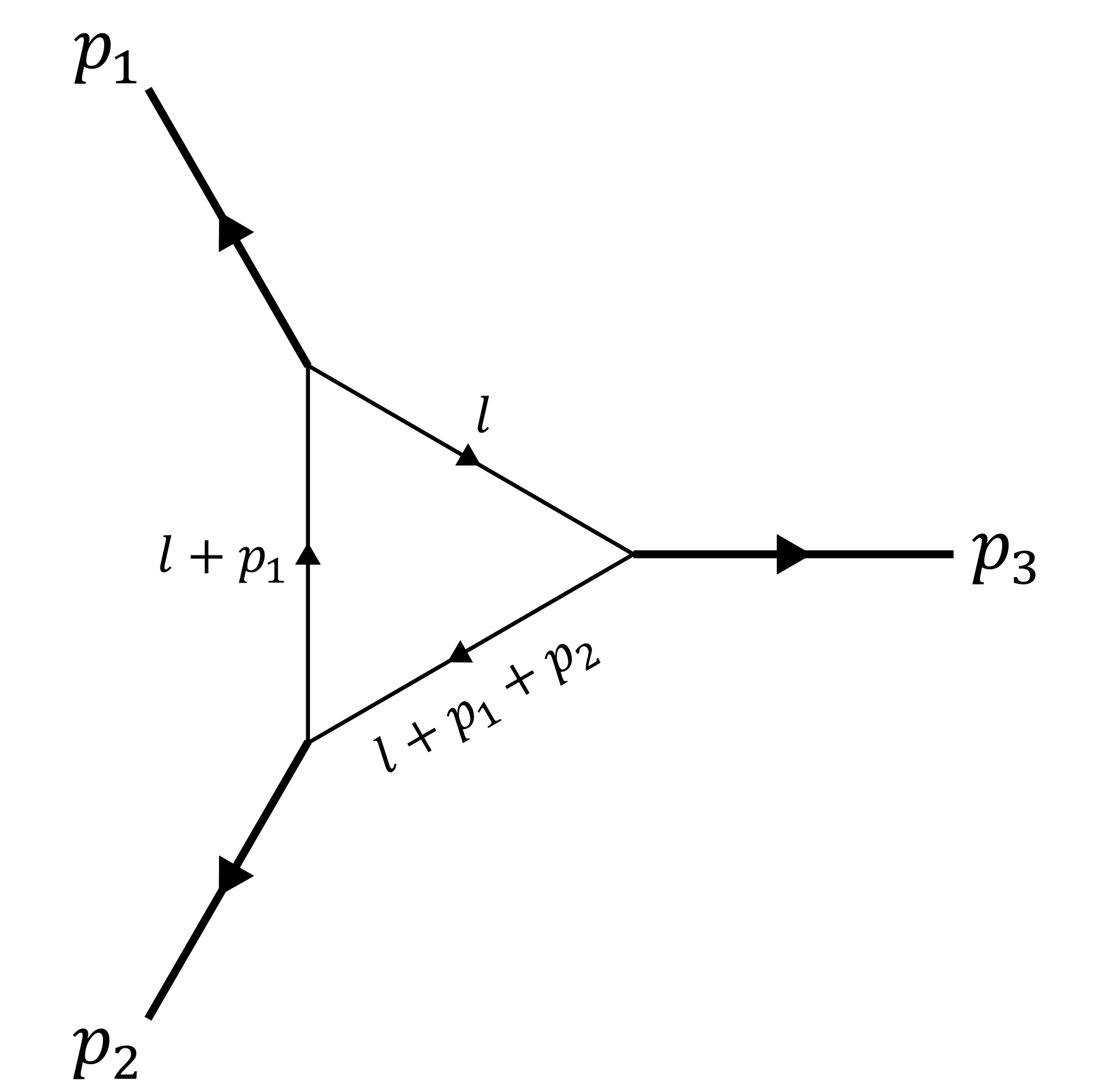}\label{fig:dual_a}
}\qquad
\subfloat[]{
\includegraphics[width=0.5\textwidth]{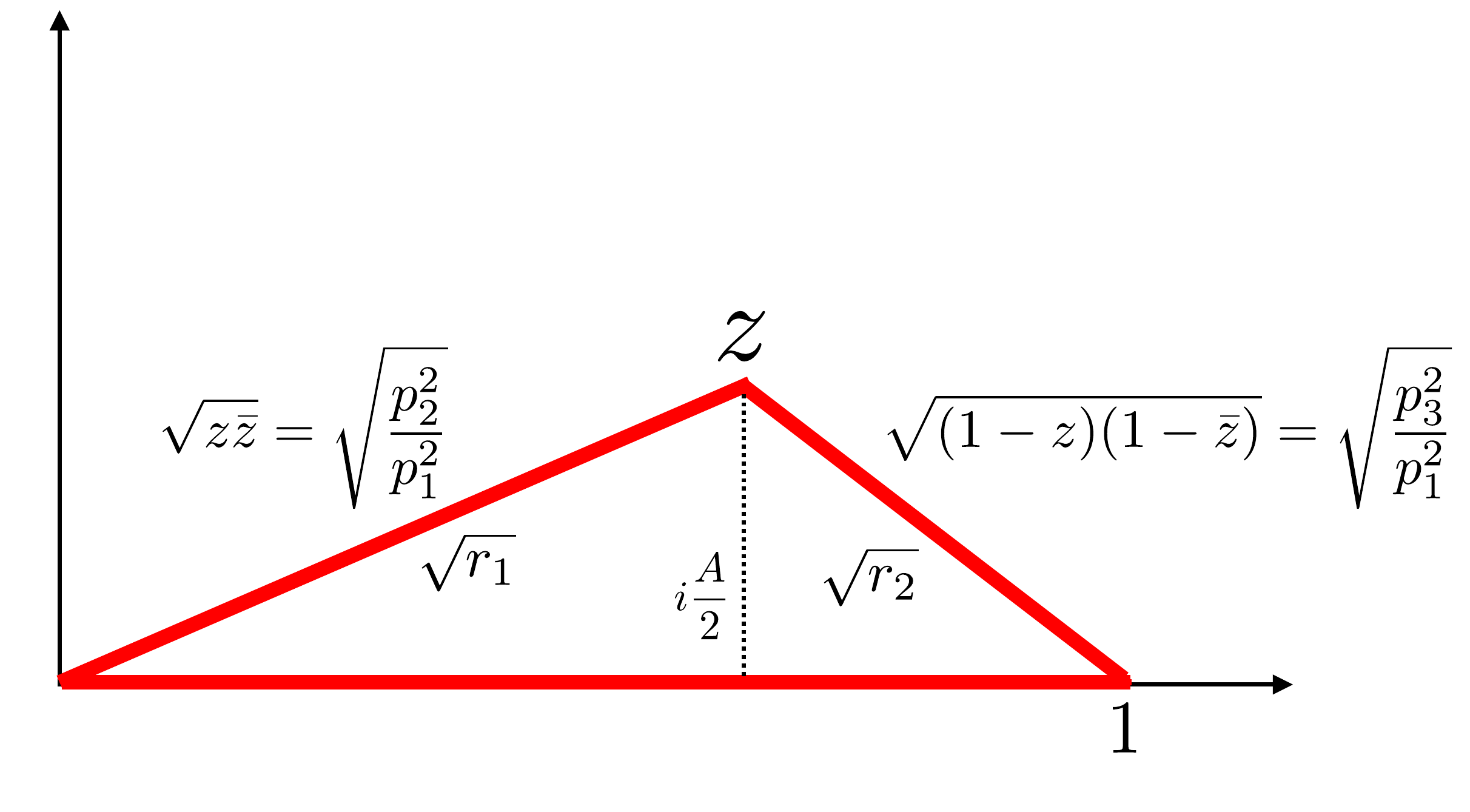}\label{fig:dual_b}
}\qquad
\end{center}
\caption{(a) Three mass triangle. (b). Corresponding dual geometry of the EEEC. The angles of the EEEC are mapped to the external masses of the three mass triangle.}
\label{fig:dual}
\end{figure}

We therefore find an interesting equivalence between the integrals for the energy correlators in the collinear limit, which are integrals of squared amplitude over phase space, and Feynman integrals for loop amplitudes. Note that this rewrite is different from the usual application of optical theorem. This correspondence can be made even more clear if we use dual coordinates \cite{Drummond:2006rz} $x_i^\mu-x_{i+1}^\mu=p_i^\mu$\,, $x_{ij}^2=(x_i-x_j)^2=(p_i+\cdots p_{j-1})^2$ in 2D Euclidean space. In this case, we simply have a correspondence between the dual coordinates and the coordinates on the celestial sphere
\begin{align}
x_{ij}^2 \leftrightarrow |z_{ij}|^2\,,
\end{align}
and between energy fractions and the Feynman parameters $\alpha_i$
This equivalence is shown schematically in \Fig{fig:dual}. We find this equivalence to be quite striking, and deserving of further investigation.

From this equivalence we can already begin to easily understand the functions that will appear in the result for the triple correlator.
 In \cite{Abreu:2017ptx} it is advocated that the $n$-point integral should be considered in $D_n$, where $D_n=n-2\epsilon$ if $n$ is even and $D_n=n+1-2\epsilon$ if $n$ is odd. These integrals are conjecturally of uniform transcendentality. They also have a correspondence with hyperbolic geometry. We therefore look at the three mass triangle integral in $D=4-2\epsilon$ as a base integral. The three mass triangle integral is given by (here we follow closely the notation of \cite{Chavez:2012kn})
\begin{align}
-e^{\gamma_E \epsilon}\int \frac{d^D k}{\pi^{D/2}}  \frac{1}{k^2 (p_2-k)^2 (p_3+k)^2} = -\frac{i}{p_1^2}  \frac{2}{z- \bar z} 2i D_2^-(z)\,,
\end{align}
where $D^-(z)$ is the Bloch-Wigner function, as defined above, and
\begin{align}
z\bar z=\frac{p_2^2}{p_1^2}\,, \qquad (1-z)(1-\bar z)=\frac{p_3^2}{p_1^2}\,.
\end{align}
Therefore, we arrive at the Bloch-Wigner function from a completely different perspective.

For this particular term in the splitting function, the result of the integral for the EEEC can be expressed in terms of a three mass triangle. Naively one might have anticipated a dependence on the fourth leg that initiated the splitting. To understand how this physically occurs, we can consider the ``eikonal" terms in the splitting function, taking as an example one that involves $(\omega_1+\omega_2)$ in the denominator. For a jet along the direction $n^\mu=(1,0,0,1)$, these terms can be thought of as originating from a Wilson line along the direction $\bar n^\mu=(1,0,0,-1)$, and therefore should introduce a dependence on the fourth direction. 

\begin{figure}[ht!]
  \centering
  \includegraphics[width=0.3\textwidth]{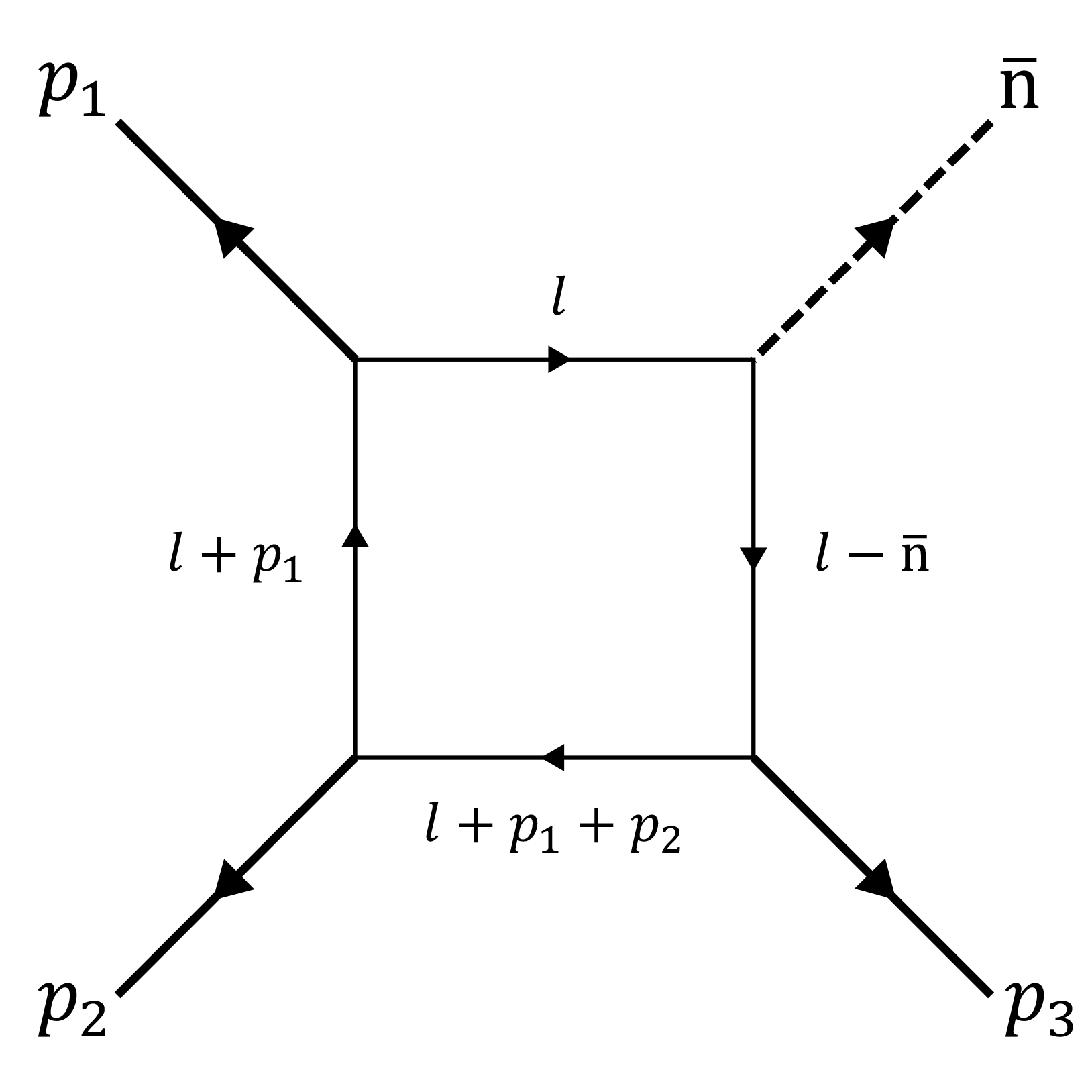}
  \caption{One-loop box with an eikonal propagator.}
  \label{fig:ebox}
\end{figure}

To illustrate the mapping of these eikonal terms into Feynman integrals, we consider the particular contribution to the triple correlator 
\begin{equation}
	F_2=\mathcal{N} \frac{1}{2 |z_{12}|^2}\times \int d\omega_1d\omega_2d\omega_3 \delta(1-\omega_1-\omega_2-\omega_3)\frac{\omega_1\omega_2\omega_3}{(\omega_1+\omega_2)s_{123}}\,.
\label{f2}
\end{equation}
Now, we consider the form of a box integral with an eikonal propagator, whose Feynman parametrization is given by~(an implicit $+i 0^+$ for the denominators is assumed)
\begin{align}\label{f2fpi}
	\mathcal{J}^{(d)}(\nu_1,\nu_2,\nu_3,\widetilde{\nu_4})=&  \int \frac{d^d l}{i\pi^{\frac{d}{2}}} \frac{1}{(l^2)^{\nu_1}((l+p_1)^2)^{\nu_2}((l+p_1+p_2)^2)^{\nu_3}(-\bar{n}\cdot l)^{\nu_4}}\\
	&\hspace{-2cm}=(-1)^{\frac{d}{2} }\frac{\Gamma(\sum_i\nu_i-\frac{d}{2})}{\prod_{i=1}^{4} \Gamma(\nu_i)} \int_0^1 d\omega_1d\omega_2d\omega_3\delta(1-\omega_1-\omega_2-\omega_3) \left(\omega_1^{\nu_3-1} \omega_2^{\nu_1-1} \omega_3^{\nu_2-1} \right) \nn \\
	&\hspace{-2cm}\times \int _{0}^{\infty} ds s^{\nu_4-1} \bigg[\omega_1\omega_2(p_1+p_2)^2+\omega_1\omega_3 p_2^2+\omega_2\omega_3 p_1^2 + s(\omega_1\bar{n}\cdot(p_1+p_2)+\omega_3\bar{n}\cdot p_1)\bigg]^{\frac{d}{2}-\sum_i\nu_i}\,.\nn
\end{align}
The corresponding Feynman diagram is depicted in Fig.~\ref{fig:ebox}.
We can now rewrite Eq. (\ref{f2}) in this way
\begin{equation}
\begin{aligned}
	F_2=\mathcal{N}\frac{1}{2 |z_{12}|^2} &\int d\omega_1d\omega_2d\omega_3 \delta(1-\omega_1-\omega_2-\omega_3)\\ &\times \int_{0}^{\infty} \! ds \, \frac{\omega_1 \omega_2 \omega_3}{\left[\omega_1\omega_2|z_{12}|^2+\omega_1\omega_3|z_{13}|^2+\omega_2\omega_3|z_{23}|^2+s(\omega_1+\omega_3)\right]^2}\,.
\label{f2new}
\end{aligned}
\end{equation}
We can therefore make the identification
\begin{equation}
	F_2 = \mathcal{N}\frac{1}{2 |z_{12}|^2}\times \mathcal{J}^{(d=10)}(2,2,2,\widetilde{1})\,.
\end{equation}
where the external momenta of the eikonal box integral are given by 
\begin{equation}
\begin{aligned}
	&\bar{n}\cdot(p_1+p_2)=\bar{n}\cdot p_1 = 1 \,,\\
	&p_1^2\rightarrow |z_{23}|^2=x_1\text{,}\qquad p_2^2\rightarrow |z_{13}|^2=x_2\text{,}\qquad (p_1+p_2)^2\rightarrow |z_{12}|^2=x_3\,.\\
\end{aligned}
\end{equation}
This eikonal box integral gives rise to the dependence on the function
\begin{align}
D^+_2(z)=\left(\text{Li}_2\left(1-|z|^2\right)+\frac{1}{2} \log \left(|1-z|^2\right) \log \left(|z|^2\right)\right)\,,
\end{align}
described previously, among other terms.

Based on this result, we believe that it would be interesting to investigate the behavior of the multipoint energy correlators in the conformal fishnet theory \cite{Gurdogan:2015csr}.\footnote{The conformal fishnet theory has the disadvantage that it is not unitary, however, we do not believe that this is a problem for studying the structure of these integrals. For an interesting recent study of the symmetry properties of such integrals, that are closely related to those appearing here, see \cite{Loebbert:2019vcj}.} We have seen that one source of complexity for the transcendental functions involved arises from the ``eikonal" terms in the splitting functions (those that involve the momentum fractions $z_i$ in the denominators). These terms are typically coming from the perturbative expansion of collinear Wilson lines. In the conformal fishnet theory, the gauge bosons are decoupled, and the splitting functions are purely functions of Mandelstam invariants. In this case there may be a simpler relation between Feynman diagrams and energy correlators in the collinear limit. Indeed, in the conformal fishnet theory, the $1\to 3$ splitting function is simply given by $P_{1\to 3}\sim 1/(s_{123})^2$, and so one finds that the result for the EEEC can be written purely in terms of the three mass triangle (Bloch-Wigner function), and does not involve the eikonal box integral. It would be interesting to understand what happens for higher point energy correlators.

The mapping between the integrals appearing in the EEEC, and one loop Feynman integrals, in addition to providing a technical simplification for computing the result, also provides a significant number of insights into the structure of the results. In particular
\begin{itemize}
\item It provides an understanding of the functions that will appear in the result, namely polylogarithms.
\item It provides a first entry condition \cite{Gaiotto:2011dt}, namely that the first entry in the $(1,n-1)$ coproduct must be a Mandelstam. In the case of the three point correlator, the Mandelstams are mapped to $z\bar z$ and $(1-z)(1-\bar z)$, and we find that this condition is indeed satisfied by our result. This condition is not something that is immediately clear from the event shape perspective. It would also be interesting to understand how this condition arises from the perspective of correlation functions \cite{Henn:2019gkr,Belitsky:2013ofa,Belitsky:2013bja,Belitsky:2013xxa} or from the lightray OPE \cite{Kravchuk:2018htv,Kologlu:2019bco,Kologlu:2019mfz}.
\item The factor of $1/(z-\bar z)$ appearing in front of the Bloch-Wigner function is identified as the leading singularity \cite{Cachazo:2008vp} of the three mass triangle diagram. The physical significance of this factor is not clear from the perspective of the EEEC. 
\item The result for the three mass triangle in $d=4$, apart from the $1/(z-\bar z)$ prefactor, is a pure function. From this perspective, the polynomials that appear in the result for the EEEC arise from integration by parts or dimension shift identities arising from factors of $\omega_i$ in the splitting functions. This may provide a convenient way of understanding these polynomials. Alternatively, it may indicate that the result for the energy correlators can be most compactly written in terms of Feynman parameter integrals in shifted dimensions.
\end{itemize}

More speculatively, it is well known that one-loop Feynman diagrams have an interpretation as volumes \cite{Davydychev:1997wa,Davydychev:1998fk,Gorsky:2009nv,Schnetz:2010pd,Hodges:2010kq,Paulos:2012qa,Nandan:2013ip,Davydychev:2017bbl}. It seems that at least in this simple case that we have illustrated here, the EEEC is directly realizing this dual geometry associated with the Feynman diagrams to which it can be associated. We find this is particularly interesting, and believe that it merits investigation at higher points.

\subsection{Possible Generalization to Higher Points} \label{sec:higher_point_duality}

\begin{figure}
\begin{center}
\includegraphics[scale=0.25]{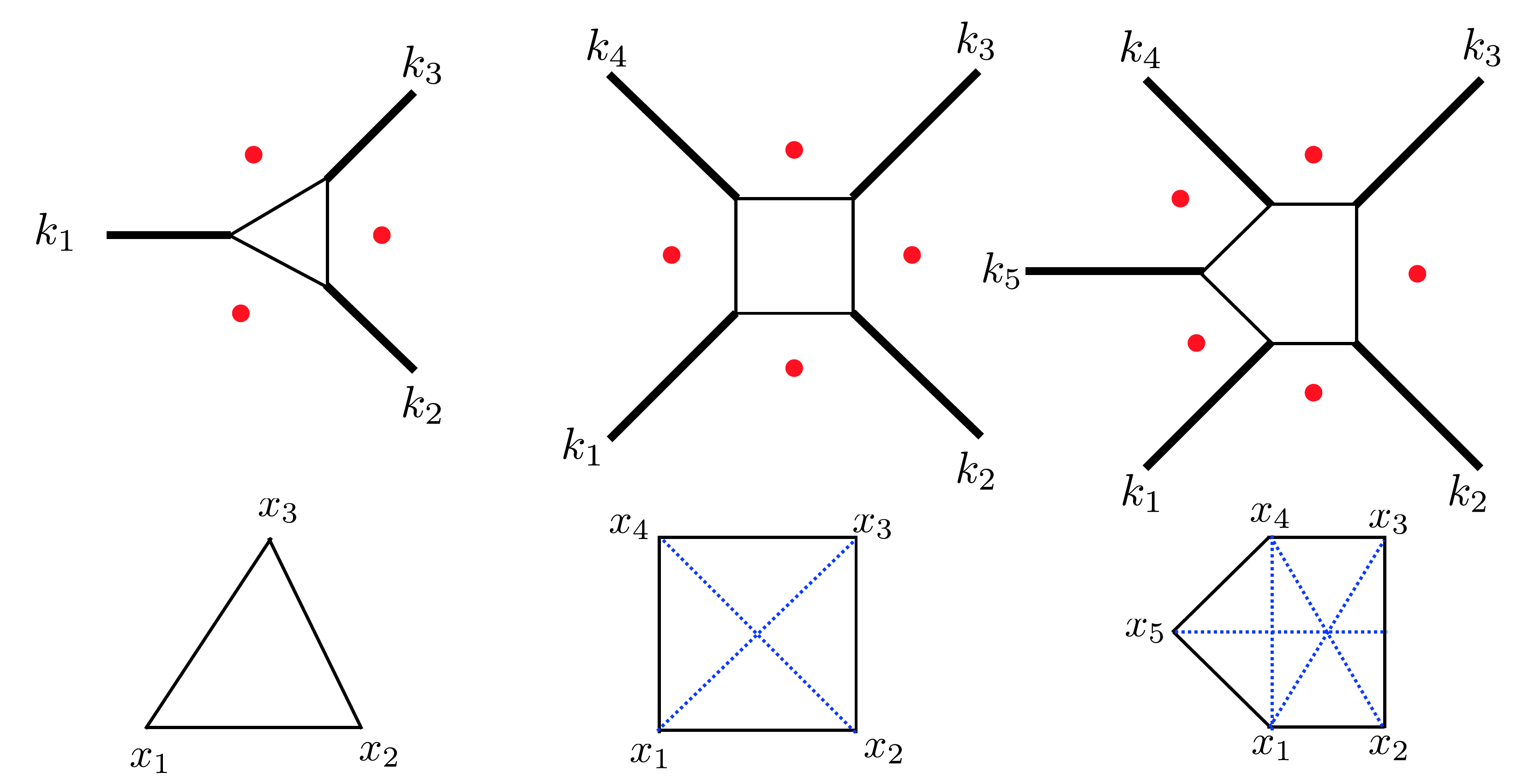}
\end{center}
\caption{Naive correspondence between multipoint energy correlators and dual coordinates for Feynman graphs. This correspondence is broken by eikonal terms in the splitting functions that introduce eikonal propagators into the dual Feynman graphs.}
\label{fig:intdual}
\end{figure}

Although we will not consider higher point correlators in this paper, it is interesting to speculate whether this correspondence can provide a guide for higher point correlators. First, we must note that it may not be true that at higher points one can always map the energy correlator integrals to one-loop Feynman diagrams. However, it is generically true that one will have integrals of the form
\begin{align}
\int d \omega_1 d\omega_2 \ldots d\omega_N \delta(1-\omega_1- \omega_2-\cdots- \omega_N) \frac{(\omega_1 \omega_2 \cdots \omega_N)^2}{s_{12\cdots N}^2} P_{1\to N}\,,
\end{align}
perhaps with some Gram determinant constraints. Even if the integrals do not take precisely the form of Feynman diagrams due to the structure of the splitting function $P_{1\to N}$, we believe that thinking about the integrals in relation to Feynman parameter integrals may be useful in two ways. First, and most practically, regardless of the exact form of $P_{1\to N}$, these integrals will have similar form to Feynman parameter integrals, and therefore many of the techniques can be carried over. Second, and more conceptually, in the three point case, the geometry, and its simple parameterization were suggested by the dual Feynman graph, and could be understand largely just from the $1/{s_{12\cdots N}^2}$ factor. It is therefore possible that this term will continue provide a  guide at higher points for what variables are convenient for parametrizing the result.

If we assume $P_{1\to N}=1$, as would occur if one had a scalar theory with an $n+1$ point interaction, then, as mentioned above, we have a correspondence between the distances on the sphere between the points of the energy correlator and the dual coordinates
\begin{align}
x_{ij}^2=(x_i-x_j)^2=(p_i+\cdots p_{j-1})^2 \leftrightarrow |x_i-x_j|^2\,.
\end{align}
In particular, this identification gives the sides of the energy correlator shapes as the masses of the external legs, and the chords as Mandelstam invariants. We note that there are of course $3n-10$ invariants plus the masses, and there are $3n-10$ chords. Unlike the case of a Wilson loop that has been more thoroughly studied, these are not null polygons. However, higher point integrals with external masses have been studied, see e.g. \cite{DelDuca:2011wh,DelDuca:2011jm}. It is also interesting to note that the dilatation symmetry for the energy correlators would correspond to a dual conformal symmetry \cite{Drummond:2006rz,Drummond:2007aua,Henn:2011xk} of the corresponding Feynman diagram.

An intriguing test case will be for the four point EEEC. The one loop four mass box integral can be written in terms of the Bloch-Wigner function \cite{Usyukina:1992jd} (see also \cite{Abreu:2014cla} for a detailed discussion) if one uses the variables defined as
\begin{align}
Z \bar Z=U=\frac{p_2^2 p_4^2}{s t} \leftrightarrow \frac{|x_3-x_2|^2 |x_4-x_1|^2}{|x_1-x_3|^2 |x_2-x_4|^2}\,, \qquad   (1-Z)(1-\bar Z)=V=\frac{p_1^2 p_3^2}{s t} \leftrightarrow \frac{|x_2-x_1|^2 |x_4-x_3|^2}{|x_1-x_3|^2 |x_2-x_4|^2}\,,
\end{align}
where
\begin{align}
Z=\frac{1}{2}\left( 1+U-V + \sqrt{\lambda(1,U,V)}  \right)\,, \qquad \bar Z=\frac{1}{2}\left( 1+U-V - \sqrt{\lambda(1,U,V)}  \right)\,,
\end{align}
and
\begin{align}
\lambda(a,b,c)=a^2+b^2+c^2-2ab-2ac-2bc\,,
\end{align}
is the Kallen function. It will be interesting to see to what extent this holds true for the four point energy correlator, and whether this duality determines the geometry. A first step in this direction will be interesting to understand the structure of the $1\to 4$ splitting functions in QCD or $\cN=4$ to see to what extent different terms can be manipulated into the form of Feynman diagrams. The $1\to 4$ quark splitting function appeared recently in \cite{DelDuca:2019ggv}.

\section{Analytic Results for the Three-Point Correlator} \label{sec:results}

In this section we give the result for the non-trivial shape dependence of the EEEC for $\cN=4$ SYM, as well as for quark and gluon jets in QCD. We then briefly discuss the differences between the QCD and $\cN=4$ results.

\subsection{$\cN=4$ SYM} \label{sec:N4}

\subsubsection{Result in Terms of Loop Integrals} \label{sec:N4_loop}

Using the relation with Feynman parameter integrals, we can write down a compact representation of the result for the EEEC in $\cN=4$ in terms of three-mass triangle integral, the three-mass box with an eikonal propagator integral, and a rational term that incorporates the OPE limit. Explicitly, we have
\begin{align}
  \label{eq:feynman_rep}
  \frac{1}{\sigma_{\rm tot}}\frac{d^3 \Sigma}{dx_L\, d\text{Re}(z)\, d\text{Im}(z)} = \frac{g^4 N_c^2}{64 \pi^5} 
\left( {\cal J}^{(d=8)}(2,2,1) + {\cal J}^{(d=10)} (2,2,2, \widetilde{1}) + {\cal R}(z)\right) + (\text{permutations})  \,,
\end{align}
where
\[
{\cal R}(z) = \frac{\zeta_2 - 1}{2 x_L (1-z) (1-\bar z)} \,,
\]
and the permutations refer to the $S_3$ action in Eq.~\eqref{eq:S3_action}, including the corresponding Jacobian~(see Eq.~\eqref{eq:quark_dis} for an explicit example).

The triangle and box integrals appearing here are UV and IR finite, and can be defined in integer dimension, $d=8$ and $d=10$. One can also use the powerful toolkit of Feynman integral calculation, such as Integration-By-Parts~(IBP) reduction~\cite{Chetyrkin:1981qh}. To do that, it is necessary to continue the dimension to $d = 8 - 2 \e$ and $d = 10 - 2 \e$, respectively. The reason is the master integrals obtained from IBP will generically contain UV or IR divergences, but the physical sum do not. We give the explicit results for these integrals in the appendix~\ref{sec:feyn_int}. 

We find the simplicity of this result to be quite remarkable. The unusual dimensions of the integrals in this result mean that when they are expressed in terms of four dimensional integrals, which are pure functions, complex polynomials are generated. This also leads support to the idea that the result can be easily bootstrapped if the correct basis of functions is chosen. In particular, in \Eq{eq:feynman_rep}, with the exception of the OPE term, ${\cal R}(z)$, which is fixed by the behavior in the OPE limit, the coefficients of the other functions are rational. 

Due to the remarkable simplicity of this result (compare with the expanded result in \Eq{eq:N4_result}), we believe that it is worth better understanding this relation between the integrals for the energy correlator and Feynman parameter integrals in shifted dimensions. Large polynomials also appear in the expression for the two-point correlator in QCD \cite{Dixon:2018qgp,Luo:2019nig} (and to some extent in $\cN=4$ \cite{Belitsky:2013ofa,Henn:2019gkr}), and it would be interesting to understand if they can also be expressed in terms of Feynman parameter integrals in shifted dimensions. It will also be extremely interesting to understand to what extent this relationship extends to higher point correlators, or beyond the collinear limit.

Finally, we wish to emphasize another feature of the integrals appearing in the result, namely that due to the fact that the observable is infrared and collinear safe, the loop integrals appearing in its result must be infrared finite. Such integrals are much simpler than their infrared divergent counterparts, and many powerful techniques exist for their study. In particular, finite integrals involving similar functions to those appearing here obey a number of interesting differential equations (see e.g. 
\cite{Drummond:2006rz,Drummond:2010cz,Drummond:2012bg,Caron-Huot:2018dsv}), and it would be interesting to understand if these could be applied in the current case. Either way, any relationship between event shape observables and infrared finite loop integrals is extremely intriguing and worthy of further study.

\subsubsection{Result in Terms of Polylogarithms} \label{sec:N4_poly}

The $\cN=4$ result can also be written directly in terms of the $z$ variable, and the transcendental functions appearing in \Sec{sec:sym}. We have
\begin{align}\label{eq:N4_result}
G(z)&=\frac{(1+ |z|^2+|1-z|^2)}{2 |z|^2|1-z|^2} (1+\zeta_2)+\frac{(-1+|z|^2+|z|^4-|z|^6-|1-z|^4-|z|^2|1-z|^4+2|1-z|^6  )}{2 |z|^2|1-z|^2   (z-\bar z)^2} \log |1-z|^2  \nn \\
&+\frac{(-1-|z|^4+2|z|^6+|1-z|^2-|z|^4|1-z|^2 +|1-z|^4-|1-z|^6  )}{2 |z|^2|1-z|^2   (z-\bar z)^2} \log |z|^2 \nn \\
&+\frac{|z|^4-1}{2 |z|^2|1-z|^4}D_2^+(z)+\frac{|1-z|^4-1}{2 |z|^4|1-z|^2}D_2^+(1-z)  +\frac{( |z|^2-|1-z|^2)( |z|^2+|1-z|^2)}{2 |z|^2|1-z|^2} D_2^+ \left( \frac{z}{z - 1}   \right) \nn \\
&+\frac{2 i D_2^-(z)}{2|1-z|^4|z|^4 (z-\bar z)^3}p_3(|z|^2,|1-z|^2)\,.
\end{align}

The transcendental functions $D_2^{\pm}(z)$ are as defined in \Sec{sec:transcendental}.
The polynomial appearing in front of the Bloch-Wigner function is given by
\begin{align}
p_3(|z|^2, |1-z|^2)=&(-1+|z|^2-|1-z|^2)(1+|z|^2-|1-z|^2)(-1+|z|^2+|1-z|^2)\nn \\
&\times\Big[(-1+|1-z|^2)^2|1-z|^2+|z|^6(1+|1-z|^2) \nn \\
& -2|z|^4(1+|1-z|^2)^2 +|z|^2(1+|1-z|^2)(1+(-5+|1-z|^2)|1-z|^2) \Big ]\,.
\end{align}
One can easily verify that this result obeys all the constraints discussed in \Sec{sec:sym}. We will discuss the OPE limit in more detail in \Sec{sec:squeeze-limit-three}.

We see that the complexity of the result resides in the rational prefactors of the different transcendental functions, which are automatically accounted for by the representation in terms of Feynman integrals in shifted dimensions. This illustrates that these standard transcendental functions do not seem to be the optimal basis for expressing the result. Nevertheless, we still find it interesting to understand the structure of the polynomials \Eq{eq:N4_result} without reference to the Feynman integral perspective, so we explore their properties a bit more in this section. 

We have chosen to write them in an explicit form in terms of the $z$ and $\bar z$ variables at the expense of the $S_3$ symmetry. Here we show that taking into account this $S_3$ symmetry, these polynomials are in fact quite minimal. To understand how to write in terms of $S_3$ invariants, we recall that we defined the function $G(z,\bar z)$ as
\beq
\frac{1}{\sigma_{\rm tot}} \frac{d \Sigma}{d x_1 d x_2 d x_3}=\frac{g^4}{32 \pi ^5 \sqrt{2 x_1 x_2+ 2 x_2 x_3 +2 x_3 x_1-x_1^2-x_2^2-x_3^2}}\times \frac{1}{x_3^2}\times G(z,\bar z)\,.
\label{eq:Sigma}
 \eeq
We can re-write this in a more symmetric way as
\begin{align}
\frac{1}{\sigma_{\rm tot}}   \frac{d \Sigma}{d x_1 d x_2 d x_3}=\frac{g^4}{32 \pi ^5 x_1 x_2 x_3}\times \frac{|z|^2|1-z|^2}{A}\times G(z,\bar z)\,,
\end{align}
where we recall from \Eq{eq:def_A} that we defined 
\begin{align}
A&=\sqrt{-r_1^2-(-1+r_2)^2+2r_1(1+r_2)}\,, 
\end{align}
and
\begin{align}
r_1=|z|^2\,, \qquad r_2=|1-z|^2\,.
\end{align}
This shows that the function that has trivial transformation properties under the $S_3$ group is
\begin{align}\label{eq:convertFG}
F(z)= \frac{|z|^2|1-z|^2}{A}\times G(z).
\end{align}
However, due to the non-trivial action of the $S_3$ symmetry, given in \Eq{eq:S3_action}, $S_3$ invariant polynomials can take a non-trivial form.

First consider the OPE limit. In this limit, we have
\begin{align}
F(z) \to  \frac{1}{2A}\,.
\end{align}
The completion of this into an $S_3$ symmetric polynomial is
\begin{align}
\frac{1}{2A}\to\frac{1}{2A} \left( 1+|z|^2+|1-z|^2 \right)\,.
\end{align}
Interestingly, this is exactly the weight 0 term in the full result in \Eq{eq:N4_result}, once we account for the conversion between $F(z,\bar z)$ and $G(z,\bar z)$, as given in \Eq{eq:convertFG}. 

We also find that the weight 1 terms can be expressed as a sum of two $S_3$ invariant terms
\begin{align}
f_1^{(1)}&=\frac{1}{A^3}\left(  r_2 \log r_1 + r_1 \log r_2   +r_1^2 r_2 \log \frac{1}{r_1} +r_1 r_2^2 \log \frac{1}{r_2} +r_1^2 \log \frac{r_2}{r_1} +r_2^2 \log \frac{r_1}{r_2} \right) \,, \nn \\
f_2^{(1)}&=\frac{1}{A^3}\left(  \log r_1 +\log r_2   +r_1^3 \log \frac{1}{r_1} +r_2^3 \log \frac{1}{r_2} +r_1^3 \log \frac{r_2}{r_1} +r_2^3 \log \frac{r_1}{r_2} \right)\,,
\end{align}
where we have written $r_1 =|z|^2$ and $r_2=|1-z|^2$ to shorten the expression. 
In particular, we have
\begin{align}
f_2^{(1)}-f_1^{(1)}&=\frac{(-1+|z|^2+|z|^4-|z|^6-|1-z|^4-|z|^2|1-z|^4+2|1-z|^6  )}{2 |z|^2|1-z|^2   (z-\bar z)^2} \log |1-z|^2  \nn \\
&+\frac{(-1-|z|^4+2|z|^6+|1-z|^2-|z|^4|1-z|^2 +|1-z|^4-|1-z|^6  )}{2 |z|^2|1-z|^2   (z-\bar z)^2} \log |z|^2 \,,
\end{align}
which is the weight 1 term in our result for the EEEC, showing that this polynomial is in fact fixed by only two coefficients.

The only polynomial that we are not able to write in completely simple manner is the coefficient of the Bloch-Wigner. It seems that it can be most simply expressed when written as a power series in $A^2=-(z-\bar z)^2$, 
\begin{align}
p_3(r_1,r_2)=-8 r_1^2 r_2^2 (1+r_1 +r_2) +2A^2 r_1 r_2 (1+r_1^2 +r_2^2) +A^4(r_1+r_1^2+r_2+r_1^2 r_2 +r_2^2 +r_2^2 r_1 )\,.
\end{align}
Here we see that we need one $S_3$ invariant at each order in the $A$ expansion.

We also note that while the result for the EEEC is not of uniform transcendental weight, the weight 0 and weigh 1 terms are minimal, in that they are fixed in terms of the weight 2 terms by the requirement of satisfying the correct behavior in the OPE limit,  as well as in the collapsed triangle limit. We believe that this makes it promising that this result could be entirely bootstrapped. However, to completely bootstrap the result would require a better understanding of the polynomials that can appear in the result, in particular, it would be nice to have an a priori prediction for their order.

\subsection{QCD Jets} \label{sec:QCD}

Having presented, and discussed in some detail the $\cN=4$ result, which takes a particularly simple form, we now present the results for both quark and gluon jets in QCD. These take a more complicated form, however, this complexity is a feature of the rational prefactors, not the transcendental functions. Despite some effort, we were unable to simplify the structure of the polynomials, which we believe in QCD are inherently complex and encode the structure of power corrections in the QCD splitting functions. Calculations in this section were performed using the program HyperInt \cite{Panzer:2014caa}.

To simplify the structure of our results, we write the rational functions in terms of the variables
\begin{align*}
	r = z  \bar{z}\text{,}\qquad s = z+\bar{z}\text{,} \qquad t=z-\bar{z}\,,
\end{align*}
and we use the following shorthand for the transcendental functions
\begin{align*}
	g_{1}^{(1)}&=\log\bigg((1-z)(1-\bar{z})\bigg) \,,\qquad \qquad	g_{1}^{(2)}=\log\left(z\bar{z}\right) \,,\\
	g_{2}^{(1)}&=\frac{1}{2}\bigg(\log(1-z)-\log(1-\bar{z})\bigg)\log(z\bar{z})+\text{Li}_2(z)-\text{Li}_2(\bar{z}) \,, \\
	g_{2}^{(2)}&=\frac{1}{2}\log\bigg(\left(1-z\right)\left(1-\bar{z}\right)\bigg)\log\left(z\bar{z}\right)+\text{Li}_2\left(1-z\bar{z}\right)\,,\\
	g_{2}^{(3)}&=\frac{1}{2}\log\left(\frac{1}{\left(1-z\right)\left(1-\bar{z}\right)}\right)\log\left(\frac{z\bar{z}}{\left(1-z\right)\left(1-\bar{z}\right)}\right)+\text{Li}_2\left(1-\frac{z\bar{z}}{\left(1-z\right)\left(1-\bar{z}\right)}\right)\,,\\
	g_{2}^{(4)}&=\frac{1}{2}\log\bigg(\left(1-z\right)\left(1-\bar{z}\right)\bigg)\log\left(z\bar{z}\right)+\text{Li}_2\bigg(1-\left(1-z\right)\left(1-\bar{z}\right)\bigg)\,, \\
	g_{2}^{(5)}&=\pi^2\,.
\end{align*}
Note that the functions $g_{2}^{(2)}$, $g_{2}^{(3)}$ and $g_{2}^{(4)}$ are related by the $S_3$ permutation symmetry. The results of this section can equally well be written in terms of loop integrals, however, unlike the case of $\cN=4$, this does not provide significant insight, and so we will write the results in terms of the transcendental functions given above.

\subsubsection{Quark Jets} \label{sec:Quark}

In this subsection we give the analytical results for quark jet. It can be written as
\begin{multline}
\frac{1}{\sigma_{\rm tot}}\frac{d\Sigma_{q}}{dx_L d\text{Re}(z) d\text{Im}(z)}=  \frac{g^4}{16 \pi^5}\frac{1}{x_L}\bigg[ G_{ q}(z)
+ G_{ q}(1 - z)
\\
 + \frac{1}{|1-z|^4} \left( G_{ q}\left(\frac{z}{z-1} \right)
+ G_{ q} \left(\frac{1}{1-z} \right) \right) + \frac{1}{|z|^4} \left( 
G_{ q}\left(\frac{1}{z}\right) + G_{ q} \left(\frac{z-1}{z} \right)
\right) 
\bigg]
 \,,
\label{eq:quark_dis}
\end{multline}
where we use that $dz d \bar{z} = d \mathrm{Re}(z) d \mathrm{Im}(z)$. We have written out the complete $S_3$ action, including the Jacobian, explicitly. $G_{q}(z)$ can be further decomposed into different color structures,
\begin{align}
G_{q} (z) = G_{\bar{q}^{'}q^{'}q}(z)+ G_{\bar{q}qq}^{(id)}(z)+ G_{ggq}(z) \,.
\end{align}
Here we present the results for each term separately. For $G_{\bar{q}^{'}q^{'}q}(z)$, we have
\begin{align*}
	&G_{\bar{q}^{'}q^{'}q}(z)=C_F T_F n_F\times \bigg\{\frac{1}{1920t^8}\bigg[-80640 r^5+16 r^4 \left(7897 s^2-3194 s+2232\right)-8 r^3 \left(11631 s^4 \right.\\
	&\left.-14936 s^3+19657 s^2-13290 s+4412\right)+4 r^2 \left(7160 s^6-8067 s^5+1978 s^4+13063 s^3-18082 s^2\right.\\
	&\left.+8860 s-1080\right)-2 r s^2 \left(2100 s^6-2480 s^5+1835 s^4-1620 s^3+4810 s^2-6610 s+2952\right)\\
	&+s^4 \left(240 s^6-300 s^5+260 s^4-67 s^3-83 s^2+150 s-144\right)\bigg]-\frac{r-s+1}{960 t^{10}}\bigg[r^5 (96256-136576 s)\\
	&+8 r^4 \left(25961 s^3-30314 s^2+26808 s-14400\right)-4 r^3\left(21989 s^5-3502 s^4-33955 s^3+64966 s^2\right. \\
	&\left.-46404 s+7552\right)+2 r^2 s \left(10240 s^6 -3732 s^5+9214 s^4-29779 s^3+53618 s^2-40220 s+8136\right)\\
	&+r s^3 \left(-2460 s^6+1100 s^5-1963 s^4-1289 s^3+6814 s^2-11866 s+6984\right)\\
	&+s^5 \left(120 s^6-60 s^5+100 s^4+50 s^3+34 s^2-39 s+72\right)\bigg]g_{1}^{(1)}+\frac{r}{960 t^{10}}\bigg[r^5 \left(176896-136576 s\right)\\
	&+8 r^4 \left(25961 s^3-40092 s^2+21164 s-15088\right)-4 r^3 \left(21989 s^5-8700 s^4-52275 s^3+78860 s^2\right.\\
	&\left.-62880 s+12360\right)+2 r^2\left(10240 s^7-7694 s^6+17392 s^5-75875 s^4+119500 s^3-97680 s^2\right.\\
	&\left.+28800 s-2160\right)-r s^2 \left(2460 s^7-2220 s^6+2999 s^5+4541 s^4-36410 s^3+61890 s^2-48900 s\right.\\
	&\left.+12960\right)+s^4 \left(120 s^7-120 s^6+140 s^5+73 s^4+380 s^3-2730 s^2+4020 s-2160\right)\bigg]g_{1}^{(2)}-\frac{1}{8 t^{11}}\\
	&\bigg[1344 r^7+r^6 \left(-5628 s^2+6552 s-3296\right)+6 r^5 \left(1232 s^4-2114 s^3+1803 s^2-1006 s+380\right)\\
	&+r^4 \left(-4158 s^6+6468 s^5-1795 s^4-5375 s^3+8200 s^2-4912 s+848\right)+2 r^3\left(660 s^8-1122 s^7\right.\\
	&\left. +924 s^6-1187 s^5+2260 s^4-2885 s^3+1901 s^2-498 s+36\right)+r^2\left(-242 s^8+440 s^7-396 s^6\right. \\
	&\left.+198 s^5+330 s^4-1080 s^3+1433 s^2-923 s+216\right) s^2+r\left(24 s^8-46 s^7+44 s^6-22 s^5-25 s^3\right. \\
	&\left.+76 s^2-85 s+36\right) s^4-\left(s^3-2 s^2+2 s-1\right) s^{11}\bigg]g_{2}^{(1)}+\frac{1}{8} \bigg[-2 r (s-1)+s^3-2 s^2+2 s-1\bigg]g_{2}^{(3)}\bigg\}\,.
\end{align*}
For $G_{\bar{q}qq}^{(id)}(z)$, we have
\begin{align*}
	&G_{\bar{q}qq}^{(id)}(z)=(C_A-2 C_F)C_F\times\bigg\{\frac{1}{11520 t^8 (r-s+1)^4}\bigg[2952 r^8 (63 s-58)-24 r^7 \left(6148 s^3+27246 s^2\right.\\
	&\left.-62853 s+31810\right)+r^6 \left(63582 s^5+494010 s^4+251712 s^3-3234312 s^2+3346592 s-518944\right)\\
	&+r^5 \left(-11160 s^7-239787 s^6-242838 s^5+792780 s^4+2344672 s^3-5448528 s^2+2740176 s\right.\\
	&\left.-567392\right)+r^4 \left(720 s^9+43740 s^8+278667 s^7-699957 s^6+112182 s^5-761156 s^4+3562352 s^3\right.\\
	&\left.-3080832 s^2+963544 s+67568\right)+r^3 \left(-2880 s^{10}-60480 s^9-37044 s^8+789742 s^7\right.\\
	&\left.-1541669 s^6+1762726 s^5-2439712 s^4+2379488 s^3-1304096 s^2+272184 s-29040\right)+r^2 s \\
	&\left(4320 s^{10}+30600 s^9-114070 s^8-77940 s^7+729201 s^6-1306849 s^5+1470360 s^4-1150278 s^3\right.\\
	&\left.+621840 s^2-180024 s+24720\right)+r s^3 \left(-2880 s^9+1080 s^8+40877 s^7-101984 s^6+69207 s^5\right.\\
	&\left.+59426 s^4-157328 s^3+136848 s^2-63996 s+12000\right)+s^5 \left(720 s^8-3780 s^7+7415 s^6-6731 s^5\right.\\
	&\left.+3355 s^4-1849 s^3+1932 s^2-882 s+180\right)\bigg]+\frac{1}{1920 t^{10} (r-s+1)^4}\bigg[32256 r^{10}-4 r^9 \left(11909 s^2\right.\\
	&\left.+37826 s-43136\right)+4 r^8 \left(14155 s^4+33716 s^3+35198 s^2-170608 s+56064\right)+r^7 \left(-16043 s^6\right.\\
	&\left.-274300 s^5+355114 s^4-384612 s^3+1346556 s^2-1134536 s+497664\right)+r^6 \left(2220 s^8+78997 s^7\right.\\
	&\left.+442073 s^6-1802686 s^5+2690374 s^4-3239252 s^3+2476568 s^2-1303360 s+124416\right)+r^5 \\
	&\left(-120 s^{10}-11040 s^9-143233 s^8-164574 s^7+2504176 s^6-5907364 s^5+7687884 s^4\right.\\
	&\left.-6057800 s^3+3293732 s^2-992024 s+217600\right)+r^4 \left(600 s^{11}+21300 s^{10}+104147 s^9-239951 s^8\right.\\
	&\left.-1238636 s^7+5128180 s^6-8773052 s^5+8432296 s^4-5077368 s^3+1781128 s^2-374016 s\right.\\
	&\left.+20480\right)+r^3 \left(-1200 s^{12}-19200 s^{11}-3678 s^{10}+206332 s^9-2900 s^8-1652064 s^7+4235853 s^6\right.\\
	&\left.-5319852 s^5+4040306 s^4-1880724 s^3+541556 s^2-81944 s+5120\right)+r^2 s^2 \left(1200 s^{11}\right.\\
	&\left.+6900 s^{10}-29250 s^9-1320 s^8+71288 s^7+85636 s^6-602283 s^5+1018501 s^4-916982 s^3\right.\\
	&\left.+475910 s^2-141380 s+18920\right)+r s^4 \left(-600 s^{10}+480 s^9+7640 s^8-24520 s^7+38363 s^6\right.\\
	&\left.-49712 s^5+71911 s^4-84326 s^3+63342 s^2-25508 s+4280\right)+s^6 \left(120 s^9-660 s^8+1480 s^7\right.\\
	&\left.-1710 s^6+913 s^5+93 s^4-557 s^3+393 s^2-162 s+30\right)\bigg]g_{1}^{(1)}+\frac{1}{1920 t^{10} (r-s+1)^4}\\
	&\bigg[-32256 r^{10}+r^9 \left(47636 s^2+73184 s-99344\right)+r^8 \left(-56620 s^4-46364 s^3+87148 s^2\right.\\
	&\left.+69472 s+41264\right)+r^7 \left(16043 s^6+212500 s^5-579364 s^4+463512 s^3-135376 s^2-227408 s\right.\\
	&\left.+26656\right)+r^6 \left(-2220 s^8-61957 s^7-221453 s^6+1556576 s^5-2797984 s^4+2347632 s^3\right.\\
	&\left.-571352 s^2-155776 s+141984\right)+r^5 \left(120 s^{10}+8760 s^9+78553 s^8-32546 s^7-1397671 s^6\right.\\
	&\left.+4161836 s^5-5541244 s^4+3775936 s^3-1383964 s^2+221760 s-34320\right)+r^4 \left(-480 s^{11}\right.\\
	&\left.-12360 s^{10}-24907 s^9+140321 s^8+408799 s^7-2358813 s^6+4396600 s^5-4238920 s^4\right.\\
	&\left.+2353060 s^3-733140 s^2+133920 s-12240\right)+2 r^3 s \left(360 s^{11}+3360 s^{10}-8561 s^9-8786 s^8\right.\\
	&\left.-26256 s^7+293094 s^6-710896 s^5+853220 s^4-589160 s^3+234740 s^2-52020 s+5400\right)\\
	&-10 r^2 s^3 \left(48 s^{10}+30 s^9-813 s^8+2248 s^7-5386 s^6+13816 s^5-25504 s^4+28729 s^3-19312 s^2\right.\\
	&\left.+7158 s-1140\right)+20 r s^5 \left(6 s^9-30 s^8+60 s^7-43 s^6-122 s^5+480 s^4-788 s^3+692 s^2-318 s\right.\\
	&\left.+60\right)\bigg]g_{1}^{(2)}-\frac{1}{32 t^{11} (r-s+1)^5}\bigg[r^{11} (2440-2604 s)+r^{10} \left(4998 s^3+5360 s^2-21420 s+10056\right)\\
	&-4 r^9 \left(1155 s^5+3615 s^4-6755 s^3-6512 s^2+12573 s-6530\right)+2 r^8 \left(924 s^7+9702 s^6-12481 s^5\right.\\
	&\left.-21516 s^4+26037 s^3+7588 s^2-26702 s+3012\right)-4 r^7 \left(99 s^9+2112 s^8+4719 s^7-31384 s^6\right.\\
	&\left.+40666 s^5-17272 s^4-3317 s^3-5698 s^2+3375 s-2934\right)+2 r^6 \left(22 s^{11}+946 s^{10}+6292 s^9\right.\\
	&\left.-10890 s^8-60868 s^7+206316 s^6-280018 s^5+197026 s^4-86265 s^3+12864 s^2+1150 s\right.\\
	&\left.-3636\right)-2 r^5 \left(s^{13}+108 s^{12}+1651 s^{11}+1650 s^{10}-22858 s^9+7666 s^8+136020 s^7-348348 s^6\right.\\
	&\left.+419956 s^5-309054 s^4+139950 s^3-41416 s^2+5662 s-844\right)+2 r^4 \left(5 s^{14}+205 s^{13}+1115 s^{12}\right.\\
	&\left.-3964 s^{11}-6468 s^{10}+28553 s^9+1588 s^8-129649 s^7+260030 s^6-272931 s^5+174884 s^4\right.\\
	&\left.-72641 s^3+18084 s^2-2778 s+204\right)-2 r^3 s \left(10 s^{14}+180 s^{13}-50 s^{12}-2855 s^{11}+6205 s^{10}\right.\\
	&\left.-854 s^9-6854 s^8-10256 s^7+51400 s^6-77864 s^5+64878 s^4-33808 s^3+10754 s^2-2004 s\right.\\
	&\left.+180\right)+r^2 s^3 \left(20 s^{13}+120 s^{12}-724 s^{11}+329 s^{10}+4018 s^9-11098 s^8+15952 s^7-19508 s^6\right.\\
	&\left.+24576 s^5-26846 s^4+20520 s^3-10228 s^2+2956 s-380\right)-2 r s^5 \left(5 s^{12}-8 s^{11}-79 s^{10}\right.\\
	&\left.+362 s^9-694 s^8+713 s^7-324 s^6-224 s^5+606 s^4-652 s^3+398 s^2-136 s+20\right)+s^{11} \left(2 s^7\right.\\
	&\left.-14 s^6+42 s^5-70 s^4+70 s^3-43 s^2+16 s-4\right)\bigg]g_{2}^{(1)}-\frac{1}{32 (r-s+1)^5}\bigg[2 r^4-2 r^3 (s+2)\\
	&+r^2 \left(s^2+2 s+4\right)-2 r \left(s^2-s+2\right)+s^2-2 s+2\bigg]g_{2}^{(2)}+\frac{1}{16} (s-1)^2g_{2}^{(3)}-\frac{g_{2}^{(5)}}{192 (r-s+1)}\bigg\}\,.
\end{align*}
For $G_{ggq}(z)$ we have
\begin{align*}
	&G_{ggq}(z)=C_F^2\times\bigg\{\frac{1}{1920 r t^8 (r-s+1)}\bigg[4320 r^6+48 r^5 \left(123 s^2-544 s+421\right)+8 r^4 \left(18 s^4\right.\\
	&\left.-1602 s^3+6263 s^2-8495 s+3992\right)+4 r^3 \left(-51 s^5+2070 s^4-7280 s^3+11598 s^2-8480 s\right.\\
	&\left.+1132\right)+2 r^2 \left(37 s^5-970 s^4+2697 s^3-3546 s^2+3320 s-644\right) s+2 r \left(10 s^4-37 s^3-101 s^2\right.\\
	&\left.+131 s-154\right) s^3+13 s^8\bigg]+\frac{1}{960 t^{10}}\bigg[48 r^5 (339 s-448)+8 r^4 \left(873 s^3-8238 s^2+14561 s\right.\\
	&\left.-8704\right)+4 r^3 \left(18 s^5-3039 s^4+20363 s^3-40627 s^2+37080 s-8448\right)-2 r^2 \left(33 s^6-2920 s^5\right.\\
	&\left.+15741 s^4-30520 s^3+33116 s^2-9492 s+512\right)+2 r s^2 \left(8 s^5-509 s^4+1810 s^3-2534 s^2\right.\\
	&\left.+4074 s-1214\right)+s^4 \left(13 s^4+50 s^3-208 s^2+84 s-274\right)\bigg]g_{1}^{(1)}+\frac{1}{960t^{10}(r-s+1)}\bigg[48 r^6 (358\\
	&-339 s)-8 r^5 \left(873 s^3-8652 s^2+13793 s-7642\right)-4 r^4 \left(18 s^5-4245 s^4+27695 s^3-52145 s^2\right.\\
	&\left.+43770 s-11160\right)+2 r^3 \left(69 s^6-6874 s^5+37015 s^4-75160 s^3+75300 s^2-29580 s+2520\right)\\
	&-2 r^2 s \left(41 s^6-2172 s^5+8771 s^4-17455 s^3+18790 s^2-8010 s+900\right)+r s^3 \left(3 s^5-664 s^4\right.\\
	&\left.+1340 s^3-2300 s^2+3120 s-1200\right)+s^5 \left(13 s^4+35 s^3-50 s^2+30 s-60\right)\bigg]g_{1}^{(2)}-\frac{1}{8t^{11}}\\
	&\bigg[-72 r^6+r^5 \left(-216 s^2+696 s-452\right)-2 r^4 \left(18 s^4-309 s^3+896 s^2-995 s+344\right)+2 r^3 \left(27 s^5\right.\\
	&\left.-277 s^4+726 s^3-952 s^2+462 s-42\right)+2 r^2 \left(-11 s^5+87 s^4-177 s^3+246 s^2-126 s+15\right) s\\
	&+2 r \left(s^4-9 s^3+10 s^2-21 s+10\right) s^3+s^7+s^5\bigg]g_{2}^{(1)}\bigg\}+C_F C_A \times\bigg\{\frac{1}{11520rt^{8}(r-s+1)}\\
	&\bigg[241920 r^7-48 r^6 \left(7897 s^2-3389 s+8757\right)+8 r^5 \left(34893 s^4-43872 s^3+114654 s^2-55760 s\right.\\
	&\left.+24616\right)+r^4 \left(-85920 s^6+23472 s^5+62208 s^4-807932 s^3+985072 s^2-503784 s+42576\right)\\
	&+6 r^3 \left(2100 s^8+2800 s^7-5925 s^6+19077 s^5+14154 s^4-54072 s^3+39244 s^2-3664 s-464\right)\\
	&-2 r^2 \left(360 s^9+2340 s^8-5010 s^7+14083 s^6-11630 s^5+13365 s^4-23487 s^3+15924 s^2+2580 s\right.\\
	&\left.-1932\right) s+2 r \left(180 s^8-405 s^7+955 s^6-530 s^5-126 s^4+336 s^3-57 s^2-213 s+462\right) s^3\\
	&-39 s^8\bigg]+\frac{1}{1920t^{10}}\bigg[128 r^6 (992-1067 s)+8 r^5 \left(25961 s^3-34733 s^2+56757 s-39936\right)-4 r^4 \\
	&\left(21989 s^5+10184 s^4-54302 s^3+187766 s^2-181048 s+32256\right)+r^3 \left(20480 s^7+20994 s^6\right.\\
	&\left.+32298 s^5-77116 s^4+386928 s^3-473164 s^2+115144 s+2048\right)-2 r^2 \left(1230 s^9+3470 s^8\right.\\
	&\left.-1878 s^7+22026 s^6-33050 s^5+64563 s^4-71304 s^3+14688 s^2+2712 s-512\right)+2 r s^2 \left(60 s^9\right.\\
	&\left.+525 s^8-625 s^7+2584 s^6+222 s^5-3078 s^4+6237 s^3-6078 s^2-1164 s+1214\right)-s^4 \left(60 s^8\right.\\
	&\left.-90 s^7+260 s^6+5 s^5+35 s^4+10 s^3-118 s^2+24 s-274\right)\bigg]g_{1}^{(1)}+\frac{1}{1920t^{10}(r-s+1)}\bigg[128 r^7 \\
	&(1067 s-1622)-8 r^6 \left(25961 s^3-44511 s^2+49583 s-51754\right)+4 r^5 \left(21989 s^5+4986 s^4\right.\\
	&\left.-77470 s^3+193146 s^2-237744 s+52776\right)-2 r^4 \left(10240 s^7+6535 s^6+25479 s^5-120500 s^4\right.\\
	&\left.+300490 s^3-401970 s^2+141240 s-8280\right)+2 r^3 \left(1230 s^9+2910 s^8-2581 s^7+31058 s^6\right.\\
	&\left.-93827 s^5+169085 s^4-189540 s^3+69600 s^2-1920 s-720\right)-2 r^2 s \left(60 s^{10}+495 s^9-845 s^8\right.\\
	&\left.+3201 s^7+1151 s^6-17753 s^5+37139 s^4-38495 s^3+10610 s^2+2610 s-900\right)+r s^3 \left(60 s^9\right.\\
	&\left.-120 s^8+320 s^7-47 s^6+367 s^5-2176 s^4+4300 s^3-3160 s^2-1320 s+1200\right)+s^5 \left(-13 s^4\right.\\
	&\left.-35 s^3+50 s^2-30 s+60\right)\bigg]g_{1}^{(2)}-\frac{1}{32t^{11}(r-s+1)}\bigg[-2688 r^8+56 r^7 \left(201 s^2-245 s+189\right)\\
	&-4 r^6 \left(3696 s^4-6475 s^3+9443 s^2-7898 s+3522\right)+4 r^5 \left(2079 s^6-2541 s^5+1190 s^4+6838 s^3\right.\\
	&\left.-14437 s^2+10065 s-1866\right)-2 r^4 \left(1320 s^8-1089 s^7+1155 s^6+474 s^5+4040 s^4-16766 s^3\right.\\
	&\left.+16344 s^2-4874 s+268\right)+4 r^3 \left(121 s^{10}+11 s^9-198 s^8+1221 s^7-2422 s^6+3195 s^5\right.\\
	&\left.-4520 s^4+3801 s^3-1165 s^2+21 s+12\right)+2 r^2 s \left(-24 s^{11}-53 s^{10}+187 s^9-605 s^8+693 s^7\right.\\
	&\left.+124 s^6-1116 s^5+1866 s^4-1462 s^3+306 s^2+102 s-30\right)+2 r (s-1)^2 s^3 \left(s^9+11 s^8-7 s^7\right.\\
	&\left.+48 s^6+26 s^5+26 s^4+3 s^3+47 s^2-8 s-20\right)-s^5 \left(s^{10}-3 s^9+7 s^8-7 s^7+2 s^6-2 s^3+2 s^2\right.\\
	&\left.-2 s+2\right)\bigg]g_{2}^{(1)}+\frac{1}{32(r-s+1)}\bigg[4 r^2 (s-1)-2 r \left(s^3-2 s^2+5 s-4\right)+s^4-3 s^3+7 s^2\\
	&-7 s+2\bigg]g_{2}^{(3)}\bigg\}\,.
\end{align*}

\subsubsection{Gluon Jets}\label{sec:Gluon}

For gluon jet we similarly write
\begin{multline}
\frac{1}{\sigma_{\rm tot}}\frac{d\Sigma_{g}}{dx_L d\text{Re}(z) d\text{Im}(z)}=  \frac{g^4}{16 \pi^5}\frac{1}{x_L}\bigg[ G_{g}(z)
+ G_{g}(1 - z)
\\
 + \frac{1}{|1-z|^4} \left( G_{g}\left(\frac{z}{z-1} \right)
+ G_{g} \left(\frac{1}{1-z} \right) \right) + \frac{1}{|z|^4} \left( 
G_{g}\left(\frac{1}{z}\right) + G_{g} \left(\frac{z-1}{z} \right)
\right) 
\bigg]
 \,.
\end{multline}
The color decomposition is
\begin{align*}
G_g(z) = G_{gq\bar{q}}^{(ab)}(z)+2 G_{gq\bar{q}}^{(nab)}(z)+G_{ggg}(z) \,.
\end{align*}
Here the superscript $(ab)$ denotes the abelian contribution, while $(nab)$ denotes the non-abelian contribution.
We again present results for each of the terms separately. For the abelian $q\bar q g$ term, $G_{gq\bar{q}}^{(ab)}(z)$, we have
\begin{align*}
	&G_{gq\bar{q}}^{(ab)}(z)=\frac{C_FT_Fn_F}{1920t^{10}}\times\bigg\{\frac{1}{r-s+1}\bigg[96 r^6 (706-511 s)+8 r^5 \left(777 s^3+20430 s^2-46648 s+21680\right)\\
	&+16 r^4 \left(105 s^5-1968 s^4-10351 s^3+39298 s^2-34898 s+9284\right)+r^3 \left(-42 s^7-3702 s^6\right.\\
	&\left.+55248 s^5-3704 s^4-348088 s^3+488464 s^2-231168 s+40320\right)+r^2 s^2 \left(75 s^6+2498 s^5\right.\\
	&\left.-38602 s^4+80736 s^3-4488 s^2-85344 s+45024\right)-8 r s^4 \left(4 s^5+80 s^4-1273 s^3+3784 s^2\right.\\
	&\left.-4137 s+1554\right)+s^6 \left(5 s^4+28 s^3-364 s^2+672 s-336\right)\bigg]+\bigg[21504 r^6+24 r^5 \left(1729 s^2-8158 s\right.\\
	&\left.+5760\right)+48 r^4 \left(35 s^4-2615 s^3+11392 s^2-13046 s+4224\right)-6 r^3 \left(7 s^6+484 s^5-23370 s^4\right.\\
	&\left.+105084 s^3-151356 s^2+82856 s-14336\right)+6 r^2 s \left(9 s^6+237 s^5-11154 s^4+51546 s^3\right.\\
	&\left.-81660 s^2+52976 s-12656\right)-12 r s^3 \left(s^5+19 s^4-966 s^3+4236 s^2-5992 s+2716\right)\\
	&-168 s^5 \left(s^2-3 s+2\right)\bigg]g_{1}^{(1)}+\frac{1}{r-s+1}\bigg[-21504 r^7-24 r^6 \left(1729 s^2-7584 s+4796\right)-48 r^5 \\
	&\left(35 s^4-3112 s^3+12233 s^2-12810 s+3904\right)+6 r^4 \left(7 s^6+764 s^5-35600 s^4+144280 s^3\right.\\
	&\left.-194180 s^2+104000 s-18800\right)-12 r^3 \left(8 s^7+357 s^6-12424 s^5+51820 s^4-79540 s^3\right.\\
	&\left.+54320 s^2-15960 s+1680\right)+6 r^2 s^2 \left(11 s^6+266 s^5-8178 s^4+33680 s^3-52720 s^2+36960 s\right.\\
	&\left.-10080\right)-12 r s^4 \left(s^5+18 s^4-480 s^3+1720 s^2-2100 s+840\right)\bigg]g_{1}^{(2)}-\frac{1}{t}\bigg[1440 r^6 (62-49 s)\\
	&-720 r^5 \left(49 s^3-498 s^2+832 s-344\right)+720 r^4 \left(125 s^4-949 s^3+1850 s^2-1278 s+276\right)\\
	&-720 r^3 \left(115 s^5-806 s^4+1704 s^3-1456 s^2+504 s-56\right)+720 r^2 s^2 \left(45 s^4-297 s^3+622 s^2\right.\\
	&\left.-532 s+168\right)-1440 r s^4 \left(3 s^3-17 s^2+28 s-14\right)\bigg]g_{2}^{(1)}\bigg\}\,.
\end{align*}
For the non-abelian $q\bar q g$ term, $G_{gq\bar{q}}^{(nab)}(z)$, we have
\begin{align*}
	&G_{gq\bar{q}}^{(nab)}(z)=C_A T_F n_F\times \bigg\{\frac{1}{11520r^5t^8(r-s+1)}\bigg[-12960 r^{10}-48 r^9 \left(369 s^2-2311 s+1840\right)\\
	&-4 r^8 \left(108 s^4-14757 s^3+83880 s^2-113064 s+92336\right)+r^7 \left(990 s^5-74964 s^4+462552 s^3\right.\\
	&\left.-640000 s^2+1094480 s+247488\right)+r^6 \left(-453 s^6+37440 s^5-358404 s^4+135508 s^3\right.\\
	&\left.-1542992 s^2-683520 s-211680\right)+r^5 \left(-9 s^6+24 s^5+11465 s^4+237198 s^3+531872 s^2\right.\\
	&\left.+1540512 s+731712\right) s^2-r^4 \left(917 s^4+34820 s^3+211638 s^2+793968 s+874368\right) s^4+5 r^3 \\
	&\left(427 s^3+6808 s^2+43470 s+86796\right) s^6-20 r^2 \left(103 s^2+1488 s+5496\right) s^8+540 r (3 s+26) s^{10}\\
	&-720 s^{12}\bigg]+\frac{1}{1920r^5t^{10}}\bigg[r^{10} (27136-16272 s)+r^9 \left(-6984 s^3+93496 s^2-199424 s+92160\right)\\
	&+r^8 \left(-72 s^5+19774 s^4-202600 s^3+524000 s^2-422368 s+211456\right)+r^7 \left(129 s^6-19858 s^5\right.\\
	&\left.+198616 s^4-625560 s^3+555992 s^2-514208 s-160768\right)+r^6 \left(-47 s^6+6892 s^5-70742 s^4\right.\\
	&\left.+377014 s^3-187208 s^2+737232 s+202192\right) s-r^5 \left(4 s^5+541 s^4+26879 s^3+136212 s^2\right.\\
	&\left.+345532 s+490576\right) s^3+2 r^4 \left(17 s^4+1869 s^3+16644 s^2+74249 s+168636\right) s^5-10 r^3 \left(20 s^3\right.\\
	&\left.+426 s^2+3315 s+12343\right) s^7+20 r^2 \left(11 s^2+191 s+1256\right) s^9-180 r (s+15) s^{11}+120 s^{13}\bigg]g_{1}^{(1)}\\
	&+\frac{1}{1920r^5t^{10}(r-s+1)}\bigg[r^{11} (16272 s-22816)+8 r^{10} \left(873 s^3-12101 s^2+23154 s-9752\right)+r^9 \\
	&\left(72 s^5-24598 s^4+241580 s^3-566960 s^2+422112 s-233472\right)+r^8 \left(-201 s^6+33164 s^5\right.\\
	&\left.-302340 s^4+848800 s^3-703720 s^2+742160 s+178560\right)+r^7 \left(176 s^7-19889 s^6+179816 s^5\right.\\
	&\left.-705970 s^4+357900 s^3-1210160 s^2-426960 s-70560\right)+r^6 \left(-43 s^6+4700 s^5-7504 s^4\right.\\
	&\left.+388380 s^3+401700 s^2+1224480 s+362640\right) s^2-2 r^5 \left(19 s^5+2109 s^4+32185 s^3+141590 s^2\right.\\
	&\left.+380610 s+298860\right) s^4+2 r^4 \left(117 s^4+4080 s^3+33730 s^2+143400 s+193080\right) s^6-60 r^3\\
	& \left(7 s^3+136 s^2+999 s+2242\right) s^8+200 r^2 \left(2 s^2+33 s+132\right) s^{10}-60 r (5 s+46) s^{12}+120 s^{14}\bigg]\\
	&g_{1}^{(2)}-\frac{1}{32r^6t^{11}(r-s+1)}\bigg[144 r^{13}+8 r^{12} \left(54 s^2-249 s+178\right)+r^{11} \left(72 s^4-2266 s^3+9424 s^2\right.\\
	&\left.-11032 s+2784\right)+r^{10} \left(-254 s^5+4968 s^4-21402 s^3+33808 s^2-15512 s+10368\right)+r^9\\
	& \left(336 s^6-5532 s^5+25304 s^4-53122 s^3+23672 s^2-36424 s-9264\right)+r^8 \left(-194 s^7+3060 s^6\right.\\
	&\left.-13156 s^5+49356 s^4-22 s^3+67568 s^2+21248 s+2352\right)+2 r^7 \left(18 s^6-513 s^5-1528 s^4\right.\\
	&\left.-17641 s^3-23216 s^2-35818 s-8092\right) s^2+4 r^6 \left(12 s^5+339 s^4+2530 s^3+8399 s^2+15501 s\right.\\
	&\left.+8309\right) s^4-2 r^5 \left(s^5+77 s^4+1034 s^3+5544 s^2+15642 s+13860\right) s^6+r^4 \left(7 s^4+224 s^3\right.\\
	&\left.+2090 s^2+9196 s+12804\right) s^8-2 r^3 \left(5 s^3+106 s^2+794 s+1760\right) s^{10}+3 r^2 \left(3 s^2+50 s+192\right) \\
	&s^{12}-2 r (3 s+26) s^{14}+2 s^{16}\bigg]g_{2}^{(1)}+\frac{1}{32r^6 (r-s+1)}\bigg[-2 r^5+r^4 (7 s+4)-2 r^3 \left(5 s^2+7 s+2\right)\\
	&+r^2 s \left(9 s^2+18 s+4\right)-2 r s^3 (3 s+4)+2 s^5\bigg]g_{2}^{(4)}\bigg\}\,.
\end{align*}
For the pure gluon term, $G_{ggg}(z)$, we have
\begin{align*}
	&G_{ggg}(z)=C_A^2\times\bigg\{\frac{1}{57600r^5t^8(r-s+1)}\bigg[1036800 r^{11}-240 r^{10} \left(15796 s^2-16472 s+14071\right)\\
	&+240 r^9 \left(15594 s^4-14988 s^3+381 s^2+22011 s-10877\right)-8 r^8 \left(173400 s^6+129150 s^5\right.\\
	&\left.-1084380 s^4+2117595 s^3-1814325 s^2+1203028 s-345478\right)+2 r^7 \left(115200 s^8+465600 s^7\right.\\
	&\left.-1821945 s^6+2524230 s^5-682980 s^4+78832 s^3+3335968 s^2-1820040 s+205080\right)+r^6 \\
	&\left(-14400 s^{10}-201600 s^9+657000 s^8-902565 s^7+533715 s^6-1591014 s^5-3976216 s^4\right.\\
	&\left.-2400040 s^3+2851720 s^2-145440 s+21600\right)+r^5 \left(14400 s^9-43200 s^8+58200 s^7-33420 s^6\right.\\
	&\left.+223904 s^5+1882921 s^4+3835750 s^3-593320 s^2-1723680 s+132480\right) s^2-r^4 \left(13544 s^5\right.\\
	&\left.+331901 s^4+1698540 s^3+1551030 s^2-1242960 s-629280\right) s^4+25 r^3 \left(853 s^4+12307 s^3\right.\\
	&\left.+34826 s^2+1662 s-31308\right) s^6-100 r^2 \left(202 s^3+1709 s^2+1920 s-3192\right) s^8+900 r \left(13 s^2\right.\\
	&\left.+53 s-62\right) s^{10}-3600 (s-1) s^{12}\bigg]+\frac{1}{1920r^5t^{10}}\bigg[-256 r^{11} (823 s-688)+8 r^{10} \left(57428 s^3\right.\\
	&\left.-69816 s^2+53943 s-27200\right)-8 r^9 \left(31606 s^5+19440 s^4-144750 s^3+253940 s^2-200085 s\right.\\
	&\left.+63104\right)+2 r^8 \left(34240 s^7+83488 s^6-236066 s^5+155705 s^4+289210 s^3-439120 s^2+316524 s\right.\\
	&\left.-57088\right)+r^7 \left(-9120 s^9-57000 s^8+143177 s^7-173825 s^6+148336 s^5-363000 s^4+33200 s^3\right.\\
	&\left.-242464 s^2+144920 s-3072\right)+r^6 \left(480 s^{10}+8520 s^9-20020 s^8+22915 s^7-6698 s^6+15636 s^5\right.\\
	&\left.+231236 s^4+286130 s^3-123060 s^2-71904 s+8496\right) s-r^5 \left(480 s^9-1080 s^8+1180 s^7-236 s^6\right.\\
	&\left.+3158 s^5+57031 s^4+212027 s^3+111780 s^2-132108 s-31152\right) s^3+2 r^4 \left(97 s^5+3720 s^4\right.\\
	&\left.+31250 s^3+56000 s^2-5625 s-42212\right) s^5-10 r^3 \left(39 s^4+860 s^3+3690 s^2+2880 s-5495\right) s^7\\
	&+20 r^2 \left(23 s^3+270 s^2+627 s-800\right) s^9-60 r \left(5 s^2+34 s-37\right) s^{11}+120 (s-1) s^{13}\bigg]g_{1}^{(1)}\\
	&+\frac{1}{1920r^5t^{10}}\bigg[256 r^{11} (823 s-958)+r^{10} \left(-459424 s^3+933888 s^2-974904 s+615280\right)+8 r^9 \\
	&\left(31606 s^5-48360 s^4+9150 s^3+62435 s^2-79350 s+26096\right)+r^8 \left(-68480 s^7+114304 s^6\right.\\
	&\left.-130748 s^5+201610 s^4-387220 s^3+214360 s^2-324840 s+27120\right)+r^7 \left(9120 s^9-15720 s^8\right.\\
	&\left.+16663 s^7+2865 s^6-45226 s^5+213710 s^4+282540 s^3+263840 s^2-13680 s+1440\right)-r^6\\
	& \left(480 s^9-840 s^8+860 s^7+195 s^6+742 s^5+25492 s^4+221280 s^3+439180 s^2+150240 s\right.\\
	&\left.-12720\right) s^2+2 r^5 \left(42 s^5+1783 s^4+29175 s^3+123970 s^2+140610 s+27660\right) s^4-2 r^4 \left(97 s^4\right.\\
	&\left.+3780 s^3+33530 s^2+80520 s+52440\right) s^6+10 r^3 \left(39 s^3+884 s^2+4362 s+6180\right) s^8-20 r^2 \\
	&\left(23 s^2+288 s+852\right) s^{10}+60 r (5 s+38) s^{12}-120 s^{14}\bigg]g_{1}^{(2)}+\frac{1}{16r^6}\bigg[r^5-3 r^4 s+5 r^3 s^2-5 r^2 s^3\\
	&+3 r s^4-s^5\bigg]g_{2}^{(4)}-\frac{1}{16r^6t^{11}}\bigg[-1152 r^{13}+12 r^{12} \left(692 s^2-1048 s+691\right)-2 r^{11} \left(7848 s^4\right.\\
	&\left.-18936 s^3+23663 s^2-15880 s+5066\right)+r^{10} \left(11088 s^6-27288 s^5+29145 s^4-5825 s^3\right.\\
	&\left.-16236 s^2+9936 s-3628\right)+2 r^9 \left(-2046 s^8+5280 s^7-6448 s^6+4119 s^5-3016 s^4+6082 s^3\right.\\
	&\left.+269 s^2+2856 s-230\right)+r^8 \left(836 s^{10}-2222 s^9+2772 s^8-1416 s^7-450 s^6+696 s^5-9543 s^4\right.\\
	&\left.-11969 s^3-4580 s^2+240 s-24\right)+r^7 \left(-90 s^{10}+244 s^9-308 s^8+154 s^7+12 s^6+219 s^5\right.\\
	&\left.+2688 s^4+11595 s^3+13970 s^2+2610 s-212\right) s^2+r^6 \left(4 s^{10}-11 s^9+14 s^8-7 s^7-24 s^5\right.\\
	&\left.-588 s^4-4630 s^3-11540 s^2-8322 s-922\right) s^4+r^5 \left(s^5+66 s^4+990 s^3+4620 s^2+6930 s\right.\\
	&\left.+2772\right) s^6-r^4 \left(3 s^4+110 s^3+990 s^2+2772 s+2310\right) s^8+r^3 \left(5 s^3+110 s^2+594 s+924\right) s^{10}\\
	&-r^2 \left(5 s^2+66 s+198\right) s^{12}+r (3 s+22) s^{14}-s^{16}\bigg]g_{2}^{(1)}+\frac{1}{16}\bigg[2 r (s-1)-4 s^3+11 s^2-14 s\\
	&+7\bigg]g_{2}^{(3)}+\frac{g_{2}^{(5)}}{192 (r-s+1)}\bigg\}\,.
\end{align*}

\subsection{Discussion} \label{sec:discuss}

Having presented the results for both QCD and $\cN=4$, we see that there is an enormous difference in the complexity of the rational prefactors. This can also observed by comparing the analytic calculations of the two-point EEC at generic angles for Higgs decays \cite{Luo:2019nig} and $e^+e^-$ annihilation \cite{Dixon:2018qgp} with the $\cN=4$ result \cite{Belitsky:2013ofa}. However, the situation is even worse for the EEEC. In particular, in the case of quark and gluon jets in QCD, we find factors of $1/(z-\bar z)^{11}$, as compared with $1/(z-\bar z)^{3}$ in $\cN=4$.

It is interesting to speculate on the reason for this remarkable simplicity in $\cN=4$. There are several features of our results that we find of particular interest in this regard. First, even the pure gluon result in QCD is very complicated, despite the fact that we are working at tree level, so that the underlying amplitude is equivalent to the pure gluon amplitude in $\cN=4$. Second, we can look at the result in $\cN=1$ with an adjoint gluino by setting $C_A\to C_F$, and $n_f \to C_A$ in the QCD result. In this case, we are also unable to observe any particular form of simplicity in the result.  We are therefore led to believe that this simplicity is associated with a particular feature of the splitting functions in $\cN=4$, when occurs when the amplitudes are squared and summed over particle species (i.e. at the cross section as opposed to amplitude level). Indeed, it is well known \cite{Dokshitzer:2005bf,Dokshitzer:2006nm} that for the $1\to 2$ splitting functions in $\cN=4$, the particular splitting functions for a given particle species are not any simpler than in QCD, but when summed over all final states, all the quantum corrections cancel, and one gets precisely the classical result. To our knowledge, there has been limited study of the summed splitting functions for higher point splittings in $\cN=4$. However, for the $1\to 3$ splitting function used for the calculation of the EEEC, we find a particularly simple result
\begin{align}
P_{\cN=4} =&\ \frac{8}{\xi_3 s_{123} s_{12}} \left(\frac{1}{\xi_1} + \frac{1}{1-\xi_1}\right) + \frac{4}{s_{13} s_{23}} \left( \frac{1}{(1- \xi_1) (1-\xi_2)} + \frac{1}{\xi_1 \xi_2} \right) 
\nn\\
&\ 
+ (\text{permutation of } 1\,,2\,,3) \,.
\end{align}   
We believe that it would be interesting to understand this in more detail, in particular, if there is some emergent classicality if one is measuring energy type observables, as opposed to considering quantum scattering amplitudes. We also believe that it would be particularly interesting to understand the consequences of the integrability of $\cN=4$ at the level of the cross section, which has received relatively little attention compared to its study at the level of amplitudes. 

As a final comment on the structure of the results, we note that there does not seem to be a naive form of the principle of maximum transcendentality for the multi-point energy correlators. While the principle of maximal transcendentality holds for the twist-two anomalous dimensions at least through to three loops \cite{Kotikov:2001sc,Kotikov:2002ab,Kotikov:2004er,Kotikov:2007cy}, and therefore controls the scaling of the energy correlators, it does not seem to persist for the non-trivial cross ratio dependence. This is perhaps not surprising since these contributions are in no sense eikonal, but it could be interesting to understand better the extent to which it is broken.

\section{The Squeezed Limit of the Three-Point Correlator}
\label{sec:squeeze-limit-three}

\begin{figure}
\begin{center}
\includegraphics[scale=0.25]{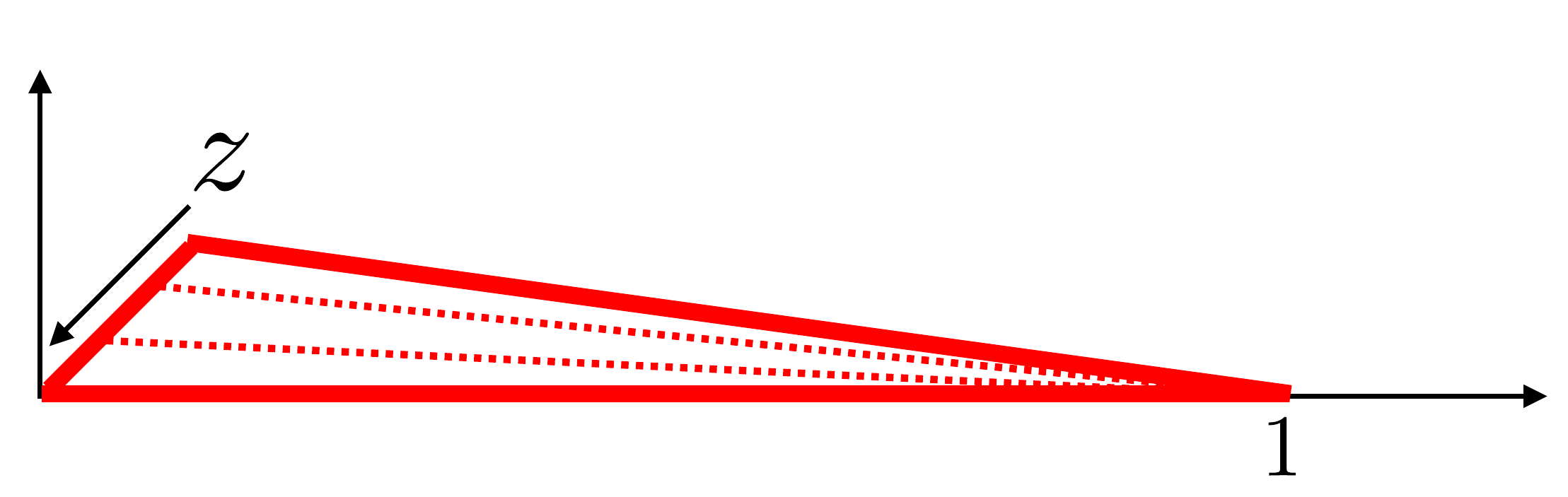}
\end{center}
\caption{The squeezed limit of the EEEC. We let $z\to0$ at a generic angle.}
\label{fig:squeeze}
\end{figure}

In this section we briefly study the squeezed (or OPE) limit of the collinear three-point correlator, which is illustrated in \Fig{fig:squeeze}. For a generic collinear three-point correlator $\frac{d^3 \Sigma}{dx_1 dx_2 dx_3}$ where $x_1, x_2, x_3 \ll 1$, the squeezed limit is defined as further taking the limit $x_1 \ll x_2 = x_3 = x_L + \Ord(x_1)$. One can also take the limit in any other ordering, but the result is the same due to bosonic symmetry of the energy measurement operator. Using the analytical results computed in this paper, one can directly verify that the triple correlator in the collinear squeezed limit scales as
\begin{align}
  \label{eq:squeeze_schematic}
 x_3^2 \frac{1}{\sigma_{\rm tot}} \frac{d^3\Sigma}{dx_3 dz d \bar{z}} \overset{\rm c.s.}{\simeq} \frac{C}{|z|^2 x_3} + \text{power corrections} \,,
\end{align}
where the constant $C$ depends on the theory under consideration. In the language of ANECs, the collinear squeezed limit corresponds to sequential light-ray OPE limits \cite{Kologlu:2019mfz}. It can also be easily understood in terms of the momentum space factorization picture for the EEC proposed in \cite{Dixon:2019uzg}, which is the approach we take here. 

Since the squeezed limit is not the primary focus of this paper, to keep this section relatively brief we restrict ourselves to considering the angularly averaged OPE limit. More precisely, in the OPE limit, we can parametrize $z=re^{i\phi}$, $\bar z=r e^{-i\phi}$, and in general the terms in the OPE expansion can have a dependence on $\phi$. In this section we average over this dependence on $\phi$, keeping only the dependence on $r^2=|z|^2$. We leave a discussion of the angular structure of the OPE, which deserves a more detailed study, to future work \cite{Chen:2020adz}.

We start from a generic triple correlator, without imposing the collinear limit. Then the limit of $x_1 \to 0$, $x_2 \to x_3$ is captured by the factorization formula,
\begin{align}
  \label{eq:x1to0}
    \frac{1}{\sigma_{\rm tot}} \frac{d^3\Sigma}{dx_3 dz d \bar{z}}\overset{|z| \to 0}{\simeq} \sum_{f_i} \int_0^1 du \, u^2 H_{f_i}(x_3, u) j_{f_i}(u^2 |z|^2 ) + \ldots \,,
\end{align} 
The jet function $j_{f_i}(u^2 |z|^2)$ is the two-point energy correlator at an angle $|z|^2$, in a jet initiated by a parton with flavor $f_i$, and momentum fraction $u = 2 p_{f_i} \cdot q/q^2$, where $q$ is, e.g., the center-of-mass momentum of $e^+e^-$ collision, and $p_{f_i}$ is the momentum of the jet. It is related to the cumulative jet function for the EEC by
\begin{align}
  \label{eq:differental_jet}
  j_{f_i}(y) = \frac{d}{dy} J_{f_i}(y) \,.
\end{align}
To leading logarithmic (LL) accuracy, the dependence on the momentum fraction $u$ in the jet function is irrelevant, and we can simply set it to $1$. Then the integral over the hard function can be interpreted as a two-point energy correlator, but with unequal energy weighting on the two points, 
\begin{align}
  \label{eq:unequal_hard_func}
  \int_0^1 du \, u^2 H_{f_i} (x_3, u) \equiv \frac{1}{\sigma_{\rm tot}} \sum_{f_j} \int d\sigma_{e^+e^- \to f_i + f_j + X} \frac{2 E_i^2 E_j}{Q^2} \delta\left(x_3 - \frac{1}{2}(1 - \cos\chi_{ij}) \right) \,.
\end{align}
This function is not collinear safe due to the $E_i^2$ weighting. However, the collinear divergence will be canceled by the boundary contact term in the triple correlator. For our purposes, we only need the result in the limit $x_3 \to 0$. In this limit, the hard function further factorizes into
\begin{align}
  \label{eq:second_fact}
   \int_0^1 du \, u^2 H_{f_i} (x_3, u) \overset{x_3 \to 0}{\simeq} \sum_{f_k} \int_0^1 dv \, v^3 H_{f_k} (v) j_{f_i/f_k} (x_3 v^2) + \ldots \,,
\end{align}
where $H_{f_k}(v)$ is the semi-inclusive fragmentation coefficient with momentum fraction $v$, which also appears in the context of EEC, and $j_{f_i/f_k}(x_3 v^2)$ is the two-point correlator with unequal energy weight for a jet initiated by $f_k$ with momentum fraction $v$. Again, when working at LL accuracy we can set $v=1$, and $H_{q}(v) = \delta(1-v)$.\footnote{We do not distinguish $q$ and $\bar{q}$ in $e^+e^-$ due to the charge conjugation invariance of QCD.} At the first order in $\alpha_s$, $j_{f_i/f_k}$ can be computed schematically by
\begin{align}
  \label{eq:unequal_jet}
  j_{f_i/f_k}(x_3) = \int d{\rm PS}^{(2)}(s,y) \frac{1}{s} P_{f_i, f_j \leftarrow f_k} (y) y^2 (1-y) \delta\left(x_3 - \frac{s}{y(1-y) Q^2} \right) \,,
\end{align}
where $d{\rm PS}^{(2)}$ is the collinear two-body phase space measure, $P_{f_i f_j \leftarrow f_k}$ is the splitting function for $f_k \to f_i + f_j$, $y$ is the momentum fraction of $f_i$ with respect to $f_k$, and $s = (p_{f_i} + p_{f_j})^2$. The difference as compared with the jet function for the EEC is the $y^2$ weighting, and the tagging of a parton flavor $f_i$. We therefore find that, to LL accuracy in the collinear squeezed limit, the triple energy correlator factorizes as
\begin{align}
  \label{eq:collinear_squeeze}
   \frac{1}{\sigma_{\rm tot}} \frac{d^3 \Sigma}{dx_3 dz d\bar{z}} \overset{\rm c.s.}{\simeq} 
\sum_{f_k} \sum_{f_i} j_{f_i/f_k}(x_3) j_{f_i}(|z|^2) + \ldots \,.
\end{align}
We can now use this formula to compute the squeezed limit for a quark jet, a gluon jet, and a jet in ${\cal N}=4$ SYM. 

We first recall the EEC jet function. At $\Ord(\alpha_s)$ and LL accuracy, it is simply given by
\begin{align}
  \label{eq:EECjetLL}
  (j_q(|z|^2), j_g(|z|^2)) = &\
\frac{\alpha_s}{4 \pi} (1, 1) 
\left(
\begin{array}{cc}
 \frac{25 C_F}{12} & -\frac{7 n_f}{30} \\
 -\frac{7 C_F}{12} & \frac{7 C_A}{5}+\frac{n_f}{3} \\
\end{array}
\right)
\nn\\
&\ = 
\frac{\alpha_s}{4 \pi}
\left( \frac{3 C_F}{2}, \frac{7 C_A}{5} + \frac{n_f}{10} \right) \,.
\end{align}
The evolution matrix on the right hand side of Eq.~\eqref{eq:EECjetLL} is the third Mellin moment of pure-singlet time-like splitting matrix. For the LL jet function of the two-point correlator with $E^2\times E$ weighting, we find
\begin{align}
  \label{eq:E2EjetLL}
  j_{q/q}(x_3) =&\ \frac{\alpha_s}{4 \pi} C_F \frac{16}{15} \frac{1}{x_3} \,,
\nn\\
  j_{g/q}(x_3) =&\ \frac{\alpha_s}{4 \pi} C_F \frac{13}{30} \frac{1}{x_3} \,,
\nn\\
  j_{g/g}(x_3) =&\ \frac{\alpha_s}{4 \pi} C_A \frac{7}{5} \frac{1}{x_3} \,,
\nn\\
  j_{q/g}(x_3) =&\ \frac{\alpha_s}{4 \pi} n_f \frac{1}{20} \frac{1}{x_3} \,.
\end{align}
For a quark jet, we then find the collinear squeezed limit to be
\begin{align}
  \label{eq:quarkcs}
   \frac{1}{\sigma_{\rm tot}} \frac{d^3 \Sigma_q}{dx_3 dz d\bar{z}} \overset{\rm c.s.}{\simeq} \left(\frac{\alpha_s}{4 \pi} \right)^2 \left(\frac{8}{5} C_F^2 + \frac{91}{150} C_F C_A + \frac{13}{300} C_F n_f\right) \frac{1}{x_3 |z|^2} \,.
\end{align}
For a gluon jet, we find
\begin{align}
  \label{eq:gluoncs}
     \frac{1}{\sigma_{\rm tot}} \frac{d^3 \Sigma_g}{dx_3 dz d\bar{z}} \overset{\rm c.s.}{\simeq} \left(\frac{\alpha_s}{4 \pi} \right)^2 \left(
\frac{49}{25} C_A^2 + \frac{7}{50} C_A n_f + \frac{3}{20} C_F n_f 
\right) \frac{1}{x_3 |z|^2} \,.
\end{align}
These results are in perfect agreement with expanding the results in Sec.~\ref{sec:Quark} and \ref{sec:Gluon}.
One can also convert the results above to ${\cal N}=1$ SYM, by setting $C_F = C_A = n_f = N_c$. We find
\begin{align}
  \label{eq:Neq1}
     \frac{1}{\sigma_{\rm tot}}  \frac{d^3 \Sigma_{q, {\cal N}=1}}{dx_3 dz d\bar{z}} \overset{\rm c.s.}{\simeq}  \left(\frac{\alpha_s}{4 \pi} \right)^2  \frac{9}{4} \frac{N_c^2}{x_3 |z|^2} \,,
\quad
     \frac{1}{\sigma_{\rm tot}} \frac{d^3 \Sigma_{g, {\cal N}=1}}{dx_3 dz d\bar{z}} \overset{\rm c.s.}{\simeq}  \left(\frac{\alpha_s}{4 \pi} \right)^2  \frac{9}{4} \frac{N_c^2}{x_3 |z|^2} \,,
\end{align}
which agrees with the expectation of ${\cal N}=1$ supersymmetry.

For ${\cal N}=4$ SYM, all the flavor dependence disappears, and one only needs a universal jet function for both EE and E$^2$E correlation,
\begin{align}
  \label{eq:jetNeq4}
  j_{\rm EE}(|z|^2) = \frac{\alpha_s}{4 \pi} N_c \frac{2}{|z|^2} \,,
\quad
j_{\text{E}^2\text{E}}(x_3) = \frac{\alpha_s}{4 \pi} N_c \frac{2}{x_3} \,.
\end{align}
We then obtain the collinear squeezed limit in ${\cal N}=4$ SYM,
\begin{align}
  \label{eq:Neq4}
       \frac{1}{\sigma_{\rm tot}} \frac{d^3 \Sigma_{{\cal N}=4}}{dx_3 dz d\bar{z}} \overset{\rm c.s.}{\simeq}  \left(\frac{\alpha_s}{4 \pi} \right)^2   \frac{4 N_c^2}{x_3 |z|^2} \,.
\end{align}
Again, this results is in complete agreement with the expansion of Eq.~\eqref{eq:N4_result}.
It will be interesting to extend this analysis to higher point correlators where there are multiple iterated squeezed limits, which should placed strong constraints on the structure of the result.

\section{Numerical Checks} \label{sec:numerics}

\begin{figure}
\begin{center}
\subfloat[]{
\includegraphics[scale=0.45]{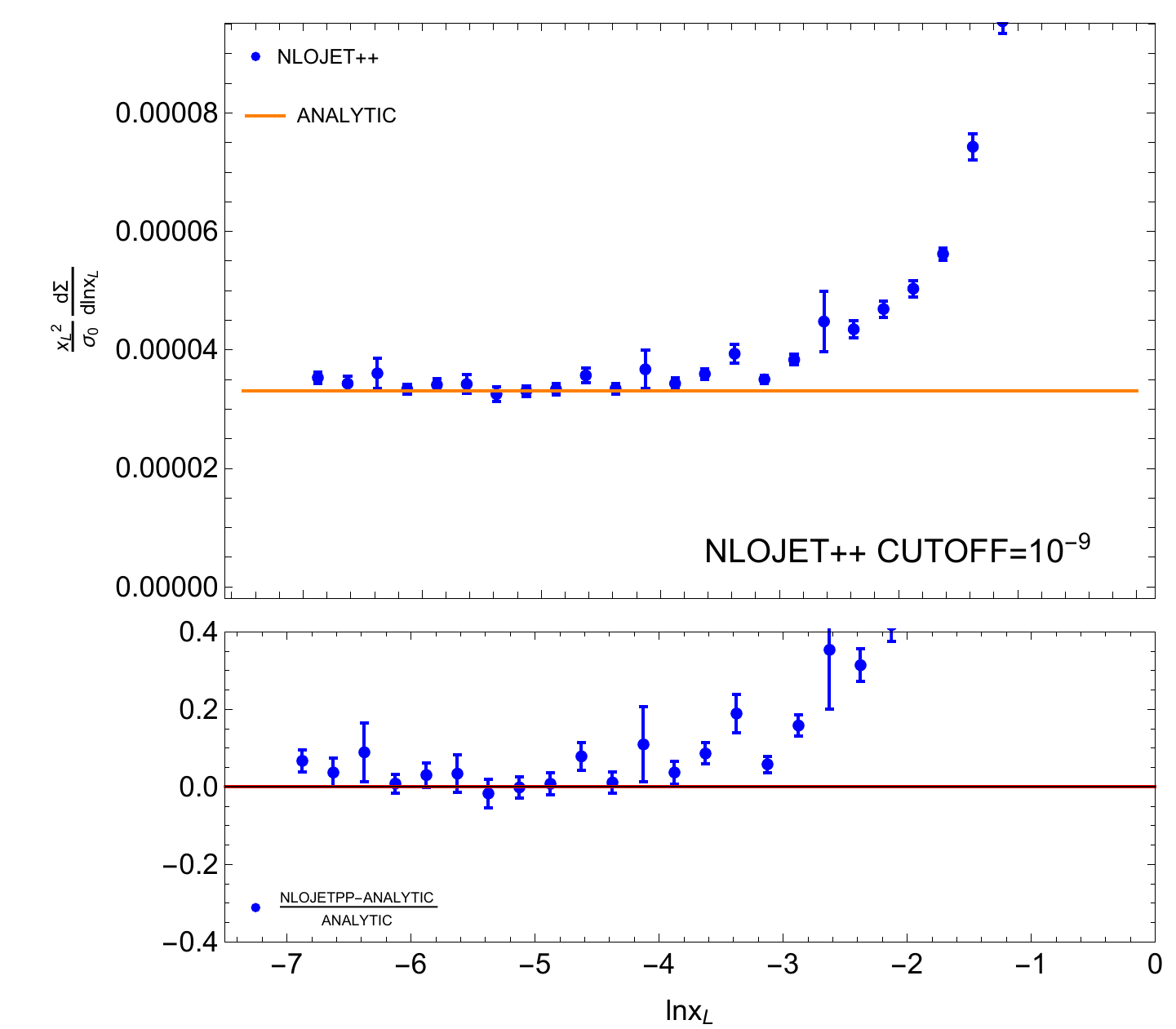}
}\qquad
\subfloat[]{
\includegraphics[scale=0.45]{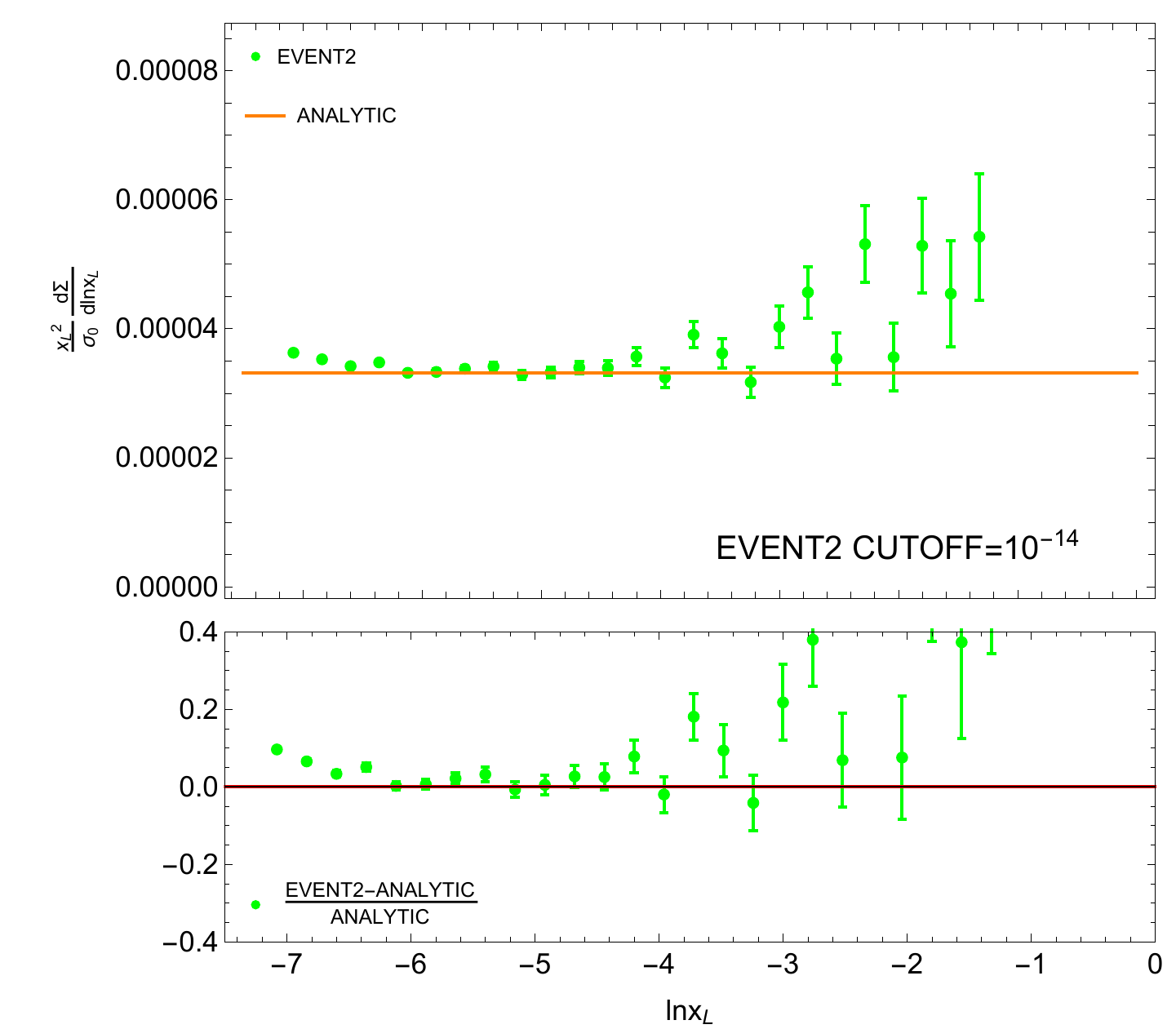}
}\qquad
\end{center}
\caption{Analytic results for the EEEC in an equilateral triangle configuration in the collinear limit, compared with numerical results from  \textsc{NLOJET++} in (a) and \textsc{Event2} in (b), with $Q=m_Z$. Excellent agreement is observed as $x_L\to 0$.}
\label{fig:label1}
\end{figure}

Particularly in QCD, our result is fairly non-trivial, and also relies on collinear factorization for a multi-particle correlation. Since this is beyond what has traditionally been studied, we can use numerical codes for the full $e^+e^- \to 4$ parton matrix elements to check that our analytic results do indeed reproduce the correct singular behavior for the EEEC observable. This check can be performed separately for each partonic channel, which provides a strong cross check on our analytic calculation. In this section we use both  \textsc{Nlojet++} \cite{Nagy:2001fj,Nagy:2003tz} and  \textsc{Event2}~\cite{Catani:1996vz} to check our results. 

We start from numerically verifying the case where the three point span an equilateral triangle, which can be obtained from our analytic calculation as~(see e.g. Eq.~\eqref{eq:Sigma} and \eqref{eq:quark_dis})
\begin{equation}
\begin{aligned}
	\frac{x_L^3}{\sigma_{\rm tot}}\frac{d\Sigma_q}{dx_L}&\,= \int\! dx_1 dx_2\,  \frac{g^4 x_L^3}{32\pi^5 \sqrt{-\Delta}} \frac{ G_q(z) }{x_L^2} \delta(x_1-x_L)\delta(x_2-x_L) \\
	&\,= \frac{i g^4}{32 \pi^5 (z - \bar{z})} G_q(z) \bigg|_{z=\frac{1}{2}+i\frac{\sqrt{3}}{2}}\,.
\end{aligned}
\end{equation}

\begin{figure}
\begin{center}
\subfloat[]{
\includegraphics[scale=0.37]{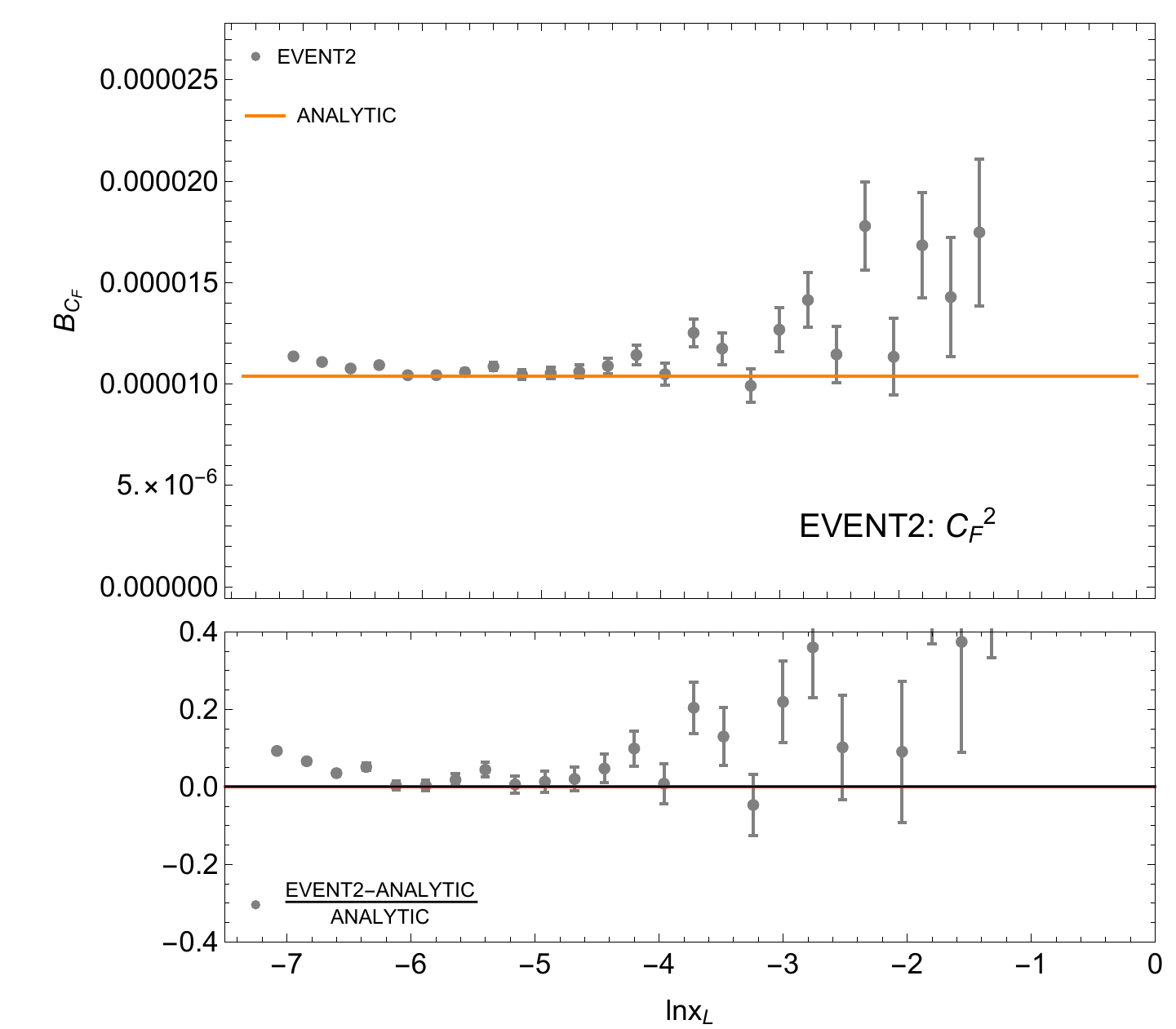}
}
\subfloat[]{\includegraphics[scale=0.37]{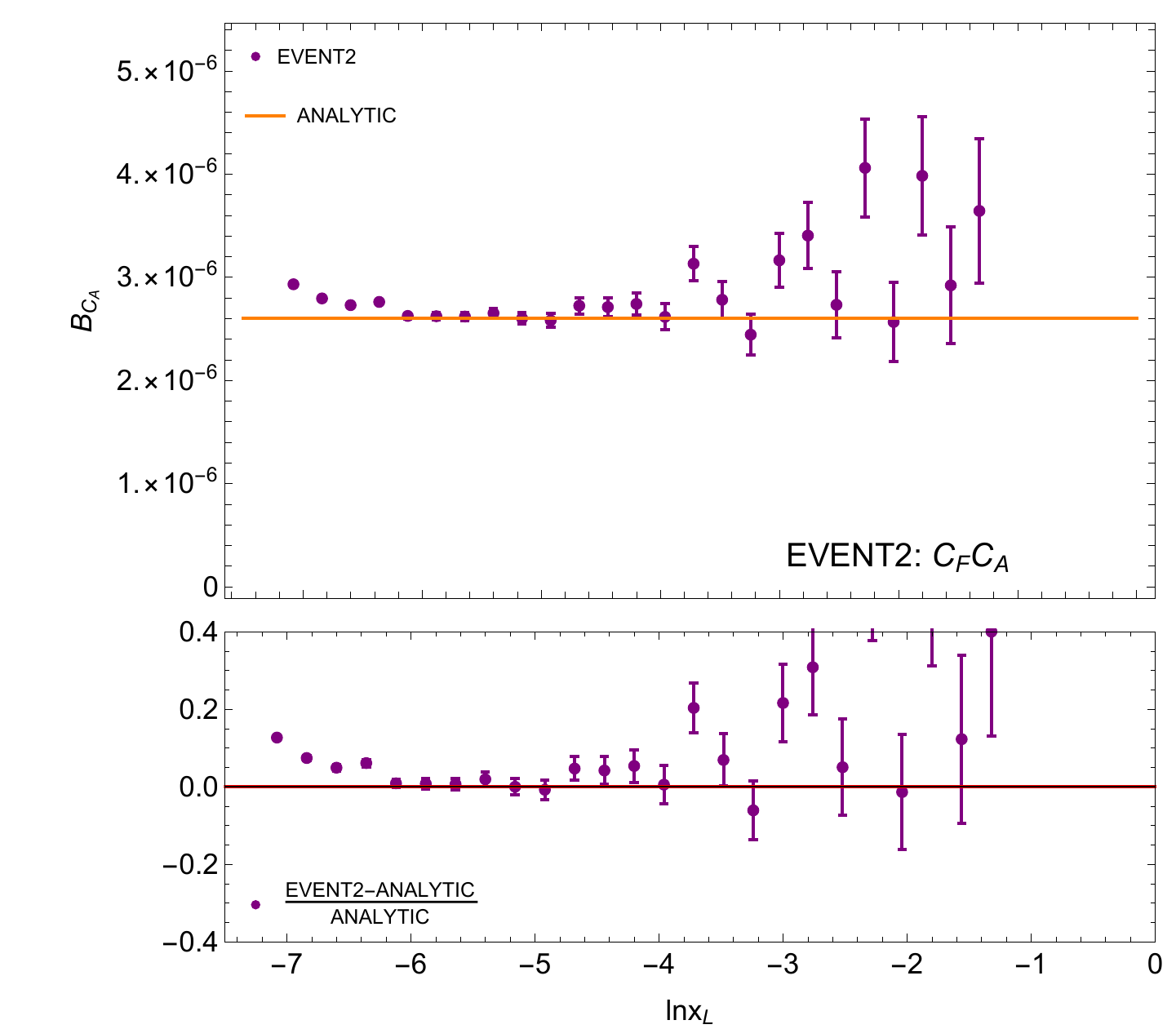}
}
\subfloat[]{\includegraphics[scale=0.37]{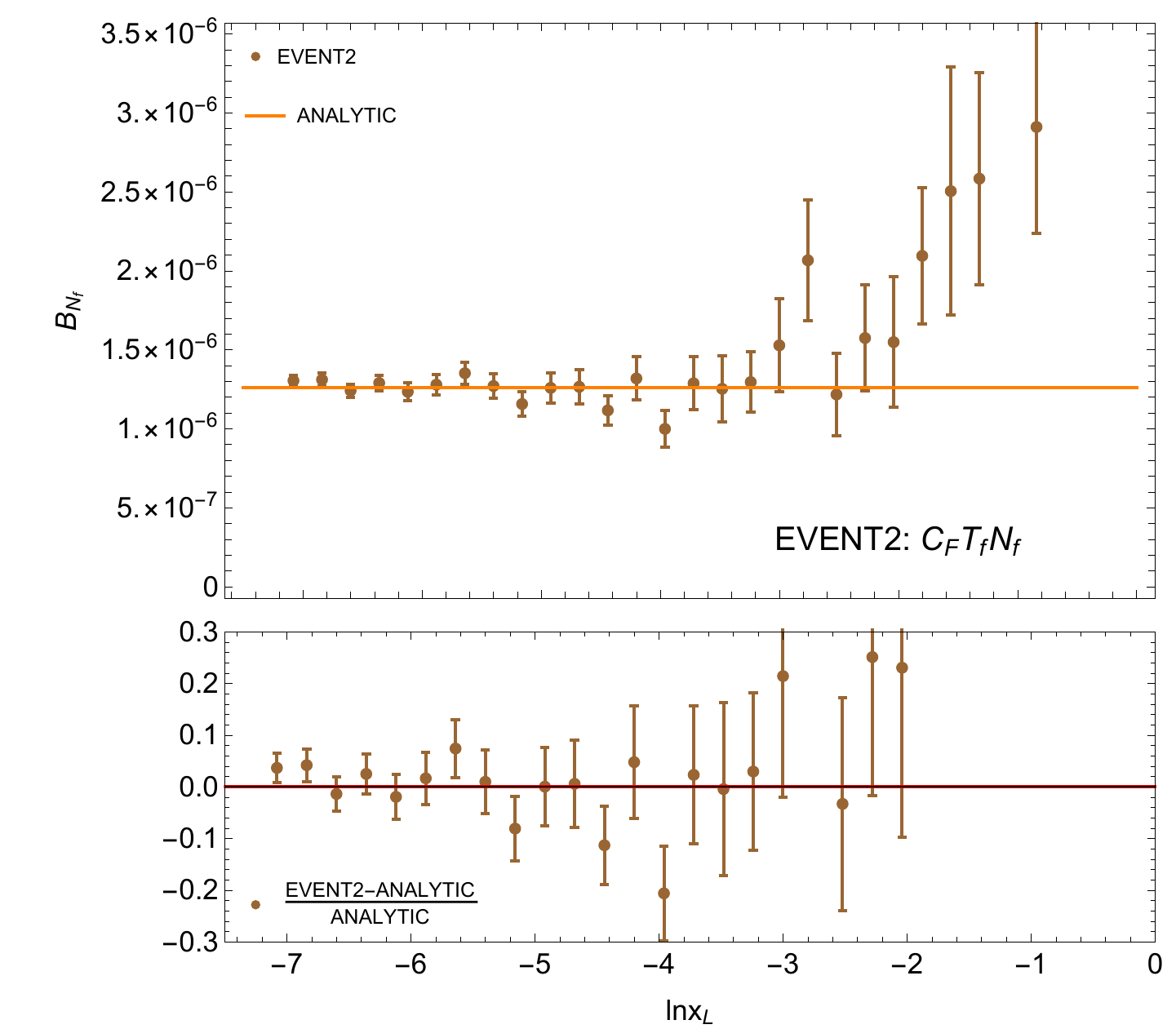}
}
\end{center}
\caption{A comparison of the analytic results for the EEEC in an equilateral triangle configuration with numerical results from \textsc{EVENT2}, separated by color channel. Results are shown for the $C_F^2$, $C_F C_A$ and $C_F N_f T_f$ channels, following the decomposition given in the text. Good agreement is observed for each channel.}
\label{fig:label2}
\end{figure}

We begin by verifying the behavior for the color summed result. In \Fig{fig:label1} we show the analytic result for the EEEC for an equilateral triangle configuration in the collinear limit, as a function of the side length, $x_L$. This is compared with numerical results from both \textsc{Nlojet++} and \textsc{Event2}. In both cases, we see agreement between the numerical result and our analytic calculation as $x_L\to 0$. This shows that our factorization formula correctly reproduces the singular behavior in the collinear limit.

To test this agreement in more detail, we can decompose the cross section by color structure
\begin{equation}
	\frac{x_L^2}{\sigma_{\rm tot}}\frac{d\Sigma_q}{d\ln x_L}=C_F^2 B_{C_F} + C_F C_A B_{C_A}+ C_FN_fT_fB_{N_f}\,,
\end{equation}
and separately compare our analytic predictions with \textsc{EVENT2} for each of the color channels $C_F^2$, $C_F C_A$ and $C_F N_f T_f$. This comparison is shown in \Fig{fig:label2}, where we again see good agreement for all the partonic channels. This provides a strong check on our result. 

\begin{figure}
\begin{center}
\includegraphics[scale=0.6]{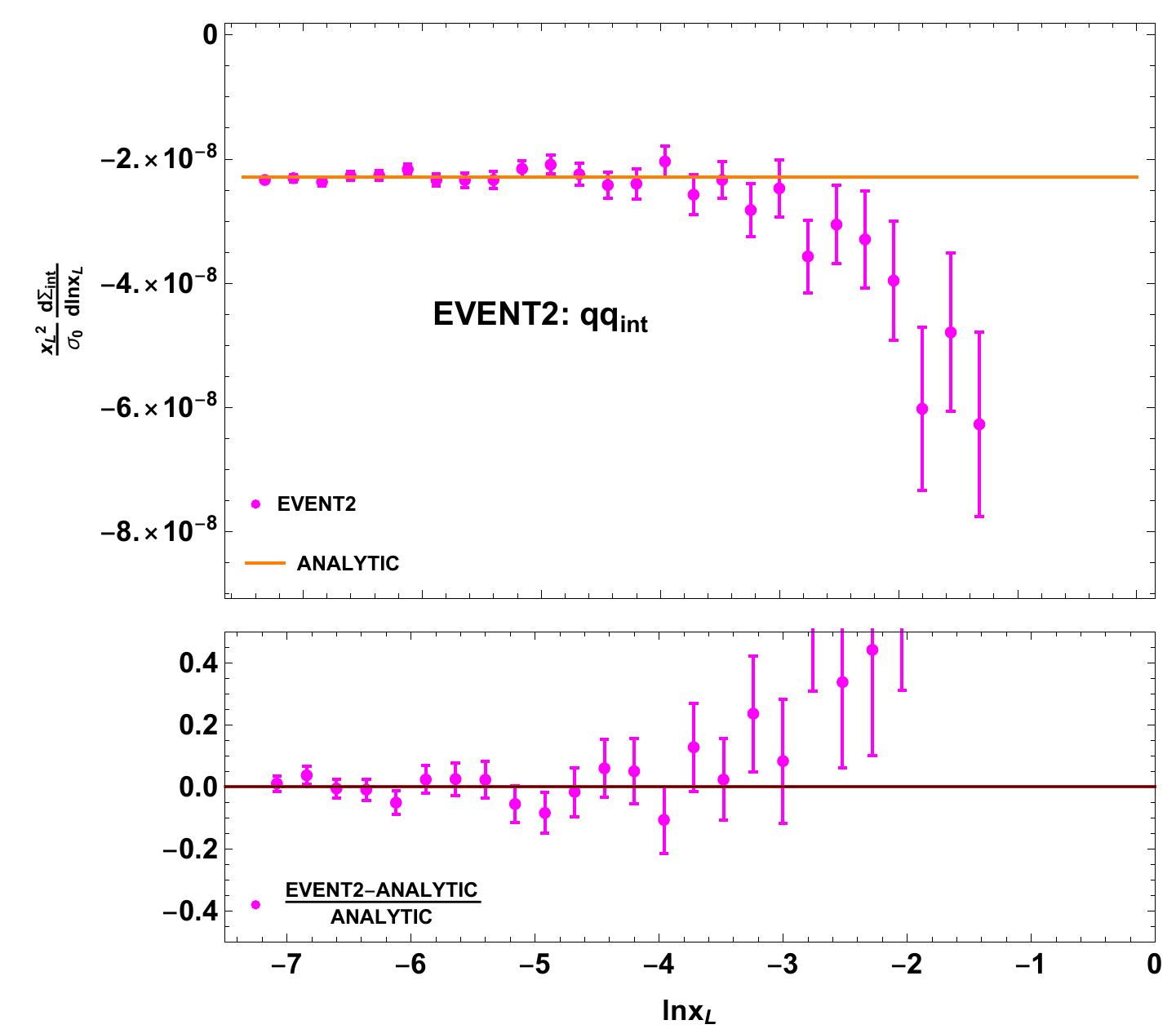}
\end{center}
\caption{The identical quark interference term compared with \textsc{Event2} for an equilateral triangle configuration.}
\label{fig:label3}
\end{figure}

Another gauge invariant subset of contributions that we can test are the identical quark ($qq \bar q \bar q$) interference terms. These terms were studied in detail for the EEC in \cite{Dixon:2018qgp}. They are a subleading color contribution, with color factor $C_F (C_A-2C_F)$. Although these terms are numerically small, they can be isolated in \textsc{Event2} and provide an additional check on our calculation. In \Fig{fig:label3}, we show a comparison of our analytic result with \textsc{Event2} for the identical quark interference term, again finding good agreement as $x_L\to 0$. 

\begin{figure}
\begin{center}
\subfloat[]{
\includegraphics[scale=0.45]{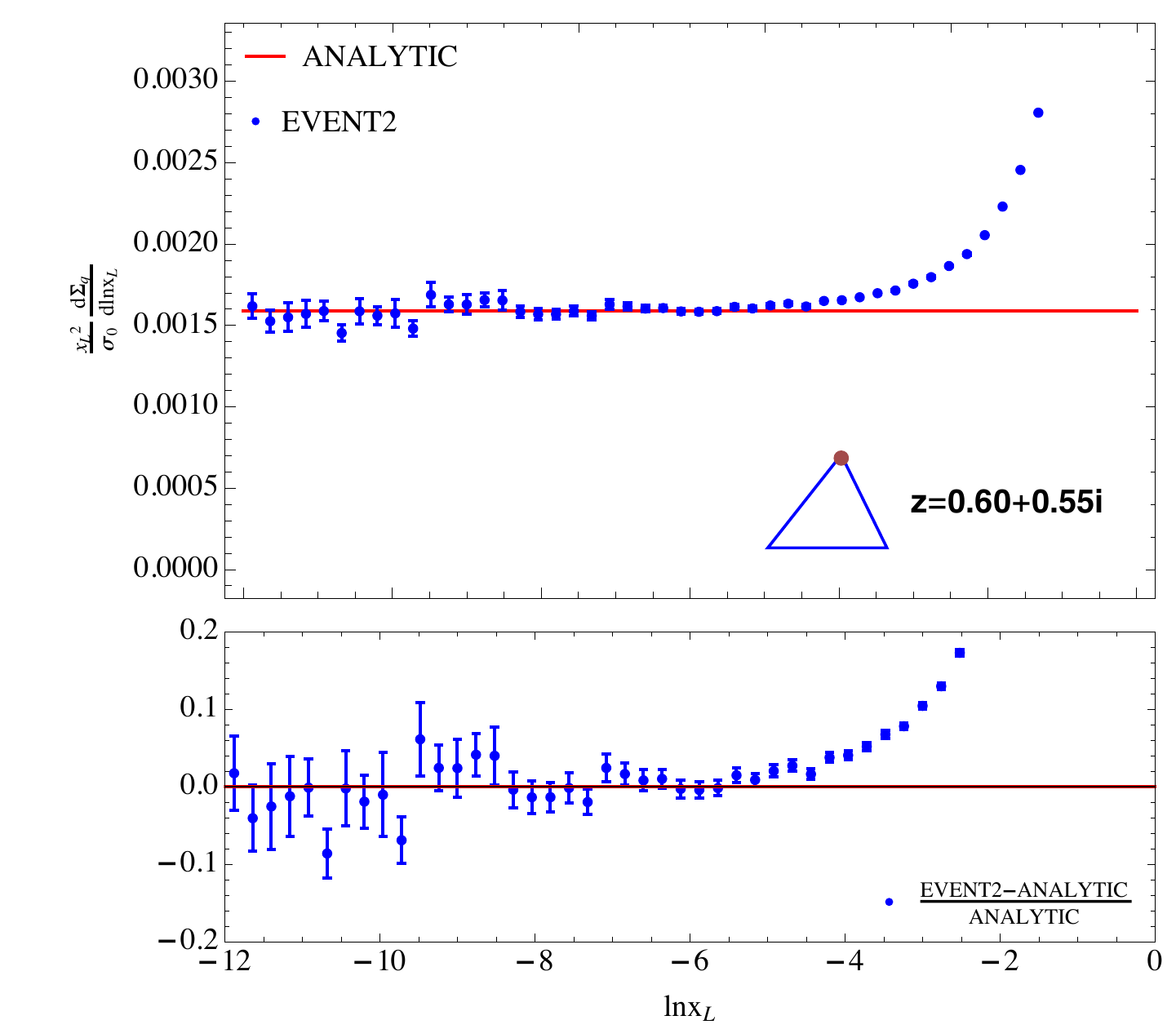}
}\qquad
\subfloat[]{
\includegraphics[scale=0.45]{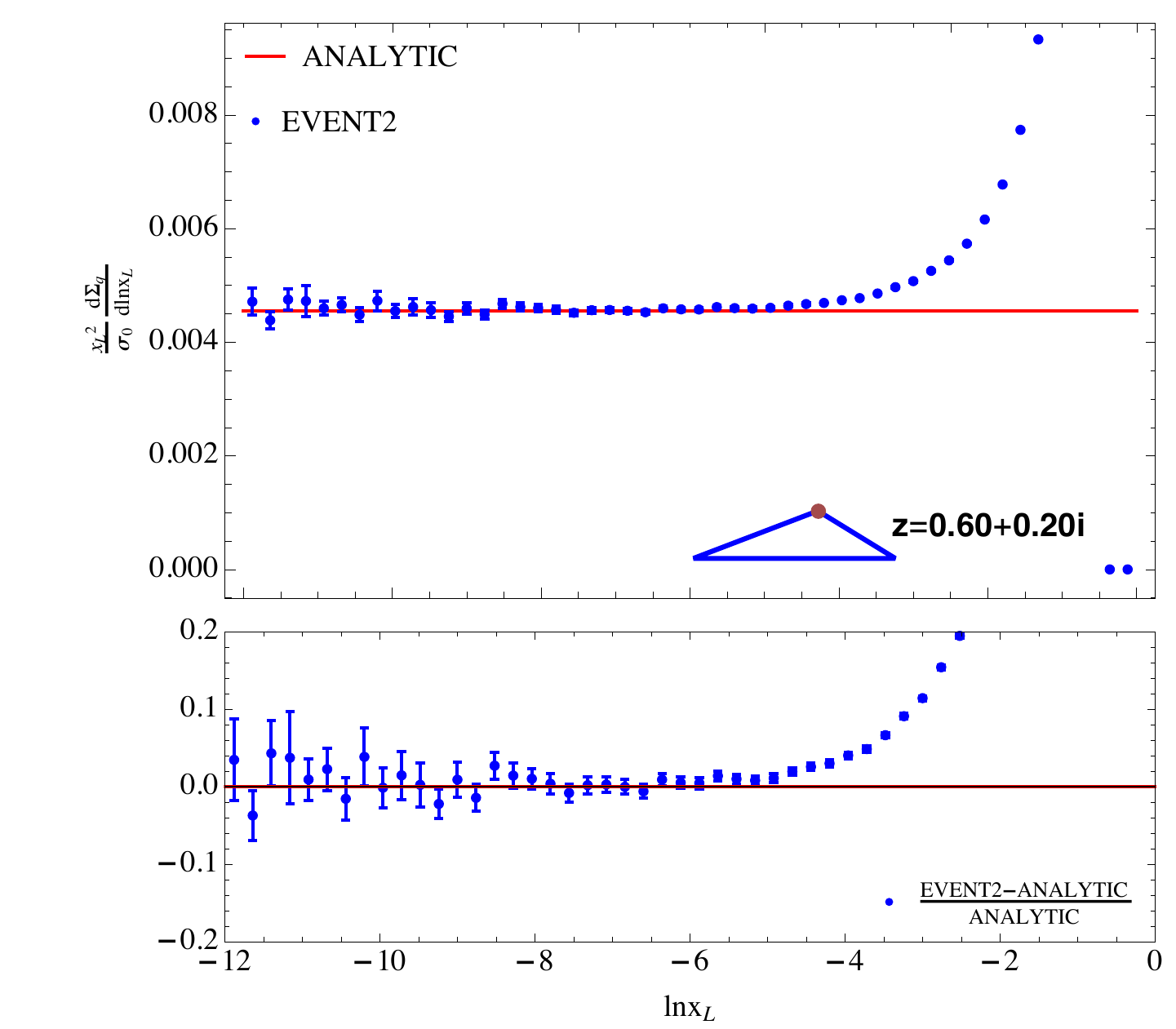}
}\qquad
\subfloat[]{
\includegraphics[scale=0.45]{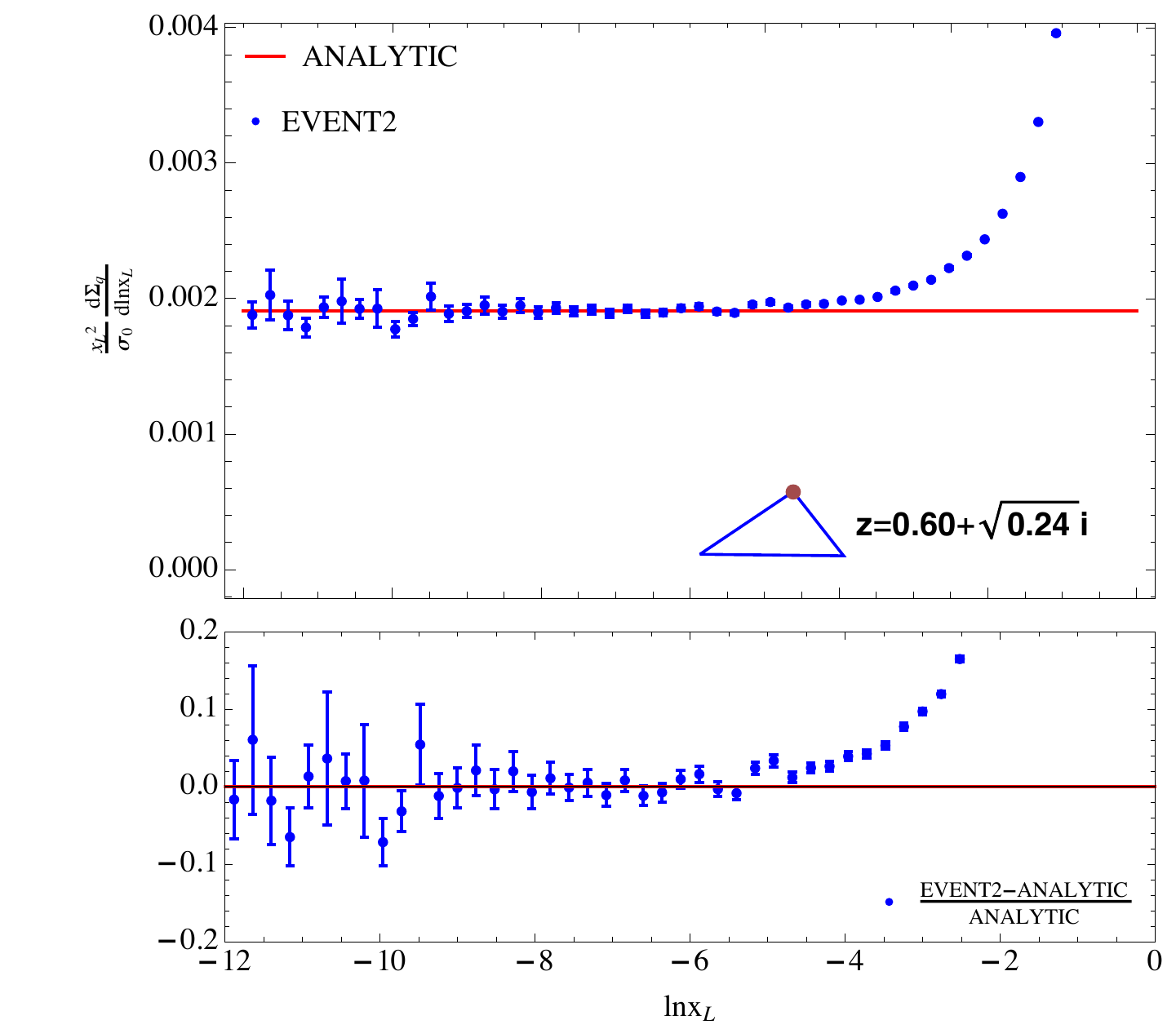}
}\qquad
\end{center}
\caption{Comparison between \textsc{Event2} in the asymptotic regime and our analytic results for different shapes of the triangle span by the three indicated points: (a) acute triangle; (b) obtuse triangle; (c) right triangle.}
\label{fig:triangle_shape}
\end{figure}

To further examine the shape dependence of the EEEC, we show in \Fig{fig:triangle_shape} three separate cases, where the three point span acute, obtuse, and right triangle, respectively. In all cases, we find good agreement between our analytical results and Monte Carlo.

The combination of all these different tests provides strong support for the correctness of our calculation, and for the factorization of the EEEC in the triple collinear limit presented in this paper. It would be interesting to extend our calculation to the next order in $\alpha_s$ to study the radiative corrections to the EEEC, and compare with the predictions of factorization.

\section{Conclusions and Future Directions} \label{sec:conclusions}

In this paper we have presented a calculation of the three point energy correlator in the collinear limit. This extends recent calculations of the two-point correlators, and is particularly interesting, since the three point correlator is the first time where non-trivial shape dependence appears. Our calculation uncovered a number of interesting properties of the energy correlators, which suggest that they can be computed to higher points and at higher loops in both QCD and $\cN=4$. Perhaps most intriguingly, we found a duality between the integrals involved in the calculation of the energy correlator, and Feynman parameter integrals. This allowed the $\cN=4$ result to be expressed in a single line, in terms of triangle and box integrals.

We presented analytic results for $\cN=4$,  and for both quark and gluon jets in QCD, and contrasted the results in the two theories. For the case of QCD quark jets, we compared our results with a  numerical calculation using the full $e^+e^-\to 4$ parton matrix elements, finding agreement for all partonic channels in the collinear limit.

Since this is the first calculation of a multipoint energy correlator, and due to the interesting nature of our results, we believe that it opens up several directions for future study, some of which we describe here:
\vspace{0.25cm}

{\bf{All Orders Resummation of Multi-Point Correlators:}} In this paper we have only considered the fixed order calculation of the non-trivial shape dependence for the triple correlator. At higher perturbative orders, one will encounter logarithms, $\log(x_L)$. As for the case of the two particle correlator, these are single collinear logarithms, and it would be interesting to understand their all orders structure for a generic multi-point correlator. This is interesting theoretically, and is also essential for phenomenological applications at the LHC.

{\bf{Jet Substructure:}} One of the primary goals of jet substructure is the understanding of the correlations of energy flow with a jet \cite{Larkoski:2017jix}. Much of the interesting information is encoded in multi-point correlators, however, these have until now evaded analytic calculation due to their apparent complexity. Here we have shown that these multiparticle correlations can indeed be analytically calculated, and that the results are much simpler than naively expected. It would be interesting to perform a more phenomenological study of these observables as relevant for the LHC. It would also be interesting to understand better their relation to more standard jet substructure observables, such as the energy correlation functions \cite{Larkoski:2013eya,Moult:2016cvt} or energy flow polynomials \cite{Komiske:2017aww}. Indeed, the moments of some of the most well used jet substructure observables, namely $D_2$ \cite{Larkoski:2014gra,Larkoski:2015kga,Larkoski:2017iuy,Larkoski:2017cqq}  and $N_2$ \cite{Moult:2016cvt} are directly related to three point energy correlators. We also believe that the multipoint correlators are interesting as precision probes of QCD at the LHC, and it would be interesting to measure them directly.

{\bf{Higher Point Correlators and Bootstrap:}} The simplicity of our results suggest that, particularly for $\cN=4$, the extension to higher point correlators is feasible. An interesting feature of the higher point energy correlator observables, is that due to their dependence on multiple variables, one can study their behavior in a number of different kinematic limits, much like amplitudes. It would be interesting to understand if one can set up a bootstrap program, similar to that which has been successful for scattering amplitudes \cite{Dixon:2011pw,Dixon:2014iba,Dixon:2015iva,Caron-Huot:2016owq,Dixon:2016nkn,Caron-Huot:2019vjl,Caron-Huot:2019bsq}. This would require a better understanding of the space of functions that describe the energy correlators perturbatively, as well as better constraints on the rational functions appearing in the results.

{\bf{Relation to Feynman Parameter Integrals:}} An interesting observation made in this paper was the relation between the calculation of the energy correlators in the collinear limit, and Feynman parameter integrals for loop amplitudes. It would be interesting to understand if this extends to higher point energy correlators, as well as to higher orders in the perturbative expansion. This relation is interesting in its own right, but is also technically extremely useful due to the wealth of knowledge regarding the structure of loop amplitudes, as well as the well developed technology for performing Feynman parameter integrals. We also believe that this relation is promising for giving a geometric interpretation to the calculation of the energy correlators, given that such an interpretation is known for amplitudes (at least to one loop) \cite{Davydychev:1997wa,Davydychev:1998fk,Gorsky:2009nv,Schnetz:2010pd,Hodges:2010kq,Paulos:2012qa,Nandan:2013ip,Davydychev:2017bbl}.

{\bf{Full Angular Dependence:}} It would be interesting to understand whether the calculation performed here in the collinear limit could be extended to derive an analytic result for the full angular dependence of the EEEC, and to understand the class of functions that appear in the result. For the EEC, there exist powerful sum rules which constrain the endpoint behavior of the result, and it would be interesting to understand the sum rules for the EEEC, and also for higher point correlators.  

{\bf{Effective Theory on the Celestial Sphere:}} More speculatively, it would be interesting to understand if there exists a field theory on the celestial sphere that can be used to compute the energy correlators directly. There has recently been much work in this direction at the amplitude level (see e.g.  \cite{Pasterski:2016qvg,Schreiber:2017jsr}). Much like the Mellin transform used in these studies, the integral over the energy weighting appearing in the definition of the energy correlator observable turns the result into a conformal primary on the sphere. One advantage of the energy correlator observables as compared with the case of amplitudes is that they are necessarily infrared finite at higher orders. By computing higher point functions, one can build up this speculative theory on the celestial sphere, and study in detail the structure of its OPE. The three point function provides the first non-trivial data in this direction.

{\bf{Consequences of Integrability for Cross Section Level Observables:}} There has been significant progress in understanding the consequences of integrability in $\cN=4$ for amplitudes and anomalous dimensions \cite{Beisert:2010jr}. However, there has been relatively little work on understanding the consequences of integrability on infrared safe event (or jet) shape observables, which are the physical observables that are measured in collider experiments. We believe that the energy correlators provide an interesting playground to study this question.

\vspace{0.25cm}
We hope that the simplicity of the results presented here motivate more work towards understanding the structure of the energy correlators, both on the more formal side, as well as in phenomenological applications to jet substructure, and we hope for a fruitful interplay between the two. We believe that there is considerable unexplored simplicity in the substructure of jets that could have a significant impact on analyses at the LHC, and in turn, we hope that the high quality jet data at the LHC can be used to better understand Lorentzian quantum field theories.

\begin{acknowledgments}
	
We thank David Simmons-Duffin, Cyuan Han Chang, Andrew McLeod, Jesse Thaler, Patrick Komiske, and Eric Metodiev for useful discussions and Lance Dixon for useful discussions and collaboration at the initial stages of this work.	
I.M was supported by the Office of High Energy Physics of the U.S. Department of Energy under Contract No. DE-AC02-76SF00515. H.X.Z. was supported in part by NSFC under contract No.~11975200.  

\end{acknowledgments}

\appendix

\section{Triple Collinear Splitting Functions}\label{app:A}

 For completeness, in this Appendix we compile the triple collinear splitting functions used for the calculation of the EEEC.

\subsection{QCD}

The triple collinear splitting amplitudes in QCD were originally computed in \cite{Campbell:1997hg,Catani:1998nv}, and are nicely summarized in \cite{Ritzmann:2014mka}. They are as follows

{\bf{Quark Splitting Functions:}}
\begin{equation}
\begin{aligned}
	P _ { \overline { q }^\prime q ^ { \prime } q } =& C _ { F } T _ { F } \frac { s _ { 123 } } { 2 s _ { 12 } } \left[ - \frac { \left[ \xi_1 \left( s _ { 12 } + 2 s _ { 23 } \right) - \xi_2 \left( s _ { 12 } + 2 s _ { 13 } \right) \right] ^ { 2 } } { \left( \xi_1 + \xi_2 \right) ^ { 2 } s _ { 12 } s _ { 123 } } + \frac { 4 \xi_3 + \left( \xi_1 - \xi_2 \right) ^ { 2 } } { \xi_1 + \xi_2 } + ( 1 - 2 \epsilon ) \left( \xi_1 + \xi_2 - \frac { s _ { 12 } } { s _ { 123 } } \right) \right]\,,\\
	P _ { \overline { q } q q } = & \left( P _ { \overline { q } ^ { \prime } q ^ { \prime } q } + 2 \leftrightarrow 3 \right) + P _ { \overline { q } q q } ^ { ( \mathrm { id } ) }  \,,\\
	P _ { \overline { q } q q } ^ { ( \mathrm { id } ) } =&  C _ { F } \left( C _ { F } - \frac { 1 } { 2 } C _ { A } \right) \left\{ ( 1 - \epsilon ) \left( \frac { 2 s _ { 23 } } { s _ { 12 } } - \epsilon \right) + \frac { s _ { 123 } } { s _ { 12 } } \left[ \frac { 1 + \xi_1 ^ { 2 } } { 1 - \xi_2 } - \frac { 2 \xi_2 } { 1 - \xi_3 } - \epsilon \left( \frac { \left( 1 - \xi_3 \right) ^ { 2 } } { 1 - \xi_2 } + 1 + \xi_1 - \frac { 2 \xi_2 } { 1 - \xi_3 } \right) \right. \right.\\ &\left. - \epsilon ^ { 2 } \left( 1 - \xi_3 \right) ] - \frac { s _ { 123 } ^ { 2 } } { 2 s _ { 12 } s _ { 13 } } \xi_1 \left[ \frac { 1 + \xi_1 ^ { 2 } } { \left( 1 - \xi_2 \right) \left( 1 - \xi_3 \right) } - \epsilon \left( 1 + 2 \frac { 1 - \xi_2 } { 1 - \xi_3 } \right) - \epsilon ^ { 2 } \right] \right\} + ( 2 \leftrightarrow 3 ) \,,\\
	P _ { g g q } =& C _ { F } ^ { 2 } \left\{ \frac { s _ { 123 } ^ { 2 } } { 2 s _ { 13 } s _ { 23 } } \xi_3 \left[ \frac { 1 + \xi_3 ^ { 2 } } { \xi_1 \xi_2 } - \epsilon \frac { \xi_1 ^ { 2 } + \xi_2 ^ { 2 } } { \xi_1 \xi_2 } - \epsilon ( 1 + \epsilon ) \right] + ( 1 - \epsilon ) \left[ \epsilon - ( 1 - \epsilon ) \frac { s _ { 23 } } { s _ { 13 } } \right] \right.\\&\left.+ \frac { s _ { 123 } } { s _ { 13 } } \left[ \frac { \xi_3 \left( 1 - \xi_1 \right) + \left( 1 - \xi_2 \right) ^ { 3 } } { \xi_1 \xi_2 } - \epsilon \left( \xi_1 ^ { 2 } + \xi_1 \xi_2 + \xi_2 ^ { 2 } \right) \frac { 1 - \xi_2 } { \xi_1 \xi_2 } + \epsilon ^ { 2 } \left( 1 + \xi_3 \right) \right] \right\}\\& + C _ { F } C _ { A } \left\{ ( 1 - \epsilon ) \left( \frac { \left[ \xi_1 \left( s _ { 12 } + 2 s _ { 23 } \right) - \xi_2 \left( s _ { 12 } + 2 s _ { 13 } \right) \right] ^ { 2 } } { 4 \left( \xi_1 + \xi_2 \right) ^ { 2 } s _ { 12 } ^ { 2 } } + \frac { 1 } { 4 } - \frac { \epsilon } { 2 } \right) \right.\\& + \frac { s _ { 123 } ^ { 2 } } { 2 s _ { 12 } s _ { 13 } } \left[ \frac { 2 \xi_3 + ( 1 - \epsilon ) \left( 1 - \xi_3 \right) ^ { 2 } } { \xi_2 } + \frac { 2 \left( 1 - \xi_2 \right) + ( 1 - \epsilon ) \xi_2 ^ { 2 } } { 1 - \xi_3 } \right]\\& - \frac { s _ { 123 } ^ { 2 } } { 4 s _ { 13 } s _ { 23 } } \xi_3 \left[ \frac { 2 \xi_3 + ( 1 - \epsilon ) \left( 1 - \xi_3 \right) ^ { 2 } } { \xi_1 \xi_2 } + \epsilon ( 1 - \epsilon ) \right]\\& + \frac { s _ { 123 } } { 2 s _ { 12 } } \left[ ( 1 - \epsilon ) \frac { \xi_1 \left( 2 - 2 \xi_1 + \xi_1 ^ { 2 } \right) - \xi_2 \left( 6 - 6 \xi_2 + \xi_2 ^ { 2 } \right) } { \xi_2 \left( 1 - \xi_3 \right) } + 2 \epsilon \frac { \xi_3 \left( \xi_1 - 2 \xi_2 \right) - \xi_2 } { \xi_2 \left( 1 - \xi_3 \right) } \right]\\&+ \frac { s _ { 123 } } { 2 s _ { 13 } } \left[ ( 1 - \epsilon ) \frac { \left( 1 - \xi_2 \right) ^ { 3 } + \xi_3 ^ { 2 } - \xi_2 } { \xi_2 \left( 1 - \xi_3 \right) } - \epsilon \left( \frac { 2 \left( 1 - \xi_2 \right) \left( \xi_2 - \xi_3 \right) } { \xi_2 \left( 1 - \xi_3 \right) } - \xi_1 + \xi_2 \right) \right.\\&\left.- \frac { \xi_3 \left( 1 - \xi_1 \right) + \left( 1 - \xi_2 \right) ^ { 3 } } { \xi_1 \xi_2 } + \epsilon \left( 1 - \xi_2 \right) \left.\left( \frac { \xi_1 ^ { 2 } + \xi_2 ^ { 2 } } { \xi_1 \xi_2 } - \epsilon \right) \right] \right\} + ( 1 \leftrightarrow 2 )\,.
\end{aligned}
\end{equation}

{\bf{Gluon Splitting Functions:}}

\begin{equation}
\begin{aligned}
	P _ { g _ { 1 } q _ { 2 } \overline { q } _ { 3 } } =& C _ { F } T _ { F } P _ { g _ { 1 } q _ { 2 } \overline { q } _ { 3 } } ^ { ( a b ) } + C _ { A } T _ { F } P _ { g _ { 1 } q _ { 2 } \overline { q } _ { 3 } } ^ { ( n a b ) }\,, \\
	P _ { g _ { 1 } q _ { 2 } \overline { q } _ { 3 } } ^ { ( a b ) }  =& - 2 - ( 1 - \epsilon ) s _ { 23 } \left( \frac { 1 } { s _ { 12 } } + \frac { 1 } { s _ { 13 } } \right) + 2 \frac { s _ { 123 } ^ { 2 } } { s _ { 12 } s _ { 13 } } \left( 1 + \xi_1 ^ { 2 } - \frac { \xi_1 + 2 \xi_2 \xi_3 } { 1 - \epsilon } \right) \\ & - \frac { s _ { 123 } } { s _ { 12 } } \left( 1 + 2 \xi_1 + \epsilon - 2 \frac { \xi_1 + \xi_2 } { 1 - \epsilon } \right) - \frac { s _ { 123 } } { s _ { 13 } } \left( 1 + 2 \xi_1 + \epsilon - 2 \frac { \xi_1 + \xi_3 } { 1 - \epsilon } \right) \,, \\
	P _ { g _ { 1 } q _ { 2 } \overline { q } _ { 3 } } ^ { ( n a b ) } =& \left\{ - \frac { t _ { 23,1 } ^ { 2 } } { 4 s _ { 23 } ^ { 2 } } + \frac { s _ { 123 } ^ { 2 } } { 2 s _ { 13 } s _ { 23 } } \xi_3 \left[ \frac { \left( 1 - \xi_1 \right) ^ { 3 } - \xi_1 ^ { 3 } } { \xi_1 \left( 1 - \xi_1 \right) } - \frac { 2 \xi_3 \left( 1 - \xi_3 - 2 \xi_1 \xi_2 \right) } { ( 1 - \epsilon ) \xi_1 \left( 1 - \xi_1 \right) } \right] \right.\\ & + \frac { s _ { 123 } } { 2 s _ { 13 } } \left( 1 - \xi_2 \right) \left[ 1 + \frac { 1 } { \xi_1 \left( 1 - \xi_1 \right) } - \frac { 2 \xi_2 \left( 1 - \xi_2 \right) } { ( 1 - \epsilon ) \xi_1 \left( 1 - \xi_1 \right) } \right] \\ & + \frac { s _ { 123 } } { 2 s _ { 23 } } \left[ \frac { 1 + \xi_1 ^ { 3 } } { \xi_1 \left( 1 - \xi_1 \right) } + \frac { \xi_1 \left( \xi_3 - \xi_2 \right) ^ { 2 } - 2 \xi_2 \xi_3 \left( 1 + \xi_1 \right) } { ( 1 - \epsilon ) \xi_1 \left( 1 - \xi_1 \right) } \right] \\ & - \frac { 1 } { 4 } + \frac { \epsilon } { 2 } - \frac { s _ { 123 } ^ { 2 } } { 2 s _ { 12 } s _ { 13 } } \left( 1 + \xi_1 ^ { 2 } - \frac { \xi_1 + 2 \xi_2 \xi_3 } { 1 - \epsilon } \right) \} + ( 2 \leftrightarrow 3 )\,,\\
	P _ { g _ { 1 } g _ { 2 } g _ { 3 } } =& C _ { A } ^ { 2 } \left\{ \frac { 1 - \epsilon } { 4 s _ { 12 } ^ { 2 } } t _ { 12,3 } ^ { 2 } + \frac { 3 } { 4 } ( 1 - \epsilon ) + \frac { s _ { 123 } } { s _ { 12 } } \left[ 4 \frac { \xi_1 \xi_2 - 1 } { 1 - \xi_3 } + \frac { \xi_1 \xi_2 - 2 } { \xi_3 } + \frac { 3 } { 2 } + \frac { 5 } { 2 } \xi_3 \right. \right.\\ & + \frac { \left( 1 - \xi_3 \left( 1 - \xi_3 \right) \right) ^ { 2 } } { \xi_3 \xi_1 \left( 1 - \xi_1 \right) } ] + \frac { s _ { 123 } ^ { 2 } } { s _ { 12 } s _ { 13 } } \left[ \frac { \xi_1 \xi_2 \left( 1 - \xi_2 \right) \left( 1 - 2 \xi_3 \right) } { \xi_3 \left( 1 - \xi_3 \right) } + \xi_2 \xi_3 - 2 + \frac { \xi_1 \left( 1 + 2 \xi_1 \right) } { 2 } \right.\\ & + \frac { 1 + 2 \xi_1 \left( 1 + \xi_1 \right) } { 2 \left( 1 - \xi_2 \right) \left( 1 - \xi_3 \right) } + \frac { 1 - 2 \xi_1 \left( 1 - \xi_1 \right) } { 2 \xi_2 \xi_3 } ] \} + ( 5 \text { permutations } )\,,\\
	t _ { i j , k } =& 2 \frac { \xi_{ i } s _ { j k } - \xi _ { j } s _ { i k } } { \xi _ { i } + \xi _ { j } } + \frac { \xi _ { i } - \xi _ { j } } { \xi _ { i } + \xi _ { j } } s _ { i j }\,.
\end{aligned}
\end{equation}

\subsection{$\cN=4$ SYM}

We extract the universal $1 \to 3$ splitting function in $\cN=4$ SYM by taking the three-particle collinear limit of the squared amplitudes for $1 \to 4$ decay for a source operator $\mathrm{tr}F^2$. We find a particularly simple answer after summing over all flavors,
\begin{align}
P_{\cN=4} =&\ \frac{8}{\xi_3 s_{123} s_{12}} \left(\frac{1}{\xi_1} + \frac{1}{1-\xi_1}\right) + \frac{4}{s_{13} s_{23}} \left( \frac{1}{(1- \xi_1) (1-\xi_2)} + \frac{1}{\xi_1 \xi_2} \right) + (\text{5 permutations}) \,.
\end{align}
It would be interesting to better understand the simplicity of this result.

\subsection{Results for Feynman Integrals}
\label{sec:feyn_int}

The one-loop integrals in Eq.~\eqref{eq:feynman_rep} can be written explicitly as
\begin{equation}
\begin{aligned}
\mathcal{J}^{(d=8)}(2,2,1)&=\frac{1}{x_L}\bigg\{ \frac{|z|^2-|1-z|^2-1}{2(z-\bar z)^2}-\frac{\log(|z|^2)}{2 (z-\bar z)^4}\bigg(1-2|z|^2+|z|^4-|1-z|^2+8|z|^2|1-z|^2\\
&+|z|^4|1-z|^2-|1-z|^4-2|z|^2|1-z|^4+|1-z|^6\bigg)+\frac{|1-z|^2\log(|1-z|^2)}{2 (z- \bar z)^4} \bigg(4|z|^2\\
&+|z|^4+4|1-z|^2-2|z|^2|1-z|^2+|1-z|^4-5\bigg)+\frac{4i |1-z|^2 D_2^{-}(z)}{ (z-\bar z)^5}\bigg(-1-|z|^2\\
&+2|z|^4+2|1-z|^2-|z|^2|1-z|^2-|1-z|^4\bigg)\bigg\}\,,
\end{aligned}
\end{equation}
\begin{equation}
\begin{aligned}
\mathcal{J}^{(d=10)}(2,2,2,\widetilde{1})&=\frac{1}{x_L}\bigg\{ \frac{2-4|z|^2+2|z|^4-3|1-z|^2-3|z|^2|1-z|^2+|1-z|^4}{2|1-z|^2(z-\bar z)^2}\\
&-\frac{|z|^2 \log(|z|^2)}{2|1-z|^2(z-\bar z)^4}\bigg(2-6|z|^2+6|z|^4-2|z|^6-7|1-z|^2+7|z|^4|1-z|^2+2|1-z|^4\\
&-8|z|^2|1-z|^4+3|1-z|^6\bigg)+\frac{\log(|1-z|^2)}{2|1-z|^2(z-\bar z)^4}\bigg(-2+8|z|^2-12|z|^4+8|z|^6-2|z|^8\\
&+7|1-z|^2-7|z|^2|1-z|^2-7|z|^4|1-z|^2+7|z|^6|1-z|^2-8|1-z|^4+4|z|^2|1-z|^4\\
&-8|z|^4|1-z|^4+3|1-z|^6+3|z|^2|1-z|^6\bigg)+\frac{(|z|^2-1)D_2^{+}(z)}{|1-z|^4}\\
&+\frac{2iD_2^{-}(z)}{|1-z|^4(z-\bar z)^5}\bigg( 1-6|z|^2+15|z|^4-20|z|^6+15|z|^8-6|z|^{10}+|z|^{12}-5|1-z|^2\\
&+15|z|^2|1-z|^2-10|z|^4|1-z|^2-10|z|^6|1-z|^2+15|z|^8|1-z|^2-5|z|^{10}|1-z|^2\\
&+10|1-z|^4-10|z|^2|1-z|^4-10|z|^6|1-z|^4+10|z|^8|1-z|^4-10|1-z|^6\\
&+4|z|^2|1-z|^6+4|z|^4|1-z|^6-10|z|^6|1-z|^6+5|1-z|^8-2|z|^2|1-z|^8\\
&+5|z|^4|1-z|^8-|1-z|^{10}-|z|^2|1-z|^{10} \bigg)\bigg\} \,.
\end{aligned}
\end{equation}

\bibliography{EEC_forward}{}
\bibliographystyle{JHEP}

\end{document}